\documentclass[aps,prb,twocolumn,floatfix,nofootinbib,superscriptaddress,longbibliography,nobibnotes]{revtex4-2}
\usepackage[utf8]{inputenc}
\usepackage{bm}
\usepackage{physics}
\usepackage{xcolor}
\usepackage[final,colorlinks,bookmarks=true,citecolor=red,linkcolor=red,urlcolor=blue]{hyperref}
\usepackage{mathbbol}
\usepackage{amsmath}
\usepackage{amssymb}
\usepackage{amsthm}
\usepackage{mathbbol}
\usepackage{multirow}
\usepackage{graphicx}
\usepackage{enumitem}
\usepackage{makecell}
\usepackage{longtable}
\usepackage{float}

\newcommand{\cyan}[1]{{\color{cyan}{#1}}}
\newcommand{\magenta}[1]{{\color{magenta}{#1}}}

\immediate\write18{latexdiff cFSPT_v1/main_v1.tex main_v2.tex > output_diff.tex}

\begin{document}
\renewcommand{\arraystretch}{1.2}
\title{Interacting Crystalline Topological Insulators in two-dimensions with Time-Reversal Symmetry}

\author{Martina O. Soldini}
\affiliation{Department of Physics, University of Zurich, Winterthurerstrasse 190, CH-8057 Z\"{u}rich, Switzerland}

\author{\"{O}mer M. Aksoy}
\affiliation{Department of Physics, Massachusetts Institute of Technology, 
Cambridge, Massachusetts 02139, USA}

\author{Titus Neupert}
\affiliation{Department of Physics, University of Zurich, 
Winterthurerstrasse 190, CH-8057 Z\"{u}rich, Switzerland}

\begin{abstract}
Topology is routinely used to understand the physics of electronic insulators. However, for strongly interacting electronic matter, such as Mott insulators, a comprehensive topological characterization is still lacking.
When their ground state only contains short range entanglement and does not break symmetries spontaneously, they generically realize crystalline fermionic symmetry protected topological phases (\mbox{cFSPTs}), supporting gapless modes at the boundaries or at the lattice defects.
Here, we provide an exhaustive classification of \mbox{cFSPTs} in two-dimensions with $\mathrm{U}(1)$ charge-conservation and spinful time-reversal symmetries, namely those generically present in spin-orbit coupled insulators, for any of the 17 wallpaper groups. 
It has been shown that the classification of \mbox{cFSPTs} can be understood from appropriate real-space decorations of lower-dimensional subspaces, and 
we expose how these relate to the Wyckoff positions of the lattice. We find that all nontrivial one-dimensional decorations require electronic interactions. 
Furthermore, we provide model Hamiltonians for various decorations, and discuss the signatures of \mbox{cFSPTs}. 
This classification paves the way to further explore topological interacting insulators, providing the backbone information in generic model systems and ultimately in experiments.
\end{abstract}

\maketitle

\section{Introduction}
A rich set of phases can be found in electronic crystalline materials when interactions are preeminent.
Among these possibilities, a prime example is that of the fermionic symmetry protected topological (\mbox{FSPT}) phases of matter.
An \mbox{FSPT} phase is characterized by the existence of a bulk gap accompanied by robust gapless boundary states, which can only be 
gapped when the protecting symmetries are broken. 
Two distinct \mbox{FSPT} phases cannot be connected by continuously
changing the parameters of the Hamiltonian, unless a 
gap closing or symmetry breaking transition occurs.
As opposed to intrinsically topologically ordered systems, ground states
of \mbox{FSPT}s contain only short range entanglement.

A classification of \emph{noninteracting} \mbox{FSPT} phases, such as band insulators and superconductors, 
has been achieved in the 10-fold way~\cite{Schnyder2009,Kitaev2009,Ryu2010}, where 10 symmetry classes are defined depending on whether or not the single-particle
Hamiltonians are symmetric under charge-conjugation symmetry, time-reversal symmetry (TRS), 
or a combination 
of the two. Over the years, weakly interacting electronic materials showcasing some of 
the boundary modes predicted by the 10-fold way have been discovered: 
the paradigmatic examples include~\footnote{
The integer quantum Hall (IQH) states~\cite{Klitzing1980},
while being the first realizations of fermionic topological phases of matter,
are not protected by any symmetries owing to the fact that their gapless chiral 
boundary modes cannot be gapped even when all the symmetries are broken.
IQH states are examples of \textit{invertible} topological phases that are not SPT phases.
These states carry long-range entanglement but do not support any nontrivial anyonic excitations~\cite{Chen2010}.
The low-energy properties of invertible topological phases (SPT or not)
are described by the so-called invertible topological field
theories~\cite{Freed2021}.} the quantum spin Hall effect~\cite{Bernevig2006,Konig2007,Roth2009}, 
and the observation of gapless surface states in three-dimensional 
topological insulators~\cite{hsieh2009,Xia2009}.

In the 10-fold way, the gapless boundary modes
are protected by internal symmetries alone. However,
the underlying lattice structure of crystalline matter naturally provides more 
symmetries such as translations, rotations, spatial inversion, or mirror operations. 
These symmetries enrich the known classification of \mbox{FSPT} phases by
providing protection to distinct gapless boundary modes, which become gappable if the symmetry is broken. 
The corresponding phases are called \emph{crystalline} \mbox{FSPT} (\mbox{cFSPT}) phases.
Early examples of noninteracting \mbox{cFSPT} have been obtained by extending 
the 10-fold way with crystalline 
symmetries~\cite{Fu2007,Fu2011,Slager2013,Zhang2013,Chiu2013,Morimoto2013,
Fang2013,Jadaun2013,Koshino2014, Liu2014,Shiozaki2014,Ando2015,Trifunovic2017},
which led to the discovery 
of novel phases such as higher order topological phases~\cite{Benalcazar2017,Schindler2018,Trifunovic2019,Fang2019,Song2017hoti}. A milestone in this direction was the development of the comprehensive frameworks of symmetry indicators~\cite{Po2017,Po2020} and topological quantum chemistry~\cite{Bradlyn2017,Kruthoff2017, Vergniory2019,Cano2021,Elcoro2021}, which allowed to characterize entire material databases. 
Since then, some of these phases have been observed in insulating 
electronic materials~\cite{Schindler2018Bismuth,Hsu2019,Yoon2020,Aggarwal2021,Shumiya2022} and insulating or superconducting metamaterials~\cite{Serra-Garcia2018,Soldini2023Shiba,wang2024}.

In the last decade, it has been realized that interaction effects
can drastically change the classification of \mbox{FSPT}s, as well as cFSPTs. 
Following the seminal works of Fidkowski and Kitaev~\cite{Fidkowski2010,Fidkowski2011},
several works have shown that the classification of \mbox{FSPT}s (and \mbox{cFSPTs}) can be reduced 
due to the presence of interactions~\cite{Ryu2012,Qi2013,Fidkowski2013,Yao2013,Metlitski2014,
Wang2014,You2014,Morimoto2015,Yoshida2015,SongXY2017,Aksoy2021a}. 
In other words, nontrivial \mbox{cFSPT} phases may be connected to the trivial one 
by interaction terms. 
Parallel to these efforts, bosonic and fermionic SPT phases have been classified
using the tools of algebraic topology such as 
cohomology and cobordism groups~\cite{Chen2011a,Chen2011b,Chen2011c,Chen2013,Chen2014,
Kapustin2014a,Kapustin2014c,Gu2014,Kapustin2015,Wang2020,Freed2021,Barkeshli2022}. 
In addition to reducing the noninteracting classification, 
interactions may also enhance the \mbox{FSPT} classification leading to 
\textit{intrinsically interacting} phases that do not admit
noninteracting counterparts~\cite{Gu2014,Wang2017,Cheng2018,Tantivasadakarn2018,
Sullivan2020,Freed2021,Cheng2022}.

A breakthrough in classifying interacting \mbox{cFSPTs} has been achieved 
when a correspondence between topological phases with crystalline symmetries
and internal symmetries was uncovered in Refs.~\onlinecite{Thorngren2018,Else2019a} 
in the form of the so-called \textit{crystalline equivalence principle}
(see also Refs.~\onlinecite{Freed2019,Debray2021}). 
This correspondence can be understood from decorating points, lines,
and planes in space with \mbox{FSPT}s through the so-called \textit{real-space construction}~\cite{Song2017,Huang2017,Zhang2020,
Zhang2022a,Zhang2022b,Rasmussen2018,Rasmussen2020,manjunath2023}.

Despite the widespread presence of interactions in real electronic materials, the experimental 
realization and observation of interacting \mbox{cFSPT} phases have so far remained elusive, 
while some of them have been realized in cold atom experiments~\cite{Potirniche2017,Sompet2022}.
A particularly relevant case is that of generic interacting electronic insulating materials
with spin-orbit coupling and TRS. These properties encompass a range of materials, 
such as nonmagnetic Mott insulators. 
In the language of the 10-fold way, such electronic insulators belong to the class AII, for which the relevant symmetries are
$\mathrm{U}(1)$ charge-conservation and spinful TRS. 

The classification of crystalline bosonic symmetry protected topological (SPT) states, 
as well as \mbox{cFSPTs} without U(1) conservation, with U(1) but without TRS, 
or with a subset of all possible spatial symmetries has been reported~\cite{Song2017,Huang2017,Zhang2020,Zhang2022a,Zhang2022b,kobayashi2024}.
However, a complete classification of interacting \mbox{cFSPT} phases with the symmetries of class AII and all space groups is so-far missing.
In this work, we take a step forward in filling this gap
by classifying all \mbox{cFSPT} phases in two-dimensional (2D)
space in class AII, enriched by any of the 17 wallpaper groups.
Using the real-space construction scheme described 
in Refs.\ \onlinecite{Song2017,Huang2017,Zhang2020,Zhang2022a,Zhang2022b}, 
for each wallpaper group we tabulate the total classification and its decomposition 
in terms of classification of lower-dimensional \mbox{FSPT}s. 
Incidentally, we realize that Wyckoff positions~\cite{Hahn2005} of a given wallpaper group
provide a natural guideline for possible decorations, 
therefore bridging a well-known notion in the context of crystalline lattices 
with the concept of `unit-cell subdivision' adopted in Refs.\ 
\onlinecite{Song2017,Huang2017,Zhang2020,Zhang2022a,Zhang2022b}. 
Interestingly, we find that all appropriate and nontrivial
one-dimensional (1D) decorations at mirror symmetric lines correspond
to intrinsically interacting \mbox{cFSPT} phases protected by the mirror symmetry. 
We provide toy-model Hamiltonians for both zero-dimensional (0D) and 1D
decorations.

The rest of the paper is organized as follows.
In Sec.~\ref{sec:Construction}, we review the classification scheme we adopted, and we
explicitly study the case of the
honeycomb lattice, i.e., wallpaper group No.\ 17 (\textit{p6mm})~\cite{Hahn2005}. 
Therein, the classification for all wallpaper groups, our main result, is summarized in 
Table~\ref{tab:wallpaper classification summary}.
Section \ref{sec:Hamiltonian realizations} presents toy model Hamiltonians 
that can be used to decorate the points and lines of the 2D lattice. 
Section \ref{sec:Signatures of cFSPT phases} discusses various signatures of 
\mbox{cFSPT} phases. Finally in Sec.~\ref{sec:conclusion} we draw the conclusion and provide an outlook.
Appendixes \ref{App:p6mm mirror equivalence} and \ref{App:All wallapers}
provide details on the classification of lower-dimensional decorations.

\section{Classification by real-space construction}\label{sec:Construction}

Following Refs.~\onlinecite{Song2017,Huang2017,Zhang2020,Zhang2022a,Zhang2022b},
we are going to construct 2D \mbox{cFSPT} phases by decorating high-symmetry points, lines,
and planes in space with appropriate \mbox{FSPT} phases, which are protected by internal symmetries only. 
The key idea behind this construction is that crystalline symmetries act as internal 
symmetries on those lower-dimensional subspaces that are invariant under their action.

In Sec.~\ref{sec:realspace construction}, we give an overview of 
the real-space construction of \mbox{cFSPT} phases in 2D. 
In Sec.~\ref{sec:block classification}, we specialize
to the case of class AII and discuss which internal symmetries 
can arise at lower-dimensional subspaces. 
Therein, we provide a classification of lower-dimensional
\mbox{FSPT} phases protected by these symmetries, and we define topological 
indices that distinguish possible decorations. 
In Sec.~\ref{sec:example 17}, we apply our scheme to classify \mbox{cFSPT} phases 
in class AII on a hexagonal lattice which has the wallpaper group \textit{p6mm}.

Our main result is summarized by Table~\ref{tab:wallpaper classification summary}, 
which lists the \mbox{cFSPT} classification in class AII for each of the 17 wallpaper groups. 
Many of the phases listed in the classification are intrinsically interacting (\cyan{cyan} entries), meaning that they can not be realized in the absence of interactions.

\begin{table}[t]
\centering
\begin{tabular}{@{\extracolsep{12 pt}}ccc}
\hline \hline
$\#$ & Wallpaper group & Classification\\
\hline
1 & $p1$ 
& $\mathbb{Z} \times \magenta{\mathbb{Z}^{\,}_2}$
\\
2 & $p2$ 
& $\mathbb{Z} \times (\mathbb{Z}^{\,}_2)^4 \times \cyan{(\mathbb{Z}^{\,}_2)^4} \times \magenta{\mathbb{Z}^{\,}_2}$
\\
3 & $p1m1$ 
& $\mathbb{Z} \times (\mathbb{Z}^{\,}_2)^2 \times \cyan{(\mathbb{Z}^{\,}_2)^4} \times \magenta{\mathbb{Z}^{\,}_2}$
\\
4 & $p1g1$ 
& $\mathbb{Z} \times \magenta{\mathbb{Z}^{\,}_2}$
\\
5 & $c1m1$ 
& $\mathbb{Z} \times \mathbb{Z}^{\,}_2 \times \cyan{(\mathbb{Z}^{\,}_2)^2} \times \magenta{\mathbb{Z}^{\,}_2}$
\\
6 &  $p2mm$ 
& $\mathbb{Z} \times (\mathbb{Z}^{\,}_2)^4 \times \cyan{(\mathbb{Z}^{\,}_2)^{12}} \times \magenta{\mathbb{Z}^{\,}_2}$
\\
7 & $p2mg$ 
& $\mathbb{Z} \times (\mathbb{Z}^{\,}_2)^3 \times \cyan{(\mathbb{Z}^{\,}_2)^4} \times \magenta{\mathbb{Z}^{\,}_2}$
\\
8 & $p2gg$ 
& $\mathbb{Z} \times (\mathbb{Z}^{\,}_2)^2 \times \cyan{(\mathbb{Z}^{\,}_2)^2} \times \magenta{\mathbb{Z}^{\,}_2}$
\\
9 & $c2mm$ 
& $\mathbb{Z} \times (\mathbb{Z}^{\,}_2)^3 \times \cyan{(\mathbb{Z}^{\,}_2)^7} \times \magenta{\mathbb{Z}^{\,}_2}$
\\
10 & $p4$ 
& $\mathbb{Z} \times (\mathbb{Z}^{\,}_4)^2 \times\mathbb{Z}^{\,}_2 \times \cyan{(\mathbb{Z}^{\,}_2)^3} \times \magenta{\mathbb{Z}^{\,}_2}$
\\
11 & $p4mm$
& $\mathbb{Z} \times (\mathbb{Z}^{\,}_4)^2 \times \mathbb{Z}^{\,}_2 \times \cyan{(\mathbb{Z}^{\,}_2)^{9}} \times \magenta{\mathbb{Z}^{\,}_2}$
\\ 
12 & $p4gm$ 
& $\mathbb{Z} \times (\mathbb{Z}^{\,}_4)^2 \times \mathbb{Z}^{\,}_2 \times \cyan{(\mathbb{Z}^{\,}_2)^{4}} \times \magenta{\mathbb{Z}^{\,}_2}$
\\
13 & $p3$ 
& $\mathbb{Z} \times (\mathbb{Z}^{\,}_3)^3 \times \magenta{\mathbb{Z}^{\,}_2}$
\\
14 & $p3m1$ 
& $\mathbb{Z} \times (\mathbb{Z}^{\,}_3)^3 \times \cyan{(\mathbb{Z}^{\,}_2)^2} \times \magenta{\mathbb{Z}^{\,}_2}$
\\
15 & $p31m$ 
& $\mathbb{Z} \times (\mathbb{Z}^{\,}_3)^2 \times \cyan{(\mathbb{Z}^{\,}_2)^2} \times \magenta{\mathbb{Z}^{\,}_2}$
\\
16 & $p6$ 
& $\mathbb{Z} \times \mathbb{Z}^{\,}_6 \times \mathbb{Z}^{\,}_3 \times \mathbb{Z}^{\,}_2 \times \cyan{(\mathbb{Z}^{\,}_2)^2} \times \magenta{\mathbb{Z}^{\,}_2}$
\\
17 & $p6mm$ 
& $\mathbb{Z} \times \mathbb{Z}^{\,}_6 \times \mathbb{Z}^{\,}_3 \times \mathbb{Z}^{\,}_2 \times \cyan{(\mathbb{Z}^{\,}_2)^6} \times \magenta{\mathbb{Z}^{\,}_2}$
\\ \hline \hline
\end{tabular}
\caption{Classification of class AII \mbox{cFSPTs} for each of the 17 wallpaper groups. Each wallpaper group classification contains a $\mathbb{Z}$ factor corresponding to the total number of Kramers' pairs in the unit-cell, and a $\mathbb{Z}^{\,}_2$ factor for the intrinsic 2D SPT phase (the 2D ``topological insulator''). The latter is colored in \magenta{magenta} to differentiate this phase which does not require crystalline symmetries nor interactions to be realized. The interaction enabled entries are highlighted in \cyan{cyan}, while the remaining entries are left in black.
}
\label{tab:wallpaper classification summary}
\end{table}

\subsection{Overview of real-space construction}
\label{sec:realspace construction}

For a given wallpaper group, we construct \mbox{cFSPTs} by following the construction based on ``decorating'' subspaces of the 2D space, 
which has been applied for classifying bosonic SPTs protected by point group symmetry, 
and cFSPTs in classes A and D of the 10-fold way
enriched by various crystalline symmetries~\cite{Song2017, Huang2017,Song2019,Zhang2020,Zhang2022a,Zhang2022b, kobayashi2024}.
In the following, we summarize the steps for the \mbox{cFSPT} construction through decoration, adapted to our case of class AII.

\paragraph{Unit-cell subdivision.} 
First, we divide the 2D space into unit-cells,
which is allowed by the translation symmetries of the underlying lattice structure. 
We then subdivide each unit-cell into $d$-dimensional subspaces, where $d=0,1,2$, which are invariant under 
some, or none, for the case of $d=2$, of the crystalline symmetries. 
Our construction differs from the previous work in Refs.~\onlinecite{Song2017, Huang2017,Song2019,Zhang2020,Zhang2022a,Zhang2022b} 
by how we subdivide the unit-cells. We obtain the relevant subspaces by 
considering the coordinates of the Wyckoff positions of the lattice (see Sec.~\ref{sec: decoration Wyckoff positions}, for a review of 
Wyckoff positions and their relevant properties).
If a given Wyckoff position has $d$ free parameters in its coordinate definition, we consider $d$-dimensional finite subspaces in the unit-cell spanned by varying the free parameters of the coordinates. 
We refer to such subspaces as $d$-dimensional \emph{Wyckoff patches}, 
where $d$ is equal to the number of free parameters 
in the definition of the Wyckoff position's coordinates.
This
defines our $d$-dimensional subspaces, to be decorated by \mbox{FSPT} phases.

\paragraph{Decorating Wyckoff patches by FSPTs.} 
A given $d$-dimensional Wyckoff patch may be left invariant under some of the crystalline symmetries such as rotations or reflections, 
which act internally within the Wyckoff patch. 
For each such $d$-dimensional Wyckoff patch, we enumerate all the possible $d'$-dimensional \mbox{FSPT}s ($d'\leq d$) that 
can be ``glued'' to that patch, which we call \emph{decorations}.
The possible decorations are then $d'$-dimensional \mbox{FSPT} phases protected by the internal symmetries of class AII, together with the crystalline symmetries that leave
the corresponding Wyckoff patch invariant. 
We enumerate distinct decorations by assigning a 
topological index to each decoration. These indicate how the 
protecting symmetries are represented at a given Wyckoff patch.
In Sec.~\ref{sec:block classification}, we detail the classification of possible $d'$-dimensional \mbox{FSPT} decorations. We remark that decorating a Wyckoff position with an \mbox{FSPT} does not imply that a physical degree of freedom (atom) needs to be placed there [see e.g., Fig.~\ref{fig:measuring cFSPTs}(c)]. 

\paragraph{Choosing gappable decorations.} 
Since \mbox{cFSPT} phases are gapped in the bulk by definition, 
we must consider only decorations that can be gapped out in the bulk 
while preserving the crystalline symmetries. 
We call such decorations \emph{gappable}, and otherwise they are \emph{non-gappable}. 
By definition, all 0D decorations are already gapped. 
Therefore, the only potentially non-gappable decorations are either 1D or 2D. 
Let us first consider the case of 1D decorations. If there exist any nontrivial 1D \mbox{FSPT} protected by the symmetries of a specific line in the 2D bulk, we can glue such 1D phases along such a line. Each glued 1D \mbox{FSPT} supports 0D gapless boundary modes. To ensure that this construction results into a gappable phase, multiple of these 1D \mbox{FSPT}s must meet at their endpoints in an appropriate number and arrangement, such that their gapless boundary modes can be gapped out by interactions that are compatible with all the protecting symmetries.
Likewise, for nontrivial 2D decorations 
we have to ensure that gapless 1D boundary modes are gapped out in the bulk
without breaking any crystalline symmetry.

\paragraph{Equivalences.} The last step is to identify equivalences between 
distinct decorations of lower-dimensional \mbox{FSPT} phases.
Two decorations are equivalent if they can be connected 
by symmetric and gap preserving local processes.
As we shall demonstrate in Sec.~\ref{sec:example 17} for wallpaper group
\textit{p6mm}, these equivalences
take the form of moving 0D \mbox{FSPT} decorations 
from one Wyckoff position to another.
The \mbox{cFSPT} classification is then given by enumerating gappable decorations modulo 
equivalences under adiabatic deformations. 

In practice, the classification is a collection of topological indices 
labeling all the inequivalent topological phases realized by various decorations. 
At the same time, this approach gives a prescription to construct 
many-body wave functions that realize the phase corresponding 
to a chosen set of topological indices as product states of decorations.

\subsection{Classification of possible decorations}
\label{sec:block classification}
The symmetry class AII refers to the \mbox{FSPT} phases with
charge-conservation and spinful TRS symmetries. 
More precisely, we denote the symmetry group by
\begin{align}
\label{eq:G AII}
G^{\mathrm{AII}}_{f}
:=
\frac{
\mathrm{U}(1)^{\mathrm{F}}\rtimes\mathbb{Z}^{\mathrm{FT}}_{4} 
}
{\mathbb{Z}^{\mathrm{F}}_{2}},
\end{align}
which comprises two subgroups. First, the subgroup $\mathrm{U}(1)^{\mathrm{F}}$
corresponds to the charge-conservation symmetry. The superscript $\mathrm{F}$
is to indicate that this subgroup contains the fermion parity symmetry $\mathbb{Z}^{\mathrm{F}}_{2}$.~\footnote{
If we parameterize the elements of $\mathrm{U}(1)^{\mathrm{F}}$ by $e^{\mathrm{i}\theta}$
with $\theta\in[0,2\pi)$, then fermion parity symmetry is generated by the element $\theta=\pi$.
Fermion parity is a symmetry of any Hamiltonian built out of fermions and cannot be broken explicitly
or spontaneously.
} Second, the subgroup $\mathbb{Z}^{\mathrm{FT}}_{4}$ corresponds to the 
spinful TRS. Here too the superscript indicates that this subgroup 
contains fermion parity symmetry, since for spinful fermions TRS squares to fermion parity. 
To avoid double counting the fermion parity symmetry, we coset by the subgroup 
$\mathbb{Z}^{\mathrm{F}}_{2}$.

As described in Sec.~\ref{sec:realspace construction}, we should classify
\mbox{FSPT} phases in $d'=0,\,1,\,2$ dimensions that are protected by the symmetries of 
class AII together with any crystalline symmetries that act internally on subspaces of the 2D space. In Table \ref{tab:Cohomology blocks}, we summarize the classification of all possible
\mbox{FSPT} decorations in ($d'=0,1,2$)-dimensional subspaces. 
In what follows, we explain how this table is obtained by 
enumerating possible enhancements of the symmetry group 
$G^{\mathrm{AII}}_{f}$ by crystalline symmetries, and we classify the corresponding \mbox{FSPT} phases.

\def\arraystretch{1.5}
\begin{table}[t]
\centering
\begin{tabular}{@{\extracolsep{10pt}} c c c}
\hline
\hline
$d'$ & CS & cFSPT Classification
\\
\hline
0D
& 
$\mathbb{Z}^{\mathrm{F}}_{2a}\rtimes\mathbb{Z}^{\mathrm{F}}_{2b}/\mathbb{Z}^{\mathrm{F}}_{2}$
& $\mathbb{Z}\times \cyan{\mathbb{Z}^{\,}_{\mathrm{gcd}(2,a)}}\times\cyan{\mathbb{Z}^{\,}_{b}}$ 
\\
\hline
1D
& $\mathbb{Z}^{\mathrm{F}}_{2b}$
& $\cyan{\mathbb{Z}^{\,}_{b}}$
\\
\hline
2D 
& $\mathbb{Z}^{\,}_{1}$ 
& $\magenta{\mathbb{Z}^{\,}_2}$
\\ \hline \hline
\end{tabular}
\caption{Classification of FSPTs in $d'=0,\, 1, \, 2$ with full symmetry group $G^{(d')}_{f, \text{tot}}$, obtained by appropriately combining the group in Eq.~\eqref{eq:G AII} with the group of crystalline symmetries acting internally, which is listed under the column CS. 
The total symmetry group $G^{(d')}_{f, \text{tot}}$ obtained this way is explicitly shown in Eqs.~\eqref{eq:Gftot 0D}, \eqref{eq:Gftot 1D}, and \eqref{eq:Gftot 2D} for $d'=0,\,1,\,2$, respectively. 
The third column, labeled \mbox{cFSPT} Classification, contains the resulting classification as obtained from cohomology groups. 
The color code follows the one of Tab.~\ref{tab:wallpaper classification summary}.}
\label{tab:Cohomology blocks}
\end{table}
\def\arraystretch{1}

\paragraph{Classification of 0D decorations.}
A subspace in $d'=0$ dimension can be kept invariant under the action of
point group symmetries. In 2D, there are ten possible double point groups, all of the form $(\mathbb{Z}^{\mathrm{F}}_{2a}\rtimes\mathbb{Z}^{\mathrm{F}}_{2b})/\mathbb{Z}^{\mathrm{F}}_{2}$,
where $a=1,2,3,4,6$ denotes the $a$-fold rotation while $b=1,2$ denotes the absence ($b=1$)
or presence ($b=2$) of a mirror symmetry.
Since we consider spinful fermions, both $2\pi$ rotations and 
applying mirror transformation twice must be equal to fermion parity symmetry. 
In other words, 
at a 0D subspace the total symmetry group 
$G^{(0)}_{f,\mathrm{tot}}$ is
\begin{align}\label{eq:Gftot 0D}
G^{(0)}_{f,\mathrm{tot}}
:=
G^{\mathrm{AII}}_{f}\times 
\frac{\mathbb{Z}^{\mathrm{F}}_{2a}\rtimes\mathbb{Z}^{\mathrm{F}}_{2b}}
{\mathbb{Z}^{\mathrm{F}}_{2}\times\mathbb{Z}^{\mathrm{F}}_{2}},
\end{align}
for $a=1,2,3,4,6$ and $b=1,2$. In 0D, the \mbox{FSPT} phases protected 
by $G^{(0)}_{f,\mathrm{tot}}$ are classified by the first cohomology group~\cite{Chen2013,Gu2014}
\begin{align}
H^{1}
\left(
G^{(0)}_{f,\mathrm{tot}},\,\mathrm{U}(1)
\right)
=
\mathbb{Z}
\times
\mathbb{Z}^{\,}_{\mathrm{gcd}(2,a)}
\times
\mathbb{Z}^{\,}_{b},
\label{eq:classification of 0D decorations}
\end{align}
where $\mathrm{gcd}(2,a)$ is the greatest common divisor of integers $a$ and $2$.
The first cohomology group $H^{1}$ classifies the 1D 
representations of the total symmetry group \eqref{eq:Gftot 0D}, i.e, possible 
charges of a 0D quantum state $\ket{\Psi}$ that are compatible with the group composition rule. 
Such a state $\ket{\Psi}$ is precisely the 0D decoration one can attach to a 
high-symmetry point that is invariant under the $(\mathbb{Z}^{\mathrm{F}}_{2a}\rtimes\mathbb{Z}^{\mathrm{F}}_{2b})/\mathbb{Z}^{\mathrm{F}}_{2}$
double point group symmetry. 

The $\mathbb{Z}$ term in Eq.~\eqref{eq:classification of 0D decorations}
is the classification of 0D \mbox{FSPT} phases protected only by charge-conservation and spinful TRS. 
Physically, it corresponds to the total number of Kramers' pairs localized at the high-symmetry point,
which we denote by the index $n\in\mathbb{Z}$.
The remaining two terms $\mathbb{Z}^{\,}_{\mathrm{gcd}(2,a)}\times\mathbb{Z}^{\,}_{b}$ give the classification of
0D FSPT phases protected by the point group symmetries. Because of TRS, any 0D decoration
can only carry real charges, i.\,e., $\pm1$, under rotation and mirror symmetries.
Whenever $\text{gcd}(2,a)=2$ or $b=2$, 
we define two $\mathbb{Z}^{\,}_{2}$-valued indices 
$r\in \mathbb{Z}^{\,}_{2}$ and $m\in\mathbb{Z}^{\,}_{2}$, 
which label
the rotation and mirror charges, respectively, of a 0D decoration.  
Crucially, these $\mathbb{Z}^{\,}_2$ indices are absent in noninteracting electronic systems in AII, 
where wave functions can only transform trivially under the action of rotations or mirror transformations~\cite{Yao2010}.
As we shall discuss in Sec.~\ref{sec:Hamiltonian realizations}, 
all 0D decorations with nontrivial indices $r$ or $m$
are intrinsically interacting.

\paragraph{Classification of 1D decorations.}
For 1D decorations in 2D space,
only mirror transformation can act internally. 
Therefore, the total symmetry group becomes
\begin{align}\label{eq:Gftot 1D}
G^{(1)}_{f,\mathrm{tot}}
:=
G^{\mathrm{AII}}_{f}\times 
\frac{\mathbb{Z}^{\mathrm{F}}_{2b}}
{\mathbb{Z}^{\mathrm{F}}_{2}},
\end{align}
where $b=1$ or $b=2$ denotes the absence or presence of mirror symmetry, respectively.
In 1D, the \mbox{FSPT} phases protected by $G^{(1)}_{f,\mathrm{tot}}$
are classified by the second cohomology group~\cite{Turzillo2019,Bourne2021,Aksoy2022}
\begin{align}
H^{2}
\left(
G^{(1)}_{f,\mathrm{tot}},\,\mathrm{U}(1)
\right)
=
\mathbb{Z}^{\,}_{b}.
\label{eq:classification of 1D decorations}
\end{align}
It classifies the projective representations of the 
total symmetry group \eqref{eq:Gftot 1D}. 
There are no nontrivial \mbox{FSPT} phases protected only by charge-conservation and spinful
TRS symmetries. Therefore, the classification \eqref{eq:classification of 1D decorations}
corresponds to the 1D \mbox{FSPT} decorations that are stabilized by the 
presence of the mirror symmetry. Physically, a nontrivial 1D decoration features
gapless Kramers' pairs at its 0D boundaries, protected by the total symmetry group in Eq.~\eqref{eq:Gftot 1D}.
The $\mathbb{Z}^{\,}_{b}$ group in 
Eq.~\eqref{eq:classification of 1D decorations} then 
corresponds to the parity of the number 
of Kramers' pairs at the boundary, which we denote by the index $\nu\in\mathbb{Z}^{\,}_{2}$. 
We dub the phase realized by nontrivial 1D decorations the 
\emph{AIIM phase}, since it is protected by the symmetries of class 
AII and mirror symmetry, and we refer to $\nu$ as the AIIM index.
In Sec.~\ref{sec:Hamiltonian realizations}, we demonstrate that the AIIM phase is intrinsically interacting
and we introduce a model Hamiltonian realizing it.

\paragraph{Classification of 2D decorations.}
There are no crystalline symmetries in 2D space that leave a 
2D decoration invariant. 
Therefore, the total symmetry group for ($d'=2$)-dimensional decorations is simply
\begin{align}
\label{eq:Gftot 2D}
G^{(2)}_{f,\mathrm{tot}}
:=
G^{\mathrm{AII}}_{f}.
\end{align}
In 2D, \mbox{FSPT} phases protected by $G^{\mathrm{AII}}_{f}$ symmetry 
are $\mathbb{Z}^{\,}_{2}$ classified~\cite{Kapustin2015,Freed2021,Barkeshli2022}.
The nontrivial 2D \mbox{FSPT} phase in class AII features gapless helical boundary states, and it is
such that odd number of boundary states cannot be gapped unless TRS is broken. 
We define the index $x\in\mathbb{Z}^{\,}_{2}$ that labels if a 2D decoration is trivial or not,
and physically corresponds to the parity of the 
number of helical edge states.

\subsection{Decorating Wyckoff patches}\label{sec: decoration Wyckoff positions}

\begin{figure}[t]
\centering
\includegraphics{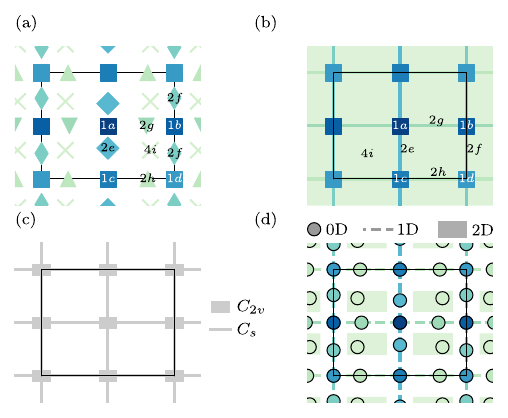}
\caption{(a) Wyckoff positions of the wallpaper group $p2mm$ (No.~6) with their labels and (c) illustration of the nontrivial site-symmetry groups of the lattice [either $C_{2v}$ or $C_s$ (marked in gray)]. (b) Unit cell subdivision based on Wyckoff patches: 0D (circles), 1D (lines), and 2D (rectangles) patches are colored and labeled according to their corresponding Wyckoff position, following the color code of (a). (d) Example of 0D (dots), 1D (dashed lines), or 2D (rectangular regions) \mbox{FSPTs} decorations of the $p2mm$ lattice. Decorations are colored according to the Wyckoff position to which they belong, following the color code in (a). In all panels, a unit-cell is marked by black lines.}
\label{fig:wyckoff_positions_6}
\end{figure}

We start this section by briefly recalling some useful concepts in the description of crystalline lattices.
Every point in the lattice is characterized by its site-symmetry group, namely the finite subgroup of the space group transformations that leave the point invariant.
A Wyckoff position is a set of points in the lattice whose site-symmetry groups are conjugate subgroups of the space group, i.e., 
the two site-symmetry groups are related by an element of the space group. Wyckoff positions are labeled by their multiplicity followed by a letter, and they are tabulated for all the space groups~\cite{Hahn2005}. The ``generic position'' on a 2D lattice is the Wyckoff position with trivial site-symmetry group. 
The coordinates of the Wyckoff positions can either be fully determined or defined up to some free parameters. 

\paragraph{Wyckoff patches.}
We now introduce the concept of \emph{Wyckoff patches}:
\textit{
Given a Wyckoff position with $d$ free parameters in the definition of its coordinates, 
we call the $d$-dimensional region of the unit-cell covered by varying the free parameters a $d$-dimensional Wyckoff patch.}

By extension, we call the \emph{site-symmetry group of the Wyckoff patch} the site-symmetry group of the Wyckoff position spanning the patch.
Wyckoff positions with nontrivial site-symmetry groups lead to either 0D or 1D patches, while the general position results into a 2D patch.  
As Wyckoff positions provide a natural way to systematically characterize the lattice, 
we employ the concept of $d$-dimensional Wyckoff patches 
as our definition of unit-cell subdivision. 

As an example, Fig.~\ref{fig:wyckoff_positions_6}(a) illustrates the Wyckoff positions of 
the unit-cell of the wallpaper group No.~6 ($p2mm$) and Fig.~\ref{fig:wyckoff_positions_6}(c) the corresponding site-symmetry group at each point. For instance, the explicit coordinates of the Wyckoff positions $1a$, $2e$ and $4i$ are
\begin{equation}\label{eq:p2mm coordinates}
\begin{split}
  1a:& \ (0, 0),\\
  2e:& \ (x, 0), (-x, 0)\\
  4i:&\ (x, y),\,(-x, y),\, (x, -y), (-x, -y)
\end{split}
\end{equation}
where we expressed coordinates in terms of lattice vectors, namely $(x, y) = x \bm{a}_x + y \bm{a}_y$ with $\bm{a}_x=a\hat{\bm{x}}, \, \bm{a}_x=b\hat{\bm{y}}$ ($a\neq b$), and the parameters $x, y$ are undetermined.
From Eq.~\eqref{eq:p2mm coordinates}, we deduce that the $1a$, $2e$, and $4i$ Wyckoff positions result into 0D, 1D, and 2D Wyckoff patches, respectively. The $4i$ Wyckoff position is the general position of the lattice, it has trivial site-symmetry group and its coordinates span the full 2D space. From the Wyckoff patches, we identify the subdivision of the unit-cell of $p2mm$, as shown in Fig.~\ref{fig:wyckoff_positions_6}(b).

\paragraph{Decorations.}
With our newly introduced notion of unit-cell subdivision, 
we can proceed and answer the question of how to decorate unit-cells.
A $d$-dimensional Wyckoff patch can be 
decorated by \mbox{cFSPTs} that: (i) are $d'$-dimensional, with $d'\leq d$, 
and (ii) are protected by the site-symmetry group of the Wyckoff patch and class AII.

The site-symmetry groups of Wyckoff patches are drawn from the same set of crystalline symmetries considered in Sec.~\ref{sec:block classification}, and they are as follows: 0D Wyckoff patches 
have $\mathbb{Z}^{\mathrm{F}}_{2a} \rtimes \mathbb{Z}^{\mathrm{F}}_{2b}$ ($a=2, 3, 4, 6$, $b=1, 2$)
symmetry, 1D have $\mathbb{Z}^{\mathrm{F}}_{2b}$ mirror symmetry, and 
2D do not have additional symmetries hence the onsite group is $\mathbb{Z}^{\,}_1$.

\renewcommand{\arraystretch}{1.5}
\begin{table}[t]
\centering
\begin{tabular}{@{\extracolsep{22 pt}} c c c c}
\hline
\hline
$d$ & SG & Wyckoff patch classification
\\
\hline
\multirow{2}{*}{0D}
& $C^{\,}_{k}$
& $\mathbb{Z} \times \cyan{\mathbb{Z}^{\,}_{\mathrm{gcd}(k,2)}}$
\\
& $C^{\,}_{kv}$
& $\mathbb{Z}\times \cyan{\mathbb{Z}^{\,}_{\mathrm{gcd}(k,2)}} \times \cyan{\mathbb{Z}^{\,}_{2}}$
\\
\hline
1D
& $C_s$
& $\mathbb{Z} \times \cyan{\mathbb{Z}^{\,}_2} \times \cyan{\mathbb{Z}^{\,}_2}$ 
\\
\hline
2D & $C_1$ & $\mathbb{Z} \times \magenta{\mathbb{Z}^{\,}_2}$
\\ \hline \hline
\end{tabular}
\caption{Classification of decorations on $d$-dimensional Wyckoff patches. 
The SG column distinguishes the site-symmetry group of the Wyckoff position, 
and the last column shows the full classification. 
The color code follows the one introduced in Tab.~\ref{tab:wallpaper classification summary}.}
\label{tab:Wyckoff position decorations}
\end{table}
\renewcommand{\arraystretch}{1}

By appropriately combining the entries of Table~\ref{tab:Cohomology blocks}, 
we can deduce the classification of the possible decorations on
Wyckoff patches fulfilling (i) and (ii).
The resulting classification is summarized in Table~\ref{tab:Wyckoff position decorations}.
The 0D Wyckoff patches can only be decorated by 0D \mbox{cFSPTs}, therefore the 0D entry of Tab.~\ref{tab:Cohomology blocks} exhausts their classification.
The 1D Wyckoff patches can either be decorated by 1D \mbox{cFSPTs}, resulting into a $\mathbb{Z}^{\,}_2$ contribution to the classification, or by 0D \mbox{cFSPTs}, 
leading to an additional $\mathbb{Z} \times \mathbb{Z}^{\,}_2$ factor corresponding to the total number of Kramers' pairs and total mirror eigenvalue of the 0D decorations, respectively. 
In fact, when multiple 0D decorations are stacked on a 1D line, they can be symmetrically and adiabatically moved along the line, 
and therefore the only relevant set of topological indices is the one counting their total charges per unit-cell. 
The 2D Wyckoff patches do not have any crystalline symmetry acting internally, 
therefore the contribution of 2D decorations is given by the $\mathbb{Z}^{\,}_2$ graded 2D AII classification. 
Decorating these Wyckoff patches by 1D \mbox{cFSPTs} does not enrich the classification, as these must be trivial in the absence of additional internal symmetries, 
while 0D decorations contribute a $\mathbb{Z}$ index enumerating the total number of Kramers' pairs.
We label decorations at a Wyckoff patch by the topological indices introduced in Sec.~\ref{sec:block classification}
with a subscript denoting the corresponding Wyckoff position. For instance, in the unit-cell subdivision
shown in Fig.~\ref{fig:wyckoff_positions_6}, 0D \mbox{FSPT}s that can be glued to the 0D Wyckoff patch $1a$ are characterized 
by the triplet $n^{\,}_{1a}$, $r^{\,}_{1a}$, and $m^{\,}_{1a}$ that indicate the total number of Kramers' pairs,
their charge under rotations, and their charge under mirror symmetry, respectively. Figure~\ref{fig:wyckoff_positions_6}(d) shows a possible set of decorations on the $p2mm$ lattice.

\subsection{Example: Wallpaper group No.\ 17}
\label{sec:example 17}

\begin{figure*}[t]
\centering
\includegraphics{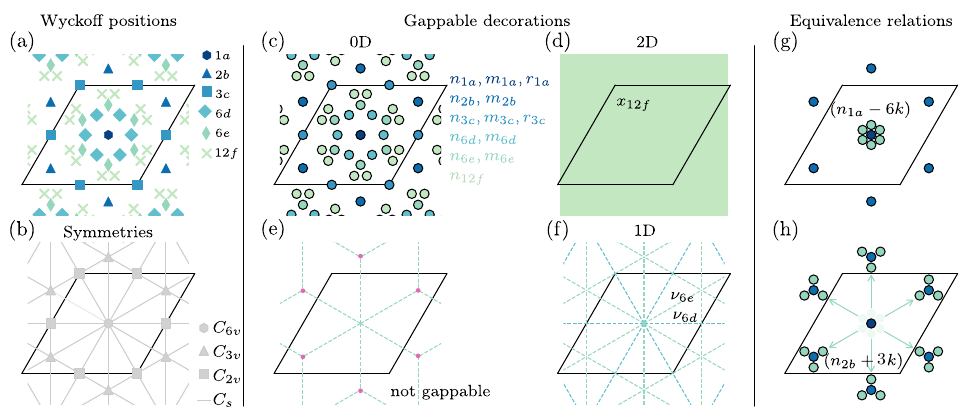}
\caption{(a) Wyckoff positions of the wallpaper group $p6mm$ (No.~17) and (b) mirror axes and rotation centers of the unit-cell, corresponding to the nontrivial site-symmetry groups in the unit-cell. (c)--(f) Representation of (c) 0D, (e), (f) 1D, and (d) 2D gappable decorations of Wyckoff patches, with the corresponding indices listed. In (e), we show an example of non-gappable decoration, and therefore with no index associated to it. (g), (h) Exemplification of the equivalence process $n^{\,}_{1a}, n^{\,}_{2b}\rightarrow n^{\,}_{1a}-6k, n^{\,}_{2b}+3k$, described around Eq.~\eqref{eq:p6mm n1a,n2b reduction} in the main text.}
\label{fig:p6mm}
\end{figure*}
Let us now discuss as an explicit example the case of wallpaper group $p6mm$ (No.~17) [Fig.~\ref{fig:p6mm}(a), (b)]. 

\setlength{\tabcolsep}{4pt}
\renewcommand{\arraystretch}{1.5}
\begin{table}[t]
\centering
\begin{tabular}{c c c c c c}
\hline\hline
$d$ & Wyckoff position & SG & Classification & indices \\
\hline
\multirow{3}{*}{0D} & $1a$ & $C_{6v}$
& $\mathbb{Z} \times \cyan{\mathbb{Z}^{\,}_2} \times \cyan{\mathbb{Z}^{\,}_2}$
& $n^{\,}_{1a},\, \cyan{r^{\,}_{1a}},\, \cyan{m^{\,}_{1a}}$
\\ 
& $2b$ & $C_{3v}$ 
& $\mathbb{Z} \times \cyan{\mathbb{Z}^{\,}_2}$
& $n^{\,}_{2b},\, \cyan{m^{\,}_{2b}}$
\\
& $3c$ & $C_{2v}$ 
& $\mathbb{Z} \times \cyan{\mathbb{Z}^{\,}_2} \times \cyan{\mathbb{Z}^{\,}_2}$
& $n^{\,}_{3c},\, \cyan{r^{\,}_{3c}},\, \cyan{m^{\,}_{3c}}$
\\ \hline
\multirow{2}{*}{1D} & $6d$ & $C_{s}$ 
& $\mathbb{Z} \times \cyan{\mathbb{Z}^{\,}_2} \times \cyan{\mathbb{Z}^{\,}_2}$
& $n^{\,}_{6d},\, \cyan{m^{\,}_{6d}},\, \cyan{\nu^{\,}_{6d}}$
\\
& $6e$ & $C_{s}$
& $\mathbb{Z} \times \cyan{\mathbb{Z}^{\,}_2} \times \cyan{\mathbb{Z}^{\,}_2} $
& $n^{\,}_{6e},\, \cyan{m^{\,}_{6e}},\, \cyan{\nu^{\,}_{6e}}$
\\ \hline
2D & $12f$ & $C_{1}$
& $\mathbb{Z} \times \magenta{\mathbb{Z}^{\,}_2}$
& $n^{\,}_{12f},\, \magenta{x^{\,}_{12f}}$
\\ \hline\hline
\end{tabular}
\caption{Decorations for the wallpaper group $p6mm$ (No.~17). The $d$, Wyckoff position, 
and SG columns list the dimension, label, and site-symmetry group for the Wyckoff positions of $p6mm$. 
The classification of the decorations and their indices are listed in the last two columns.}
\label{tab:p6mm decorations}
\end{table}
To find the classification of $p6mm$, we follow the procedure outlined in Sec.~\ref{sec:Construction}: (i) we subdivide the unit-cell into Wyckoff patches,
(ii) we enumerate the decorations for all the Wyckoff patches, (iii) we then identify the gappable decorations, (iv) and we identify equivalent decorations.

\paragraph{Unit cell subdivision.} The unit-cell subdivision is done based on the notion of Wyckoff patches. The Wyckoff positions and the site-symmetry group of the points in the unit-cell are shown in Fig.~\ref{fig:p6mm}(a),(b) and are listed in Table~\ref{tab:p6mm decorations}. The unit-cell of this space group contains three 0D Wyckoff patches, $1a$, $2b$, and $3c$, two 1D Wyckoff patches, $6d$ and $6e$, and the 2D Wyckoff patch $12f$.

\paragraph{Enumeration of decorations.} The set of possible decorations of the $p6mm$ lattice alongside the indices that classify the decorations are summarized in Table~\ref{tab:p6mm decorations} and depicted in Figs.~\ref{fig:p6mm}(c)--(f).
For the 0D Wyckoff patches, the indices describing the decorations are the total number of Kramers' pairs at each Wyckoff patch
\begin{equation}
(n^{\,}_{1a},\, n^{\,}_{2b},\, n^{\,}_{3c}) \, \in (\mathbb{Z})^3
\end{equation}
and the rotation and mirror charges
\begin{equation}
(r^{\,}_{1a},\, m^{\,}_{1a},\, m^{\,}_{2b},\, r^{\,}_{3c},\, m^{\,}_{3c}) \, \in (\mathbb{Z}^{\,}_{2})^5.
\end{equation}
The 1D Wyckoff patches are endowed with the AIIM indices
\begin{equation}
(\nu^{\,}_{6d},\, \nu^{\,}_{6e}) \, \in (\mathbb{Z}^{\,}_{2})^2,
\end{equation}
and the number of Kramers' pairs and mirror charges stemming from 0D decorations
\begin{equation}
(n^{\,}_{6d},\, n^{\,}_{6e}, \, m^{\,}_{6d},\, m^{\,}_{6e}) \, \in (\mathbb{Z})^2 \times (\mathbb{Z}^{\,}_{2})^2.
\end{equation}
Finally, 2D patches are classified by the number of Kramers' pairs of 0D decorations and the 2D AII topological phase index
\begin{equation}
(n^{\,}_{12f},\, x^{\,}_{12f})\in \mathbb{Z} \times \mathbb{Z}^{\,}_{2}.
\end{equation}

\paragraph{Gappable decorations.} We now
rule out those decorations that result in gapless modes appearing in the bulk, and we consider 0D, 1D and 2D decorations one by one.
All the 0D decorations are gappable, therefore there is no reduction of any of the indices counting the Kramers' pairs or mirror and rotation charges due to the requirement of gappability [Fig.~\ref{fig:p6mm}(c)]. 
For 1D decorations on the $6e$ Wyckoff patches, there is only one minimal way to decorate the unit-cells such that six lines meet at every crossing point, allowing to gap the edge modes in the bulk, while other decorations are not gappable: If we choose the $6e$ patches to terminate at the $2b$ patch, whose site-symmetry group contains mirror and three-fold rotation symmetries, there would be three 0D gapless boundary modes meeting at a point that are not gappable, and such a decoration has to be discarded [Fig.~\ref{fig:p6mm}(e)]. However, if the $6d$ patches are continued to the next $1a$ position, the resulting decoration is gappable [Fig.~\ref{fig:p6mm}(f)].
The 1D decorations along the $6d$ patches are symmetry constrained to meet in even number at every crossing point, therefore, they can be gapped out in the bulk [Fig.~\ref{fig:p6mm}(f)]. Likewise, the edge modes on neighboring 2D Wyckoff patches can be gapped out by covering the whole 2D space by the same phase [Fig.~\ref{fig:p6mm}(d)]. 

\paragraph{Equivalences.} We now consider the equivalences that reduce the set of indices listed above by connecting different phases through adiabatic and symmetric transformations. We start by listing the processes of peeling off charges: For example, $6k$ Kramers' pairs can be pulled away from the $1a$ Wyckoff patch, moved along the $6e$ lines, and brought to the $2b$ Wyckoff patches. This is illustrated in Figs.~\ref{fig:p6mm}(g) and \ref{fig:p6mm}(h). We write this process as
\begin{equation}\label{eq:p6mm n1a,n2b reduction}
(n^{\,}_{1a},\, n^{\,}_{2b}) \, \rightarrow \, (n^{\,}_{1a} - 6k,\, n^{\,}_{2b} + 3 k),
\end{equation}
while all the remaining indices are left unchanged.
Similarly, the following processes are allowed
\begin{equation}
\begin{split}
(n^{\,}_{1a},\, n^{\,}_{3c}) \, &\rightarrow \, (n^{\,}_{1a} - 6k,\, n^{\,}_{3c} + 2 k),\\
(n^{\,}_{1a},\, n^{\,}_{6d}) \, &\rightarrow \, (n^{\,}_{1a} - 6k,\, n^{\,}_{6d} + k),\\
(n^{\,}_{1a},\, n^{\,}_{6e}) \, &\rightarrow \, (n^{\,}_{1a} - 6k,\, n^{\,}_{6e} + k),\\
(n^{\,}_{1a},\, n^{\,}_{12f}) \, &\rightarrow \, (n^{\,}_{1a} - 12k,\, n^{\,}_{12f} + k).
\end{split}
\end{equation}
From the collection of equivalences listed above, 
we identify the following independent indices:
\begin{equation}
\begin{split}
&
(n^{\,}_{1a},\,n^{\,}_{2b},\,
n^{\,}_{3c},\,n^{\,}_{6e},\,
n^{\,}_{6d},\,n^{\,}_{12f}) \, \in \, (\mathbb{Z})^6 \\
&\rightarrow 
(n^{\,}_{1a},\,n^{\,}_{2b},\,n^{\,}_{3c}) \, \in \, \mathbb{Z}^{\,}_6 \times \mathbb{Z}^{\,}_3 \times \mathbb{Z}^{\,}_2,
\end{split}
\end{equation}
and bearing in mind that there is still a well defined index counting the total number of Kramers' pair in the unit-cell, $n \in \mathbb{Z}$.
Another set of reduction is realized by the neutralization of the mirror charges where an odd number of Wyckoff patches meet.
In particular, the $m^{\,}_{2b}$ and $m^{\,}_{6e}$ indices reduce to trivial and therefore are dropped out in the classification (Appendix~\ref{App:p6mm mirror equivalence}).

The overall classification of $p6mm$ cFSPTs finally becomes
\begin{equation}
\begin{split}
\mathbb{Z}\times \mathbb{Z}^{\,}_6 \times \mathbb{Z}^{\,}_3 \times (\mathbb{Z}^{\,}_2)^6 \times \mathbb{Z}^{\,}_2.
\end{split}
\end{equation}
In this classification, the $(\mathbb{Z}^{\,}_2)^6$ contribution, with indices $(r_{1a}, \, m_{1a}, \,r_{3c}, \, m_{3c}, \, \nu_{6d}, \,\nu_{6e}) \in (\mathbb{Z}^{\,}_2)^6$, is both intrinsically interacting and requires crystalline symmetries to exist. For this reason, they are highlighted in cyan in Table~\ref{tab:wallpaper classification summary}.

All the remaining wallpaper groups are treated similarly, and the resulting classification is contained in Table~\ref{tab:wallpaper classification summary}, where entries corresponding to intrinsically interacting phases are highlighted. The key steps followed towards filling Table~\ref{tab:wallpaper classification summary} are reported in Appendix~\ref{App:All wallapers}, where we also list the decoration indices corresponding to each wallpaper group classification explicitly. 

\section{Hamiltonian realizations}
\label{sec:Hamiltonian realizations}
We present some of the possible Hamiltonian realizations 
for the 0D and 1D decorations. 
From the latter, one can construct toy Hamiltonians for the full 2D \mbox{cFSPTs} 
by taking the transitionally invariant sums of 0D and 1D Hamiltonians 
acting on the appropriate Wyckoff patches.
We do not discuss the case of 2D Wyckoff patches, as there are no crystalline symmetries acting onsite, and the $\mathbb{Z}_2$ classification that arises without crystalline symmetries. An example of realization of the nontrivial phase is for instance the Kane-Mele model, for the case of noninteracting fermions~\cite{Kane_2005}.

\subsection{0D FSPTs: Mott atomic limits}

Noninteracting electronic gapped and non-degenerate ground states with TRS and 
particle conservation are constrained to transform trivially under the action of 
crystalline symmetries~\cite{Yao2010}
\begin{equation}\label{eq:nonint ground state transformation}
\widehat{U}(s)\ket{\Psi_{\text{nonint}}} = +\ket{\Psi_{\text{nonint}}},
\end{equation}
for any element $s$ in the space group, 
provided that $\widehat{U}(s)$ is a symmetry of both the ground state and the Hamiltonian.
However, Eq.\ \eqref{eq:nonint ground state transformation} does not necessarily 
hold in interacting systems. Therefore, for indices $r=\pm1$
and $m=\pm1$ that label the rotation and mirror charge of 
0D decorations can only be nontrivial
in the presence of interactions. 

An example of a 0D system where 
the interacting ground state transforms nontrivially 
under crystalline symmetries is the `Hubbard square', 
a four-site square molecule with a spinful orbital 
located at each corner of the square~\cite{Scalapino1996, Chakravarty2001, 
Schumann2002, Yao2010} [Fig.~\ref{fig:measuring cFSPTs}(a)]. The Hamiltonian is given by
\begin{align}
\label{eq:Hubbard square H}
\widehat{H}^{\,}_{\mathrm{HS}}
&= 
\sum_{i = 1}^4 
\sum_{\sigma=\uparrow,\downarrow} 
\left[- t^{\,}_1 
(\hat{c}^{\dagger}_{i, \sigma} \hat{c}^{\,}_{i+1, \sigma} + \mathrm{H.c.}) +
t^{\,}_2\, \hat{c}^{\dagger}_{i, \sigma} \hat{c}^{\,}_{i+2, \sigma} \right]
\nonumber\\
&\qquad 
+ 
U \sum_{i=1}^4 
\hat{n}^{\,}_{i, \uparrow}\hat{n}^{\,}_{i, \downarrow} 
- 
\mu \sum_{i = 1}^4 \sum_{\sigma} \hat{n}^{\,}_{i, \sigma},
\end{align}
where $\hat{c}^{\,}_{i, \sigma}$ ($\hat{c}^{\dagger}_{i, \sigma}$) denotes 
the annihilation (creation) operator of the electronic state 
at the $i$-th corner with spin 
$\sigma\in \{\uparrow, \downarrow\}$, 
$t_1$ and $t_2$ are the tunneling amplitude along the square edges and diagonals, respectively, 
$U$ is the Hubbard coupling, and $\mu$ the chemical potential.
In the $C^{\,}_{4}$ eigenbasis, the electronic operators are
\begin{equation}\label{eq:HS C4 transform operators}
\hat{c}^{\,}_{\ell, \sigma} 
= 
\frac{1}{2} \sum_{i=1}^4 e^{\mathrm{i} 2\pi i\ell/4}\,
\hat{c}^{\,}_{i, \sigma}, \quad \ell = 0,\frac{\pi}{2}, \pi, -\frac{\pi}{2}.
\end{equation}
At half-filling, for $t^{\,}_{1}>t^{\,}_{2}$ ($t^{\,}_{1} < t^{\,}_{2}$) and $U>0$, the ground state of Eq.~\eqref{eq:Hubbard square H} 
transforms with $-1$ ($+1$) under the action of $C^{\,}_4$ rotation symmetry. 
Expressed in the basis~\eqref{eq:HS C4 transform operators}, the ground state in the limit $U\rightarrow0$ and $t^{\,}_{1}>t^{\,}_{2}$ can be written as~\cite{Yao2010}
\begin{equation}\label{eq:Hubbard square GS}
\ket{\Psi^{\,}_{\mathrm{HS}}} = 
\frac{1}{\sqrt{2}} \hat{c}^{\dagger}_{0, \uparrow} \hat{c}^{\dagger}_{0, \downarrow}(\hat{c}^{\dagger}_{\frac{\pi}{2}, \uparrow} \hat{c}^{\dagger}_{\frac{\pi}{2}, \downarrow} - \hat{c}^{\dagger}_{-\frac{\pi}{2}, \downarrow} \hat{c}^{\dagger}_{-\frac{\pi}{2}, \uparrow}) \ket{0}.
\end{equation}
This ground state is symmetric under TRS and carries eigenvalue $-1$ under the action of $C^{\,}_{4}$ rotation symmetry. 
Therefore, it realizes an example of 0D \mbox{cFSPT} with nontrivial rotation charge, $r=-1$.
The state $\ket{\Psi^{\,}_{\mathrm{HS}}}$ also carries nontrivial $\mathrm{U}(1)$-charge since it is a linear 
combination of states with two Kramers' pairs, i.e., $n=2$. 
To tune the indices $n$ and $r$ independently, we can construct another Hamiltonian
$\widehat{H}^{'}_{\mathrm{HS}}$ that is related to $\widehat{H}^{\,}_{\mathrm{HS}}$ by the unitary particle hole symmetry
$\hat{c}^{\dagger}_{i,\sigma}\leftrightarrow\hat{c}^{\,}_{i,\sigma}$. Choosing $(t^{\,}_{1}<t^{\,}_{2})$ in Hamiltonian
$\widehat{H}^{'}_{\mathrm{HS}}$ delivers a ground state $\ket{\phi^{\,}_{\mathrm{HS}}}$
that carries $n=-2$ and trivial rotation charge $r=+1$.
Tensor product of the two ground states $\ket{\Psi^{\,}_{\mathrm{HS}}}\otimes \ket{\phi^{\,}_{\mathrm{HS}}}$
is a 0D decoration with indices $n=0$ and $r=0$.

Beyond the example of the Hubbard square, this construction can be extended to other point groups 
and used to obtain \mbox{cFSPTs} as ground states of Hubbard Hamiltonians~\cite{Muechler2014, Soldini_2023}. 
The \mbox{cFSPTs} obtained as product states of 0D \mbox{cFSPTs} realized through Hubbard like Hamiltonians, so-called Mott atomic limits (MAL),
were studied in Ref.~\onlinecite{Soldini_2023}. 
There, the \mbox{cFSPT} wave functions are written as
\begin{equation}
\ket{\mathrm{MAL}} = \prod_{\bm{r} \in \Lambda} \hat{O}^{\dagger}_{\bm{r}} \ket{0},
\end{equation}
where each $\hat{O}^{\dagger}_{\bm{r}}$ contains $N$-particle creation operators, and $\hat{O}^{\dagger}_{\bm{r}}$'s 
commute at different sites, while they describe entangled states within the $N$-particles they create. 
Similar states are studied in Ref.~\onlinecite{Herzog-Arbeitman2023}.

\subsection{1D FSPTs: The AIIM model}\label{Sec:AIIM}
\label{sec:AIIM model}
In Sec.~\ref{sec:block classification} we have pointed out that fermionic 1D systems 
with TRS, particle-number conservation, and a mirror symmetry acting as an onsite symmetry 
have $\mathbb{Z}^{\,}_{2}$ classification. 
In this section, we propose a model Hamiltonian that realizes the nontrivial phase of this classification, which we dub the AIIM model.

To better understand this $\mathbb{Z}^{\,}_{2}$ classification,
we notice that the internal symmetry group in Eq.~\eqref{eq:Gftot 1D} (with $b=2$)
has the following isomorphism:
\begin{equation}
\begin{split}
G^{(1)}_{f,\mathrm{tot}}
=
\frac{G^{\mathrm{AII}}_{f} \times 
\mathbb{Z}^{\mathrm{F}}_4}{\mathbb{Z}^{\mathrm{F}}_{2}}
&\cong
G^{\mathrm{AII}}_{f}
\rtimes
\mathbb{Z}^{\mathrm{T}'}_{2},
\label{eq:isomorphic groups in int enbld IFT}
\end{split}
\end{equation}
with $\mathbb{Z}^{\mathrm{T}'}_{2}$ being a spinless TRS
whose generator is the product of the TRS and the mirror 
symmetry generators, $t\in\mathbb{Z}^{\mathrm{FT}}_{4}$ and $m\in\mathbb{Z}^{\mathrm{FM}}_{4}$, 
i.e., $t' = t\,m$. 
Indeed, $t'$ is represented by an antiunitary operator which squares to the identity, 
and has a nontrivial action on the particle conservation $\mathrm{U}(1)^{\mathrm{F}}$ subgroup. 
We claim that the 1D nontrivial \mbox{FSPT} phase 
with symmetry group~\eqref{eq:isomorphic groups in int enbld IFT}
must be intrinsically interacting, meaning that there is no noninteracting counterpart of such a phase. 
The reason for this is as follows.
The only nontrivial projective representation of group $G^{(1)}_{f,\mathrm{tot}}$
is when the representation $\widehat{U}(t')$ of element $t'$ squares to minus identity.
In other words, the nontrivial 1D \mbox{FSPT} with $G^{(1)}_{f,\mathrm{tot}}$ symmetry must support 
gapless boundary modes that realize a projective representation of the subgroup
\begin{align}
\mathrm{U}(1)^{\mathrm{F}}\rtimes \mathbb{Z}^{\mathrm{T}'}_{2} \subset G^{(1)}_{f,\mathrm{tot}},
\end{align}
which describes the symmetry class AI of the 10-fold way. It is known that the noninteracting classification
for class AI in 1D is trivial~\cite{Schnyder2009,Kitaev2009,Ryu2010}, while
its fully interacting classification is $\mathbb{Z}^{\,}_{2}$~\cite{Freed2021}. 

\begin{figure*}[t!]
\centering
\includegraphics{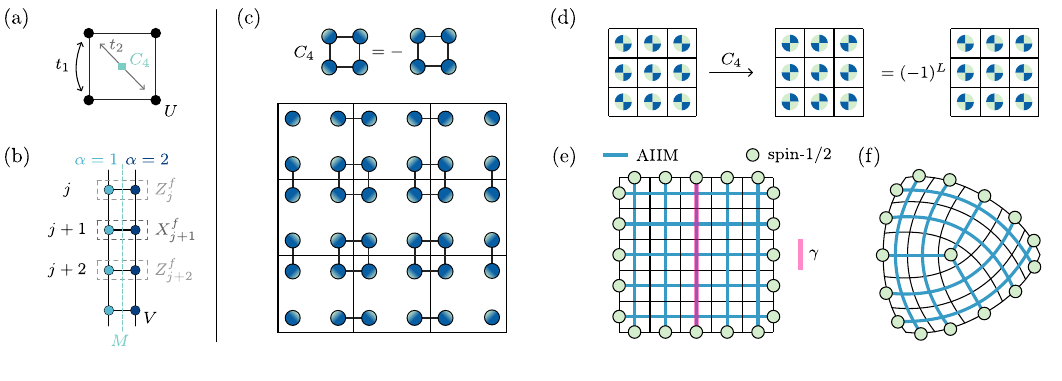}
\caption{(a), (b) Model Hamiltonians: (a) schematic of the Hubbard square with the four-fold rotation center marked by a blue square, and the two hopping terms, and Hubbard interaction indicated. (b) Schematic of the AIIM chain. The action of a single term in the Hamiltonian in Eq.~\eqref{eq:AIIM fermionic Hamiltonian} is shown, alongside the mirror-symmetry axis. In (a) and (b), the dots indicate spinful fermionic orbitals. (c)--(f) Distinguishing \mbox{cFSPTs}: (c) Square lattice ($p4mm$) decorated by the rotation-odd Hubbard square ground state of Eq.~\eqref{eq:Hubbard square GS} at the $1b$ Wyckoff patch (the corner of the unit-cell). Since there are no physical sites at the $1b$ Wyckoff patches, the lattice can be terminated by cutting through the decorations. This leads to a local two-fold degeneracy of the ground state at each corner, enforced by the combination of TRS, half-filling, and $C_4$ symmetry of the full cluster. (d) Evaluation of the $C^{\,}_{4}$ charge at the $1a$ Wyckoff position in $p4mm$. (e)~Example of string order-parameter evaluation along the path $\gamma$ (marked by the pink line) on 1D Wyckoff patches' decorations (marked by blue lines), which terminate with zero boundary modes for the nontrivial AIIM phase (marked by green dots). (f) A $\pi/2$ disclination in the $p4mm$ lattice leaves a gapless mode pinned at the disclination core.}
\label{fig:measuring cFSPTs}
\end{figure*}

We consider a 1D lattice of size $L$ with each site $j$ supporting 
two orbitals of spinful electrons~[Fig.~\ref{fig:measuring cFSPTs}(b)]. The corresponding creation and annihilation operators satisfy
\begin{align}
\left\{
\hat{c}^{\,}_{j,\alpha,\sigma},\,
\hat{c}^{\dagger}_{j',\alpha',\sigma'}
\right\}
=
\delta^{\,}_{j,j'}\,
\delta^{\,}_{\alpha,\alpha'}\,
\delta^{\,}_{\sigma,\sigma'},
\end{align}
with orbital and spin indices $\alpha=1,2$ 
and $\sigma=\uparrow,\downarrow$, respectively. 
To realize the intrinsically interacting 1D \mbox{FSPT} phase, 
we then propose the Hamiltonian
\begin{subequations}\label{eq:AIIM fermionic Hamiltonian}
\begin{equation}
\begin{split}
\widehat{H}^{\,}_{\mathrm{AIIM}} =
& - 
\sum_{j=1}^L 
\widehat{Z}^f_j\,\widehat{X}^f_{j+1}\,\widehat{Z}^f_{j+2} \\
& \quad + V \sum_{j=1}^L (\hat{n}^{\,}_{j, 1} - 1)(\hat{n}^{\,}_{j, 2} - 1),
\end{split}
\end{equation}
where 
\begin{align}
\hat{n}^{\,}_{j,\alpha} = \sum_{\sigma}\,\hat{n}^{\,}_{j,\alpha,\sigma},
\quad 
\hat{n}^{\,}_{j,\alpha,\sigma}=\hat{c}^{\dagger}_{j,\alpha,\sigma}\,\hat{c}^{\,}_{j,\alpha,\sigma}
\end{align}
are number operators, and $\widehat{Z}^{f}_{j}$ and $\widehat{X}^{f}_{j}$ are defined as
\begin{align}
\label{eq:ZXZ X Z mapping to fermions}
& \widehat{Z}^f_j 
:= 
\frac{1}{2} 
\sum_{\alpha,\sigma}
(-1)^{\alpha+1}\,
\hat{n}^{\,}_{j, \sigma, \alpha},
\\\nonumber
&
\widehat{X}^{f}_{j} 
:= 
\hat{c}^{\dagger}_{j, \uparrow, 1}\,
\hat{c}^{\dagger}_{j, \downarrow, 1}\,
\hat{c}^{\,}_{j, \downarrow, 2}\, 
\hat{c}^{\,}_{j, \uparrow, 2} 
+ 
\hat{c}^{\dagger}_{j, \uparrow, 2}\, 
\hat{c}^{\dagger}_{j, \downarrow, 2}\, 
\hat{c}^{\,}_{j, \downarrow, 1}\,
\hat{c}^{\,}_{j, \uparrow, 1}.
\end{align}
\end{subequations}
In the AIIM Hamiltonian of Eq.~\eqref{eq:AIIM fermionic Hamiltonian}, 
the first term contains four-body 
interactions while the second term is an onsite repulsive ($V>0$) interaction term. 
This Hamiltonian is symmetric under the action of the operators
\begin{subequations}
\label{eq:G1tot representation}
\begin{align}
&
\widehat{U}(\theta):=
e^{\mathrm{i}\theta \sum_{j,\alpha,\sigma}\hat{n}^{\,}_{j,\alpha,\sigma}},
\\
&
\widehat{U}(t):=
e^{\mathrm{i}\frac{\pi}{2} \sum_{j,\alpha}\sum_{\sigma,\nu}
\hat{c}^{\dagger}_{j,\alpha,\sigma}\,
\sigma^{y}_{\sigma,\nu}\,
\hat{c}^{\,}_{j,\alpha,\nu}}\,
\mathsf{K},
\\
&
\widehat{U}(m):=
e^{\mathrm{i}\frac{\pi}{2} \sum_{j}\sum_{\alpha,\rho}\sum_{\sigma,\nu}
\hat{c}^{\dagger}_{j,\alpha,\sigma}\,
\tau^{x}_{\alpha,\rho}\,
\sigma^{x}_{\sigma,\nu}\,
\hat{c}^{\,}_{j,\rho,\nu}},
\end{align}
that respectively implement $\mathrm{U}(1)$, TRS, and the mirror transformations
\begin{align}
&
\widehat{U}(\theta): \hat{c}^{\,}_{j,\alpha,\sigma} \mapsto e^{\mathrm{i}\theta}\hat{c}^{\,}_{j,\alpha,\sigma},
\\
&
\widehat{U}(t): \hat{c}^{\,}_{j,\alpha,\sigma} \mapsto \mathrm{i}\sigma^{y}_{\sigma,\nu}\hat{c}^{\,}_{j,\alpha,\nu},
\\
&
\widehat{U}(m): \hat{c}^{\,}_{j,\alpha,\sigma} \mapsto \mathrm{i}\tau^{x}_{\alpha,\rho}\,\sigma^{x}_{\sigma,\nu}\,\hat{c}^{\,}_{j,\rho,\nu},
\end{align}
\end{subequations}
where $\mathsf{K}$ is the complex conjugation and
$\sigma^{i}$ and $\tau^{i}$ ($i=x,y,z$) are Pauli matrices. 
Operators \eqref{eq:G1tot representation} 
form a global representation of symmetry group $G^{(1)}_{f,\mathrm{tot}}$.

To demonstrate that the Hamiltonian \eqref{eq:AIIM fermionic Hamiltonian} realizes a nontrivial \mbox{FSPT} phase, 
we consider the limit $V\gg 1$. The onsite repulsive interaction is then locally minimized in the two-dimensional
local subspace spanned by the states
\begin{align}
\label{eq:AIIM low-energy subspace}
\ket{1}_j:=
\hat{c}^{\dagger}_{j, \uparrow, 1} \hat{c}^{\dagger}_{j, \downarrow, 1} \ket{0},
\quad
\ket{2}_j:=
\hat{c}^{\dagger}_{j, \uparrow, 2} \hat{c}^{\dagger}_{j, \downarrow, 2} \ket{0},
\end{align}
where $\ket{0}$ is the vacuum state. Hence, at low-energies the 16-dimensional
local fermionic Fock space at each site $j$ is reduced to the 2D Hilbert space spanned
by states $\ket{1}^{\,}_{j}$ and $\ket{2}^{\,}_{j}$.
In this subspace, the effective Hamiltonian becomes 
\begin{align}
\label{eq:ZXZ}
\widehat{H}^{\,}_{\mathrm{AIIM}}\Big\vert_{V\gg 1}
\approx
\widehat{H}^{\,}_{\mathrm{ZXZ}}
:=
-
\sum_{j}
\widehat{Z}^{\,}_{j}\,
\widehat{X}^{\,}_{j+1}\,
\widehat{Z}^{\,}_{j+2},
\end{align}
where $\widehat{Z}^{\,}_{j}$ and $\widehat{X}^{\,}_{j}$ are effective degrees of freedom
acting on the low-energy two-level Hilbert space~\eqref{eq:AIIM low-energy subspace}. These operators are obtained by 
restricting the operators $\widehat{Z}^{f}_{j}$ and $\widehat{X}^{f}_{j}$
defined in Eq.\ \eqref{eq:ZXZ X Z mapping to fermions}
to the low-energy subspace. 
They satisfy the $\mathfrak{su}(2)$ algebra 
\begin{align}
\left[
\widehat{X}^{\,}_{j},\,
\widehat{Y}^{\,}_{k}
\right]
=
\delta^{\,}_{j,k}\,
2\mathrm{i}
\widehat{Z}^{\,}_{k},
\qquad
\widehat{Y}^{\,}_{j}
:=
\mathrm{i}
\widehat{X}^{\,}_{j}\,
\widehat{Z}^{\,}_{k}.
\end{align}
The Hamiltonian $\widehat{H}^{\,}_{\mathrm{ZXZ}}$ is the cluster model
introduced for spin-1/2 degrees of freedom in Ref.~\cite{Suzuki1971}.
The ground state of this Hamiltonian realizes a 1D bosonic SPT phase protected
by $\mathbb{Z}^{\mathrm{T}}_{2}\times\mathbb{Z}^{\mathrm{M}}_{2}$ symmetry~\cite{Son2011}. 
The protecting symmetries are represented by the operators
\begin{align}
\widehat{U}(t)\Big\vert_{V\gg 1}
\approx
\mathsf{K},
\qquad
\widehat{U}(m)\Big\vert_{V\gg 1}
\approx
\prod_{j}
\widehat{X}^{\,}_{j},
\end{align}
which are the representations of TRS and mirror symmetries in Eq.\ \eqref{eq:G1tot representation}
in the low-energy subspace.
The AIIM low-energy Hamiltonian~\eqref{eq:ZXZ} has a non-degenerate and gapped ground state when periodic boundary conditions are imposed. Instead,
with open boundary conditions, it has a fourfold ground state 
degeneracy owing to the existence of the operators
\begin{align}
\left\{
\widehat{Z}^{\,}_1, \, \widehat{X}^{\,}_1\widehat{Z}^{\,}_2, \, \widehat{Y}^{\,}_1\widehat{Z}^{\,}_2
\right\},
\quad
\left\{
\widehat{Z}^{\,}_{L}, \, \widehat{Z}^{\,}_{L-1}\widehat{X}^{\,}_{L}, \, \widehat{Z}^{\,}_{L-1}\widehat{Y}^{\,}_{L}
\right\}
\end{align}
localized at the boundaries and with the following properties: they commute with the Hamiltonian with open boundary conditions, form an $\mathfrak{su}(2)$ algebra, and are odd under either $\widehat{U}(t)$ or $\widehat{U}(m)$ (which prevents them from being added to the Hamiltonian as perturbations). Therefore, each boundary supports a single gapless spin-1/2
degree of freedom. 
On the boundary degrees of freedom, the $\mathbb{Z}^{\mathrm{T}}_{2}\times\mathbb{Z}^{\mathrm{M}}_{2}$
symmetry is realized projectively~\cite{Verresen2017,Aksoy2022}.

The intrinsically interacting 
1D \mbox{FSPT} phase realized by the AIIM model \eqref{eq:AIIM fermionic Hamiltonian} can be understood as
the bosonic SPT phase of the low-energy degrees of freedom that result from confinement of fermions
due to interactions. In particular, the fermionic symmetry group $G^{(1)}_{f,\mathrm{tot}}$
reduces to the group $\mathbb{Z}^{\mathrm{T}}_{2}\times\mathbb{Z}^{\mathrm{M}}_{2}$.
This is because in the low-energy subspace the $\mathrm{U}(1)^{\mathrm{F}}$
symmetry trivializes to
\begin{align}
\widehat{U}(\theta)\Big\vert_{V\gg 1}
\approx
e^{\mathrm{i}2\theta\,L},
\end{align}
which is just a complex phase. Fermion parity symmetry, $\widehat{U}(\theta=\pi)$,
then just becomes identity, which is why the TRS and mirror symmetry subgroups $\mathbb{Z}^{\mathrm{FT}}_{4}$
and $\mathbb{Z}^{\mathrm{F}}_{4}$ reduce to $\mathbb{Z}^{\mathrm{T}}_{2}$ and $\mathbb{Z}^{\mathrm{M}}_{2}$, respectively.
A gappable decoration of a 1D line by the AIIM Hamiltonian 
\eqref{eq:AIIM fermionic Hamiltonian} then corresponds to the index $\nu=1$
introduced in Sec.\ \ref{sec:block classification}, while a trivial decoration 
corresponds to $\nu=0$.

To obtain a cFSPT where 1D Wyckoff patches are decorated by the AIIM phase, 
one has to ensure that the boundary modes of each 1D decoration are gapped 
out in the bulk. This is easily achieved by ensuring that the 1D decorations terminate 
at the same point, and share the same mirror axis. 
Then, the boundary modes at a meeting point where a chain ends and another one begins 
are lifted by adding the term
\begin{equation}
\label{eq:gapping AIIM chains}
\widehat{Z}^{f}_{L-1}\,\widehat{X}^{f}_{L}\,\widehat{\mathcal{Z}}^{f}_{1} 
+
\widehat{Z}^{f}_{L}\,\widehat{\mathcal{X}}^{f}_{1}\,\widehat{\mathcal{Z}}^{f}_{2} 
\end{equation}
to the Hamiltonian. In Eq.~\eqref{eq:gapping AIIM chains}, operators with indices $j\leq L$ belong to the first chain, and with $j\geq 1$ to the second, and they are distinguished by non calligraphic and calligraphic symbols, respectively. If several pairs of 1D patches are related by rotation 
symmetry, as in the example of $6d$ and $6e$ decorations in $p6mm$ 
[Fig.~\ref{fig:p6mm}(f)], a set of rotationally symmetric copies of~\eqref{eq:gapping AIIM chains} are added to the full Hamiltonian.

\section{Signatures of cFSPT phases}\label{sec:Signatures of cFSPT phases} 

We briefly address the question of how 
\mbox{cFSPTs} can be detected and distinguished by using ground state properties.
We discuss the signatures of 0D and 1D decorations separately. 

\subsection{Signatures for 0D decorations}
As we have discussed in Sec.~\ref{sec:block classification}
nontrivial 0D decorations are characterized by symmetry charges localized at 
0D Wyckoff patches. For a given 2D lattice with $L$ 
unit-cells in total, related by translations,
the many-body ground state of the full lattice $\ket{\Psi^{\,}_{L}}$ transforms as
\begin{subequations}
\label{eq:0D decoration charges}
\begin{align}
&
\widehat{U}(\theta)\ket{\Psi^{\,}_{L}}
=
e^{\mathrm{i}2\,L\,\theta \sum_{\alpha}n^{\,}_{\alpha}}\,
\ket{\Psi^{\,}_{L}},
\\
&
\widehat{U}(r)\ket{\Psi^{\,}_{L}}
=
\prod_{\alpha}
r^{L}_{\alpha}
\ket{\Psi^{\,}_{L}},
\\
&
\widehat{U}(m)\ket{\Psi^{\,}_{L}}
=
\prod_{\alpha}
m^{L}_{\alpha}
\ket{\Psi^{\,}_{L}},
\end{align}
\end{subequations}
under $\mathrm{U}(1)$ transformations, rotations, and reflections, respectively,
provided that the latter two are the symmetries of the lattice 
[Fig.~\ref{fig:measuring cFSPTs}(d)]. 
Hereby the index $\alpha$ runs over the set of all 0D Wyckoff positions.
Therefore, a partial but by far not exhaustive characterization of the cFSPT due to 0D decorations can be extracted from the
particular sensitivity of the total symmetry charge on the system size $L$.

For \mbox{cFSPTs} that are only composed by 0D decorations, 
it was hinted in Refs.~\onlinecite{Iraola2021,Soldini_2023} 
that single-particle Green's function, often used to 
treat topological interacting systems by defining a ``topological Hamiltonian''~\cite{Gurarie2011,Wang2012,Wang2013,Lessnich2021,Iraola2021},
fail in distinguishing between trivial and nontrivial symmetry charges of 0D decorations.
However, in Ref.~\onlinecite{Soldini_2023} is was shown 
that 0D decorations where the nontrivial symmetry charge is carried 
by two-particle entangled ``clusters'' of electrons can be distinguished 
by the two-particle Green's function, and a generalization to 
$n$-particle entanglement and $n$-particle Green's function is suggested. 
In Ref.~\onlinecite{Herzog-Arbeitman2023} an alternative 
scheme based on real space invariants was proposed.

When open boundary conditions are imposed, \mbox{cFSPTs} with only 0D decorations
can support 0D topological boundary modes under certain conditions. This is possible if the 0D Wyckoff patch on which a MAL is located does not contain 
an atom itself. Then, one can define open boundary conditions
that are globally preserving the relevant spatial symmetry and cut out a $k$-th 
of the Wyckoff patch with a MAL transforming nontrivially under $C^{\,}_k$ point group elements. 
A zero-energy excitation is bound to such a corner~[Fig.~\ref{fig:measuring cFSPTs}(c)].
These modes are similar to those arising as ``filling anomalies'' in 
so-called obstructed atomic limits of noninteracting insulators~\cite{Benalcazar2019,Schindler2019,Fang2021}.

\subsection{Signatures for 1D decorations}

For \mbox{cFSPTs} with nontrivial 1D decorations, 
the ground state $\ket{\Psi}$ does not necessarily carry 
global charge under symmetries. However, the 1D \mbox{FSPT} phases acquire an
expectation value for non-local string order 
parameters~\cite{PerezGarcia2008,Pollmann2012a,Pollmann2012b}.
Namely, the analog of Eq.\ \eqref{eq:0D decoration charges} is given by
\begin{align}
\bra{\Psi^{\,}_{L}}\,
\widehat{O}^{\,}_{\gamma}\,
\ket{\Psi^{\,}_{L}}
\neq 0,
\end{align}
where $\widehat{O}^{\,}_{\gamma}$ is a non-local string operator along
a mirror symmetric line $\gamma$ [Fig.~\ref{fig:measuring cFSPTs}(e)]. 
For the AIIM model introduced in Sec.~\ref{sec:AIIM model}, the string order parameter is 
given by
\begin{align}
\widehat{O}^{\,}_{\gamma}
:=
\widehat{Z}^{f}_{1}\,
\widehat{Y}^{f}_{2}\,
\widehat{X}^{f}_{3}\,
\widehat{X}^{f}_{4}\,
\cdots
\widehat{X}^{f}_{N-2}\,
\widehat{Y}^{f}_{N-1}\,
\widehat{Z}^{f}_{N}\,
\end{align}
in terms of the effective spin-$\frac{1}{2}$ degrees of freedom introduced
in Eq.\ \eqref{eq:ZXZ X Z mapping to fermions},
where $j=1,\cdots,N$ labels the sites along the line $\gamma$.
Alternatively, ground states of \mbox{cFSPT} phases can be detected by partial crystalline 
symmetry transformations~\cite{Shapourian2017,Shiozaki2018,zhang2023complete}.

There are two paradigmatic signatures of \mbox{cFSPTs} with nontrivial 1D decorations.
First, when open boundary conditions are imposed, such \mbox{cFSPTs} feature gapless boundary states
that are protected by both internal and crystalline symmetries. More precisely, 
as shown for the AIIM model in Sec.~\ref{sec:AIIM model}, each 1D decoration features an effective 
spin-1/2 degree of freedom when open boundary conditions are imposed. 
The boundary is then described by TRS invariant spin-$\frac{1}{2}$ chains with translation and 
(if present) additional crystalline symmetries. The protected gaplessness at the boundary 
can then be understood as a direct consequence of the generalized
Lieb-Schultz-Mattis (LSM) theorems~\cite{Lieb1961,Oshikawa2000,Hastings2004,Hastings2005,Cheng2016,Cho2017,Huang2017,Tasaki2018,
Jian2018,Ogata2019,Ogata2021,Yao2021,Aksoy2021b} that apply to both effective spin-1/2 as well as the original 
fermionic degrees of freedom. Second, there are gapless boundary modes localized at lattice defects such as dislocations and disclinations (if rotation symmetry is 
present)~\cite{Thorngren2018}. This is because the otherwise gappable 1D decorations 
can no longer be generically gapped at the defect center as dislocations and disclinations
require the removal of appropriate 1D decorations. For example, in the $p4mm$ wallpaper group, where four 1D 
Wyckoff patches meet at the center of the square, a $\pi/2$ disclination, which amounts to ``cutting'' a $\pi/2$ 
angle wide slice from the lattice, would prevent the hybridization of 
one of the gapless spin-1/2 degrees of freedom, therefore leading to a gapless mode located at the disclination core [Fig.~\ref{fig:measuring cFSPTs}(f)]. We refer to the mode pinned at the defect core as gapless, in the sense that it leads to a true ground state degeneracy, arising from local degrees of freedom.

\section{Discussion}\label{sec:conclusion}
In summary, we have completed the classification of \mbox{cFSPTs} with TRS and U(1) charge-conservation for all the 17 wallpaper groups, and the result of the classification is summarized in Table~\ref{tab:wallpaper classification summary}.
We have adopted a real-space based construction first proposed in Refs.~\onlinecite{Song2017,Huang2017,Zhang2020,Zhang2022a,Zhang2022b}, and we have introduced the concept of Wyckoff patches as a natural prescription to obtain the unit-cell subdivision.

Due to the constructive nature of the classification procedure we adopted, each term appearing in the products of Table~\ref{tab:wallpaper classification summary} can be precisely associated to a topological index, and each index in turn characterizes the decorations of a specific Wyckoff patch. The indices corresponding to each entry of Table~\ref{tab:wallpaper classification summary} are listed in Appendix~\ref{App:All wallapers}. For each wallpaper group the classification summarized in Table~\ref{tab:wallpaper classification summary} always contains a $\mathbb{Z}$ factor counting the total charge in the unit-cell, and a $\mathbb{Z}_2$ factor always appearing at the end (colored in \magenta{magenta}) for the intrinsic 2D topological phase, which does not require crystalline symmetries to be protected. In addition to those, the wallpaper \mbox{cFSPTs} are classified by a series of $\mathbb{Z}_n$ ($n=2, 3, 4, 6$) terms (colored in black) that describe the charge distribution within the unit-cell, and all the remaining $\mathbb{Z}_2$ factors (colored in \cyan{cyan}) classify intrinsically interacting decorations. The latter correspond to either mirror and rotation charges, or to the AIIM phase, namely, the 1D non trivial \mbox{cFSPT} phase protected by mirror symmetry.

We remark that in our classification scheme, two Hamiltonians are in the 
same cFSPT phase only if they can be deformed to one another without
symmetry breaking or gap closing. In particular, we do not allow stacking 
with ``ancilla'' degrees of freedom. This is in contrast to the classification of 
the so-called strong topological phases where stacking with a product state does not change the topological phase. Depending on the choice of ancillas some of the entries in Table \ref{tab:wallpaper classification summary}
can be trivialized. For instance, all $\mathbb{Z}$ entries corresponding to 
the U(1) charge per unit-cell, can be changed by stacking with charged ancillas. These phases correspond to the so-called fragile topological phases~\cite{Po2018,Bradlyn2019,Bouhon2019,Else2019b}.

The ability to associate indices to real-space decorations is a rather powerful aspect of the real-space construction, as it allows to not only extract the classification of \mbox{cFSPTs}, but also to construct fixed-point wave functions that lie in each of the phases predicted by the classification. These correspond to wave functions built by gluing together decorations matching the values of the topological indices describing the phase.
We have both recalled and proposed model Hamiltonians for the realization of all the decorations necessary towards the construction of \mbox{cFSPTs} as their unique ground state, and finally we discussed possible approaches towards distinguishing these phases.

With this work, we hope to provide a tool that may support further study of interacting topological phases. In particular, we expect that this classification will be useful in characterizing more realistic Hamiltonian ground states that fall in the same symmetry classes as the \mbox{cFSPTs} analyzed here, namely, the one of interacting spin-orbit coupled interacting insulators, that do not undergo spontaneous symmetry breaking. The model Hamiltonians presented here may be a starting point to test the resilience of \mbox{cFSPTs} against disorder and perturbation away from the fixed-point wave functions. In addition, providing a classification to compare the results against, might be helpful in exploring the regime of validity of various methods and approaches used to detect and characterize \mbox{cFSPTs}.

\begin{acknowledgments}
We thank Apoorv Tiwari, Juven C.\ Wang, and Zheyan Wan for useful and extensive discussions in the early stages of this work.
The authors acknowledge useful discussions with 
Max Geier and Christopher Mudry. M.\ O.\ S.\ acknowledges funding 
from the University of Zurich Forschungskredit (Grant No.\ FK-23-110). 
\"O.~M.~A.\ is supported by Swiss National Science Foundation (SNSF)
under Grant No.\ P500PT-214429 and 
National Science Foundation (NSF) Grant No. DMR-2022428. T.N. and M.\ O.\ S.\ acknowledge support from the Swiss National Science Foundation (SNSF) through a Consolidator Grant (iTQC, TMCG-2-213805) and through Grant No.\ 200021E-198011 as part of the FOR 5249 (QUAST) lead by the Deutsche Forschungsgemeinschaft (DFG, German Research Foundation). 
\end{acknowledgments}

\section*{Data availability}
This publication is theoretical work that does not require supporting research data.

\begin{appendix}
\onecolumngrid
\renewcommand\thefigure{A\arabic{figure}}

\section{Mirror charge equivalence}\label{App:p6mm mirror equivalence}
\begin{figure}[th]

\centering
\includegraphics{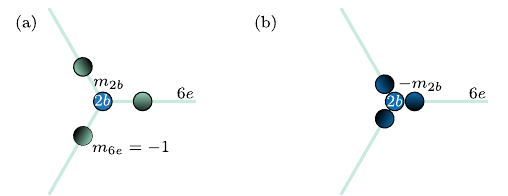}
\caption{Trivialization of the $m^{\,}_{2b}$ charge: (a) We move six mirror odd Kramers' pairs from the $1a$ to the $6e$ Wyckoff patches, which leads to a single Kramer's pair with mirror odd charge on each $6e$ Wyckoff patch, such that $m_{6e}=-1$ (assuming that initially $m_{6e}=1$, without loss of generality). (b) We transfer three Kramers' pairs to each $2b$ Wyckoff patch, which results into a change of sign of $m^{\,}_{2b}$ and trivializes $m_{6e}$.}
\label{fig:m2b}
\end{figure}

In Sec.~\ref{sec:example 17} of the main text, we derive the classification for the wallpaper group \textit{p6mm} explicitly. There, we list the equivalences that reduce the set of independent topological indices.
One of the equivalences comes from trivializing the $m^{\,}_{2b}$ mirror charge that classifies 0D decorations of the $2b$ Wyckoff patches. This trivialization is possible because an odd number of Kramers' pairs can be brought to the $2b$ patches through the $6e$ Wyckoff patches [Fig.~\ref{fig:m2b}].
More formally, let us consider each $2b$ Wyckoff patch to be decorated by a many-body state with the property
\begin{equation}
\hat{M}_x \ket{\Psi_{2b}} = m^{\,}_{2b} \ket{\Psi_{2b}},
\end{equation}
where $\hat{M}_x$ is the mirror operation with axis orthogonal to the $x$ direction.
From the $1a$ patch, we can pull six Kramers' pairs along the $6b$ lines [Fig.~\ref{fig:p6mm}(g)], that can either be even or odd under mirror symmetry, without altering the $m^{\,}_{1a}$ charge.
We choose to pull these six pairs such that they carry an odd charge under the onsite mirror of the $6e$ patches, $m^{\,}_{6e}=-1$ [Fig.~\ref{fig:m2b}(a)]. Then, we pull these pairs close to the $2b$ position, where we can write the wave function of these decorations along the $6b$ patches as
\begin{equation}
\ket{\Psi_{6b}} = \hat{O}^{\dagger}_{6e,\,1} \hat{O}^{\dagger}_{6e,\,2}\hat{O}^{\dagger}_{6e,\,3}\ket{0},
\end{equation} 
transforming as
\begin{equation}
\begin{split}
M\ket{\Psi_{6b}} &= -\hat{O}^{\dagger}_{6e,\,1} \hat{O}^{\dagger}_{6e,\,3}\hat{O}^{\dagger}_{6e,\,2}\ket{0} = - \ket{\Psi_{6b}}, \\ 
C_3 \ket{\Psi_{6b}} &= \hat{O}^{\dagger}_{6e,\,2} \hat{O}^{\dagger}_{6e,\,3}\hat{O}^{\dagger}_{6e,\,1}\ket{0} = \ket{\Psi_{6b}}.
\end{split}
\end{equation} 
The $m^{\,}_{2b}$ charge can then be neutralized by the process [Fig.~\ref{fig:m2b}(b)]
\begin{equation}
\begin{split}
\ket{\Psi_{2b}} \rightarrow \ket{\Psi_{2b}'} &= \hat{O}^{\dagger}_{6e,\,1} 
\hat{O}^{\dagger}_{6e,\,2}
\hat{O}^{\dagger}_{6e,\,3} 
\ket{\Psi_{2b}}, \\ 
M \ket{\Psi_{2b}'} &= - m^{\,}_{2b} \ket{\Psi_{2b}'}.
\end{split}
\end{equation}
At the same time, $m^{\,}_{6e}$ also changes sign, therefore returning to its original value.
Likewise, the inverse equivalence process is also allowed: the mirror charge at the $m^{\,}_{6e}$ positions can be neutralized by peeling $3k$ Kramers' pairs from the $2b$ position globally transforming oddly under the mirror symmetry of each $6e$ Wyckoff position.
Note that this particular process is possible due to the existence of an $n$-fold rotation in the site-symmetry group of $2b$, where $n$ is odd. For even $n$, the mirror neutralization is in general not possible.

\setcounter{table}{0}
\renewcommand{\thetable}{A\arabic{table}} 
\section{Classification of wallpaper group crystalline SPTs}\label{App:All wallapers}

In this appendix we provide the missing steps followed to obtain the cFSPTs classification for all the 17 wallpaper groups, namely, the various steps that lead to the entries in Table~\ref{tab:wallpaper classification summary}.
Details on the wallpaper groups, such as the coordinates of the Wyckoff positions and their site-symmetry groups, can be found for instance in Ref.~\onlinecite{Hahn2005}, or online on the Bilbao Crystallographic Server website~\cite{Aroyo2006,Aroyo2006b, Aroyo2011}.

\begin{figure}[ht!]
\centering
\includegraphics[width=\textwidth]{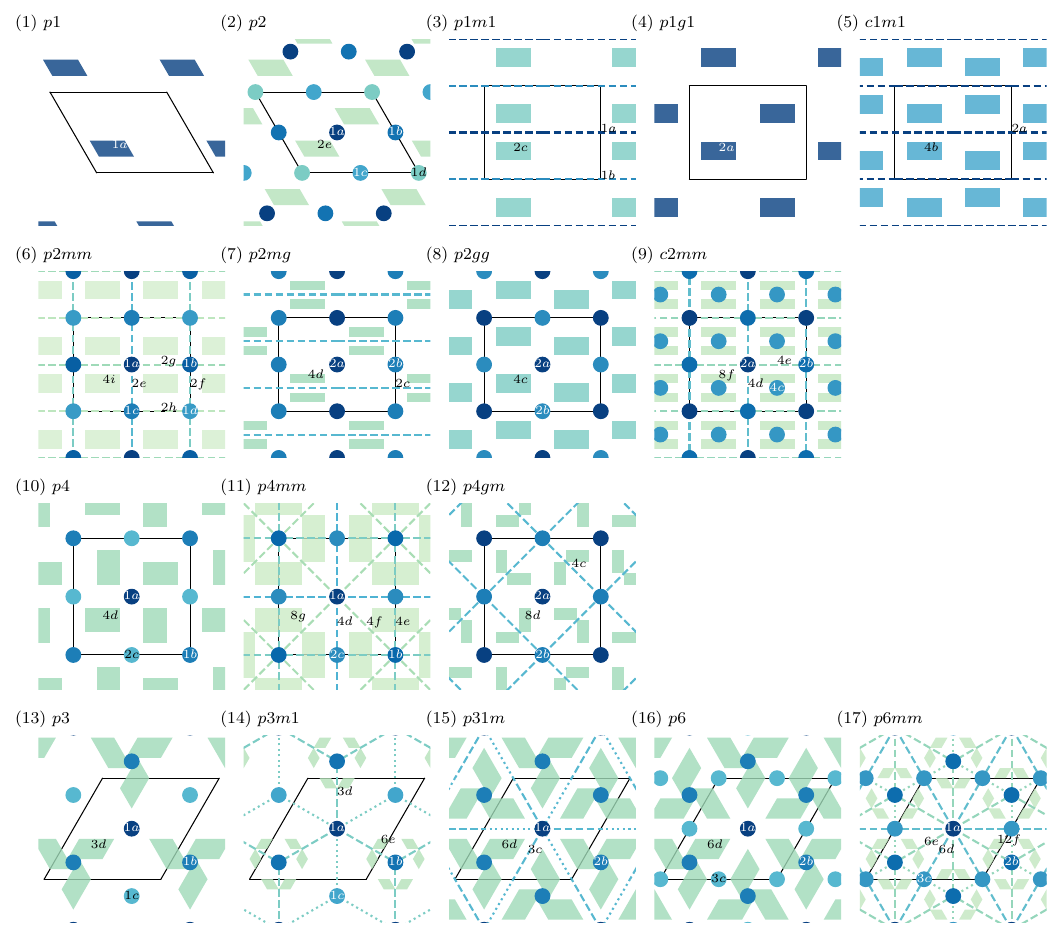}
\caption{Representation of the Wyckoff patches for the 17 wallpaper groups. The panels are labeled by the Wallpaper group's number and label. Each Wyckoff patch is labeled by the Wyckoff position from which it originates, and it is represented according to its dimension: 0D Wyckoff patches are marked by dots, 1D by dashed or dotted lines, and 2D by shaded quadrilateral regions. In each panel, a single unit-cell is marked by black lines. For the wallpaper groups Nos.\ 14, 15 and 17, dashed and dotted lines showcase how some of the 1D patches can be decorated in different ways: 1D decorations can be glued along dashed lines only, dotted lines only, or along both at the same time.}
\label{fig:wallpapers cSPT classification}
\end{figure}

\paragraph{Summary of wallpaper groups}
Figure~\ref{fig:wallpapers cSPT classification} depicts the unit-cell and Wyckoff positions for each of the 17 wallpaper groups. In all the panels, each Wyckoff position is represented as a dot, line or shaded quadrilateral for 0D, 1D or 2D Wyckoff patches, respectively. 

\paragraph{Enumeration of gappable decorations of wallpaper groups}
Tables~\ref{tab:p1}--\ref{tab:p6mm} list the classification of gappable decorations for the 17 wallpaper groups, prior to taking into account any equivalence relations. The caption of each table indicates the wallpaper group to which the table refers to. For all of the Tables ~\ref{tab:p1}--\ref{tab:p6mm}, the column labels indicate the following:
\begin{enumerate}[label=(\roman*)]
\item ``WP": list of the Wyckoff position labels for the selected wallpaper group. These labels are conventionally defined~\cite{Hahn2005}, and are composed by a number (indicating the multiplicity of the Wyckoff position) followed by a letter (the labels follow the alphabetical order, starting from the most symmetric Wyckoff position).
\item ``dim": Dimension of the Wyckoff patch corresponding to the Wyckoff position shown in the first column. In practice, the dimension is equal to the number of free parameters in the definition of the Wyckoff position coordinates.
\item ``site symm.": Site-symmetry group of the Wyckoff position expressed in the Schoenflies notation.
\item ``$G_{\text{spa}}$": Site-symmetry group of the Wyckoff position written as internal spatial symmetry group, namely as groups of the form $\mathbb{Z}^{\mathrm{F}}_{2a} \rtimes \mathbb{Z}^{\mathrm{F}}_{2b}$ ($a=1, 2, 3, 4, 6$ for $a$-fold rotation, $b=1, 2$ if mirror is not present or present, respectively). The individual factors contain the fermion parity, which motivates the superscript $\mathrm{F}$. The fermion parity $\mathbb{Z}^{\mathrm{{F}}}_2$ has to be modded out when repeated more than once. This is described more in detail in Sec.~\ref{sec:block classification} of the main text.
\item ``Classification": Classification for each Wyckoff patch, following the prescription of Table~\ref{tab:Wyckoff position decorations} in the main text, based on the entry in $G_{\text{spa}}$.
\item ``Gappable decorations": subset of gappable decorations deduced from the ``Classification" column. In particular, all 0D and 2D decorations are always gappable, while we only retain those 1D decorations whose endpoints meet in even number, and share the same mirror axis pair-wise.
\item ``Indices": List of topological indices referred to the gappable decorations. Note that the order in which the indices are listed matches the order in which the groups are written in the product of groups in the column ``Gappable decorations."
\end{enumerate}
For the topological indices, we follow the convention introduced in the main text. We summarize the convention here for convenience. For a Wyckoff patch whose Wyckoff position is labeled by $wp$, we indicate by the follwing:
\begin{enumerate}
\item $n_{wp} \in \mathbb{Z}$ (or $\mathbb{Z}_n$ with $n=2, 3, 4, 6$ after the reduction due to equivalence relations): Number of Kramers' pairs of 0D decorations at the Wyckoff patch $wp$.
\item $r_{wp} \in \mathbb{Z}^{\,}_2$: Rotation charge of 0D decorations at the 0D $wp$.
\item $m_{wp} \in \mathbb{Z}^{\,}_2$: Mirror charge of 0D decorations at the 0D or 1D Wyckoff patch $wp$.
\item $\nu_{wp} \in \mathbb{Z}^{\,}_2$: 1D topological phase of 1D decorations at the 1D Wyckoff patch $wp$. This index classifies whether the 1D decorations are in the trivial phase, or the intrinsically interacting nontrivial phase which we call AIIM phase, discussed in Sec.~\ref{sec:AIIM model} of the main text.
\item $x_{wp} \in \mathbb{Z}^{\,}_2$: 2D topological phase of 2D decorations at the 2D Wyckoff patch $wp$. This index counts whether the 2D decorations have helical edge states (nontrivial phase) or not (trivial phase).
\end{enumerate}
When a list of such indices is written next to the product of groups the indices belong to, we keep the ordering consistent: For instance, the $i$th index listed in the column ``Indices" belongs to the $i$th group of the product in ``Gappable decorations."

Note that while $n_{wp}$ and $x_{wp}$ are generically present for both interacting and noninteracting systems, instead the $r_{wp}$, $m_{wp}$, and $\nu_{wp}$ indices can assume nontrivial values only when interactions are present, i.\,e., their nontrivial values describe intrinsically interacting phases.

\paragraph{Equivalence relations}
In Appendix~\ref{App:gappable decorations and equivalences}, we list the equivalence relations for each of the 17 wallpaper groups, which allow to reduce the indices of the gappable decorations listed in Tables~\ref{tab:p1}--\ref{tab:p6mm}. Equivalences are adiabatic and symmetry respecting deformations that allow to connect phases characterized by different topological indices without a gap closing occurring. These equivalences can either partially reduce, or fully trivialize, the group to which the indices originally belonged. 

We separate equivalences into two sets:
\begin{enumerate}[label=(\roman*)]
    \item Those obtained by symmetrically peeling off 0D decorations from a Wyckoff patch to another one, which only changes the value of the indices counting the number of Kramer's pairs at the Wyckoff patches.
    \item Those obtained by symmetrically pulling 0D decorations carrying odd mirror charges from a 0D Wyckoff patch to a 1D Wyckoff patch (and vice versa). This type of equivalence results into a change of two of the Wyckoff patch mirror charges at once if the site-symmetry group of the 0D Wyckoff patch contains an odd-fold rotation, while only the 1D Wyckoff patch mirror charge is altered if the 0D Wyckoff patch only has an odd-fold rotation symmetry.
    
    Note that when two mirror charges flip sign at the same time, such an equivalence leaves out as a topological index the overall parity of the mirror charges. We call this index $m \equiv m^{\,}_{1a} m^{\,}_{1b} m^{\,}_{1c} m^{\,}_{3d}$ in the case of wallpaper group $p3m1$ (No.\ 14), and $m\equiv m^{\,}_{1a} m^{\,}_{1b}$ for the wallpaper group $p31m$ (No.\ 15).
\end{enumerate}
We indicate each process by writing the changes in the affected indices, while it is assumed that all the other indices are left unchanged.
For example, the process described in Eq.~\eqref{eq:p6mm n1a,n2b reduction} of the main text, for the wallpaper group $p6mm$, 
\begin{equation}
(n^{\,}_{1a},\, n^{\,}_{2b}) \, \rightarrow \, (n^{\,}_{1a} - 6k,\, n^{\,}_{2b} + 3 k),
\end{equation}
is listed in Appendix~\ref{App:gappable decorations and equivalences} as
\begin{equation}
n^{\,}_{1a} - 6k,\, n^{\,}_{2b} + 3 k,
\end{equation}
and a mirror charge trivialization of the type $m_{6e} \rightarrow - m_{6e}$ is listed as $ - m_{6e}$.

After listing the equivalence processes, for every wallpaper group we then explicitly show the reduction of the indices occurring due to the equivalence processes. We write on the left-hand side of the arrow ``$\rightarrow$" the original indices with their classification (omitting those indices that are left unchanged by all the equivalence relations), and on the right hand side the reduced classification. In particular, indices that are fully trivialized, namely, whose group becomes $\mathbb{Z}_1$ after taking into account equivalence relations, are omitted on the right hand side of the arrow.

\paragraph{Classification of cFSPTs in wallpaper groups with corresponding indices.}
Table~\ref{tab:final classification appendix} lists the cFSPT classification obtained for each wallpaper group, alongside the indices corresponding to each factor in the classification. The first two columns identify the wallpaper group through its number (``No.") and label (``Wallpaper"). The ``Classification" column lists the full (interacting) classification, while the ``Noninteracting classification" column shows the classification retained in the absence of interactions. The last column, ``Indices", explicitly lists the topological indices that are associated to the (interacting) classification. Once more, the ordering of the indices listed in the ``Indices" column and the groups appearing in the product in the ``Classification" column is consistent.

\clearpage
\newpage
\subsection{Enumeration of gappable decorations and equivalences }\label{App:gappable decorations and equivalences}
In this section, we list all the equivalence relations for each
993 of the wallpaper groups [Sec.~\ref{App:All wallapers}].
\begin{enumerate}[label=\textbf{No.\ \arabic*}]
    \item ($p1$)

\begin{table*}[ht!]
\centering
\begin{tabular*}{\textwidth}{@{\extracolsep{\fill}} c c c c c c c}
\hline
WP & dim & site symm. & $G_{\text{spa}}$ & Classification & Gappable decorations & Indices \\
\hline
$1a$   & 2D & $C_1$ & $\mathbb{Z}^{\,}_1$ & $\mathbb{Z}\times\mathbb{Z}^{\,}_2$ & $\mathbb{Z} \times \mathbb{Z}^{\,}_2$ & $(n^{\,}_{1a},\, x^{\,}_{1a})$ \\ \hline
\end{tabular*}
\caption{No.\ 1, $p1$.}
\label{tab:p1}
\end{table*}

    No equivalence relations.
    \item ($p2$)

\begin{table*}[h!]
\centering
\begin{tabular*}{\textwidth}{@{\extracolsep{\fill}}c c c c c c c}
\hline
WP & dim & site symm. & $G_{\mathrm{spa}}$ & Classification & Gappable decorations & Indices \\
\hline
$1a$  &  0D & $C_2$ & $\mathbb{Z}^{\mathrm{F}}_4$ & $\mathbb{Z}\times\mathbb{Z}^{\,}_2$ & $\mathbb{Z}\times\mathbb{Z}^{\,}_2$ & $(n_{1a},\, r_{1a})$\\
$1b$  &  0D & $C_2$ & $\mathbb{Z}^{\mathrm{F}}_4$ & $\mathbb{Z}\times\mathbb{Z}^{\,}_2$ & $\mathbb{Z}\times\mathbb{Z}^{\,}_2$ & $(n_{1b},\, r_{1b})$\\
$1c$  &  0D & $C_2$ & $\mathbb{Z}^{\mathrm{F}}_4$ & $\mathbb{Z}\times\mathbb{Z}^{\,}_2$ & $\mathbb{Z}\times\mathbb{Z}^{\,}_2$ & $(n_{1c},\, r_{1c})$\\
$1d$  &  0D & $C_2$ & $\mathbb{Z}^{\mathrm{F}}_4$ & $\mathbb{Z}\times\mathbb{Z}^{\,}_2$ & $\mathbb{Z}\times\mathbb{Z}^{\,}_2$ & $(n_{1d},\, r_{1d})$\\
$2e$  &  2D & $C_1$ & $\mathbb{Z}_1$ & $\mathbb{Z}\times\mathbb{Z}^{\,}_2$ & $\mathbb{Z}\times\mathbb{Z}^{\,}_2$ & $(n^{\,}_{2e},\, x^{\,}_{2e})$ \\ \hline
\end{tabular*}
\caption{No.\ 2, $p2$.}
\label{tab:p2}
\end{table*}
    Equivalence relations:
    Type (i)
    \begin{align}
        n^{\,}_{1a}-2k,\, n^{\,}_{1b}+2k \\
        n^{\,}_{1a}-2k,\, n^{\,}_{1c}+2k \\
        n^{\,}_{1a}-2k,\, n^{\,}_{1d}+2k \\
        n^{\,}_{1a}-2k,\, n^{\,}_{2e}+k,
    \end{align}
    Reduced indices:
    \begin{equation}
        \begin{split}
        (n^{\,}_{1a}, \, n^{\,}_{1b}, \, n^{\,}_{1c}, \, n^{\,}_{1d}, \, n^{\,}_{2e}) \in \mathbb{Z}^5_{\,} \rightarrow
        (n^{\,}_{1a}, \, n^{\,}_{1b}, \, n^{\,}_{1c}, \, n^{\,}_{1d}) \in (\mathbb{Z}_{2})^4
        \end{split}
    \end{equation}  
    \item ($p1m1$)

\begin{table*}[h!]
\centering
\begin{tabular*}{\textwidth}{@{\extracolsep{\fill}}c c c c c c c}
\hline
WP & dim & site symm. & $G_{\mathrm{spa}}$ & Classification & Gappable decorations & Indices \\
\hline
$1a$ & 1D & $C_s$ 
& $\mathbb{Z}^{\mathrm{F}}_4$ 
& $\mathbb{Z} \times \mathbb{Z}^{\,}_2 \times \mathbb{Z}^{\,}_2$ 
& $\mathbb{Z} \times \mathbb{Z}^{\,}_2 \times \mathbb{Z}^{\,}_2 $ 
& $(n^{\,}_{1a},\, m^{\,}_{1a}, \nu^{\,}_{1a})$ 
\\
$1b$ & 1D & $C_s$ 
& $\mathbb{Z}^{\mathrm{F}}_4$ 
& $\mathbb{Z} \times \mathbb{Z}^{\,}_2 \times \mathbb{Z}^{\,}_2$ 
& $\mathbb{Z} \times \mathbb{Z}^{\,}_2 \times \mathbb{Z}^{\,}_2$ 
& $(n^{\,}_{1b},\, m^{\,}_{1b},\, \nu^{\,}_{1b})$ 
\\
$2c$ & 2D & $C_1$ 
& $\mathbb{Z}^{\,}_1$ 
& $\mathbb{Z} \times \mathbb{Z}^{\,}_2$
& $\mathbb{Z} \times \mathbb{Z}^{\,}_2$ 
& $(n^{\,}_{2c},\, x^{\,}_{2c})$ \\ \hline
\end{tabular*}
\caption{No.\ 3,  $p1m1$.}
\label{tab:p1m1}
\end{table*}
    Equivalence relations: Type (i)
    \begin{align}
            n^{\,}_{1a}-2k,\, n^{\,}_{1b}+2k \\
            n^{\,}_{1a}-2k,\, n^{\,}_{2c}+k.
    \end{align}
    
    Reduced indices:
    \begin{equation}
        \begin{split}
            (n^{\,}_{1a},\, n^{\,}_{1b},\, n^{\,}_{2c}) \in \mathbb{Z}^3_{\,} \rightarrow (n^{\,}_{1a},\, n^{\,}_{1b}) \in (\mathbb{Z}^{\,}_2)^2
        \end{split}
    \end{equation}  
    \clearpage
    \item ($p1g1$) 
    
\begin{table*}[h!]
\centering
\begin{tabular*}{\textwidth}{@{\extracolsep{\fill}}c c c c c c c}
\hline
WP & dim & site symm. & $G_{\mathrm{spa}}$ & Classification & Gappable decorations & Indices \\
\hline
$2a$ & 2D & $C_1$ 
& $\mathbb{Z}^{\mathrm{F}}_1$ 
& $\mathbb{Z} \times \mathbb{Z}^{\,}_2$
& $\mathbb{Z} \times \mathbb{Z}^{\,}_2$ 
& $(n^{\,}_{2a},\, x^{\,}_{2a})$
\\ \hline
\end{tabular*}
\caption{No.\ 4, $p1g1$.}
\label{tab:p1g1}
\end{table*}

    No equivalence relations.
    
    \item ($c1m1$)
    
    \begin{table*}[h!]
    \centering
\begin{tabular*}{\textwidth}{@{\extracolsep{\fill}}c c c c c c c}
\hline
WP & dim & site symm. & $G_{\mathrm{spa}}$ & Classification & Gappable decorations & Indices \\
\hline
$2a$ & 1D & $C_s$ 
& $\mathbb{Z}^{\mathrm{F}}_4$
& $\mathbb{Z} \times \mathbb{Z}^{\,}_2 \times \mathbb{Z}^{\,}_2$
& $\mathbb{Z} \times \mathbb{Z}^{\,}_2 \times \mathbb{Z}^{\,}_2$
& $(n^{\,}_{2a},\,m^{\,}_{2a},\, \nu^{\,}_{2a})$
\\
$4b$ & 2D & $C_1$
& $\mathbb{Z}^{\,}_1$
& $\mathbb{Z} \times \mathbb{Z}^{\,}_2$
& $\mathbb{Z} \times \mathbb{Z}^{\,}_2$
& $(n^{\,}_{4b},\, x^{\,}_{4b})$
\\ \hline
\end{tabular*}
\caption{No.\ 5, $c1m1$.}
\label{tab:c1m1}
\end{table*}
    
    Equivalence relations: Type (i)
    \begin{align}
            n^{\,}_{2a}-2k,\, n^{\,}_{4b}+k.
    \end{align}
    
    Reduced indices:
    \begin{equation}
        \begin{split}
        (n^{\,}_{2a},\, n^{\,}_{4b}) \in \mathbb{Z}^2_{\,} \rightarrow n^{\,}_{2a} \in \mathbb{Z}^{\,}_2
        \end{split}
    \end{equation} 

    \item ($p2mm$)
    \begin{table*}[h!]
\centering
\begin{tabular*}{\textwidth}{@{\extracolsep{\fill}}c c c c c c c}
\hline
WP & dim & site symm. & $G_{\mathrm{spa}}$ & Classification & Gappable decorations & Indices \\
\hline
$1a$ & 0D & $C_{2v}$ 
& $\mathbb{Z}^{\mathrm{F}}_4 \rtimes \mathbb{Z}^{\mathrm{F}}_4 / \mathbb{Z}^{\mathrm{F}}_2$ 
& $\mathbb{Z}\times\mathbb{Z}^{\,}_2\times\mathbb{Z}^{\,}_2$ 
& $\mathbb{Z}\times\mathbb{Z}^{\,}_2\times\mathbb{Z}^{\,}_2$ 
& $(n^{\,}_{1a},\, r^{\,}_{1a}, \, m^{\,}_{1a})$ 
\\
$1b$ & 0D & $C_{2v}$
& $\mathbb{Z}^{\mathrm{F}}_4 \rtimes \mathbb{Z}^{\mathrm{F}}_4/ \mathbb{Z}^{\mathrm{F}}_2$ 
& $\mathbb{Z}\times\mathbb{Z}^{\,}_2\times\mathbb{Z}^{\,}_2$ 
& $\mathbb{Z}\times\mathbb{Z}^{\,}_2\times\mathbb{Z}^{\,}_2$ 
& $(n^{\,}_{1b},\, r^{\,}_{1b}, \, m^{\,}_{1b})$
\\
$1c$ & 0D & $C_{2v}$ 
& $\mathbb{Z}^{\mathrm{F}}_4 \rtimes \mathbb{Z}^{\mathrm{F}}_4/ \mathbb{Z}^{\mathrm{F}}_2$ 
& $\mathbb{Z}\times\mathbb{Z}^{\,}_2\times\mathbb{Z}^{\,}_2$ 
&  $\mathbb{Z}\times\mathbb{Z}^{\,}_2\times\mathbb{Z}^{\,}_2$ 
& $(n^{\,}_{1c},\, r^{\,}_{1c}, \, m^{\,}_{1c})$
\\
$1d$ & 0D & $C_{2v}$ 
& $\mathbb{Z}^{\mathrm{F}}_4 \rtimes \mathbb{Z}^{\mathrm{F}}_4/ \mathbb{Z}^{\mathrm{F}}_2$ 
& $\mathbb{Z}\times\mathbb{Z}^{\,}_2\times\mathbb{Z}^{\,}_2$ 
& $\mathbb{Z}\times\mathbb{Z}^{\,}_2\times\mathbb{Z}^{\,}_2$ 
& $(n^{\,}_{1d},\, r^{\,}_{1d}, \, m^{\,}_{1d})$
\\
$2e$ & 1D & $C_s$   
& $\mathbb{Z}^{\mathrm{F}}_4$ 
& $\mathbb{Z}^{\,} \times \mathbb{Z}^{\,}_2 \times \mathbb{Z}^{\,}_2$ 
& $\mathbb{Z}^{\,} \times \mathbb{Z}^{\,}_2 \times \mathbb{Z}^{\,}_2$ 
& $(n_{2e},\, m^{\,}_{2e},\, \nu^{\,}_{2e})$
\\
$2f$ & 1D & $C_s$   
& $\mathbb{Z}^{\mathrm{F}}_4$ 
& $\mathbb{Z}^{\,} \times\mathbb{Z}^{\,}_2 \times \mathbb{Z}^{\,}_2$ 
& $\mathbb{Z}^{\,} \times\mathbb{Z}^{\,}_2 \times \mathbb{Z}^{\,}_2$ 
& $(n_{2f},\, m^{\,}_{2f},\, \nu^{\,}_{2f})$ 
\\
$2g$ & 1D & $C_s$   
& $\mathbb{Z}^{\mathrm{F}}_4$ 
& $\mathbb{Z}^{\,} \times\mathbb{Z}^{\,}_2 \times \mathbb{Z}^{\,}_2$ 
& $\mathbb{Z}^{\,} \times\mathbb{Z}^{\,}_2 \times \mathbb{Z}^{\,}_2$ 
& $(n_{2g},\, m^{\,}_{2g},\, \nu^{\,}_{2g})$ 
\\
$2h$ & 1D & $C_s$   
& $\mathbb{Z}^{\mathrm{F}}_4$ 
& $\mathbb{Z}^{\,} \times\mathbb{Z}^{\,}_2 \times \mathbb{Z}^{\,}_2$ 
& $\mathbb{Z}^{\,} \times\mathbb{Z}^{\,}_2 \times \mathbb{Z}^{\,}_2$ 
& $(n_{2h},\, m^{\,}_{2h},\, \nu^{\,}_{2h})$
\\
$4i$ & 2D & $C_1$   
& $\mathbb{Z}^{\,}_1$         
& $\mathbb{Z}^{\,} \times \mathbb{Z}^{\,}_2$ 
& $\mathbb{Z}^{\,} \times \mathbb{Z}^{\,}_2$ 
& $(n^{\,}_{4i},\, x^{\,}_{4i})$ 
\\ \hline
\end{tabular*}
\caption{No.\ 6, $p2mm$.}
\label{tab:p2mm}
\end{table*}

    Equivalence relations: 
    Type (i)
    \begin{align}
            n^{\,}_{1a}-2k,\, n^{\,}_{1b}+2k\\ 
            n^{\,}_{1a}-2k,\, n^{\,}_{1c}+2k\\ 
            n^{\,}_{1a}-2k,\, n^{\,}_{1d}+2k\\
            n^{\,}_{1a}-2k,\, n^{\,}_{2e}+k\\
            n^{\,}_{1a}-2k,\, n^{\,}_{2f}+k\\
            n^{\,}_{1a}-2k,\, n^{\,}_{2g}+k\\
            n^{\,}_{1a}-2k,\, n^{\,}_{2h}+k\\
            n^{\,}_{1a}-2k,\, n^{\,}_{4i}+k.
    \end{align}
    Type (ii)
    \begin{align}
            -m^{\,}_{2e}\\
            -m^{\,}_{2g}\\
            -m^{\,}_{2f}\\
            -m^{\,}_{2h}.
    \end{align}
    
    Reduced indices:
    \begin{align}
        (n^{\,}_{1a},\, n^{\,}_{1b},\, n^{\,}_{1c},\, n^{\,}_{1d},\, n^{\,}_{2e},\, n^{\,}_{2f},\, n^{\,}_{2g},\, n^{\,}_{2h},\, n^{\,}_{4i},
        m^{\,}_{2e},\, m^{\,}_{2f},\, m^{\,}_{2g}\, m^{\,}_{2h}) \in \mathbb{Z}^9_{\,} \times (\mathbb{Z}^{\,}_2)^4 \\
        \rightarrow
        (n^{\,}_{1a},\, n^{\,}_{1b},\, n^{\,}_{1c},\, n^{\,}_{1d}) \in (\mathbb{Z}^{\,}_2)^4
    \end{align}
    
    \item ($p2mg$)

    \begin{table*}[h!]
\centering
\begin{tabular*}{\textwidth}{@{\extracolsep{\fill}}c c c c c c c}
\hline
WP & dim & site symm. & $G_{\mathrm{spa}}$ & Classification & Gappable decorations & Indices \\
\hline
$2a$ & 0D & $C_2$ 
& $\mathbb{Z}^{\mathrm{F}}_4$ 
& $\mathbb{Z} \times \mathbb{Z}_2$ 
& $\mathbb{Z} \times \mathbb{Z}_2$ 
& $(n^{\,}_{2a},\, r^{\,}_{2a})$ 
\\
$2b$ & 0D & $C_2$ 
& $\mathbb{Z}^{\mathrm{F}}_4$ 
& $\mathbb{Z} \times \mathbb{Z}_2$
& $\mathbb{Z} \times \mathbb{Z}_2$ 
& $(n^{\,}_{2b},\, r^{\,}_{2b})$ 
\\
$2c$ & 1D & $C_s$
& $\mathbb{Z}^{\mathrm{F}}_4$ 
& $\mathbb{Z} \times \mathbb{Z}_2 \times \mathbb{Z}_2$
& $\mathbb{Z} \times \mathbb{Z}_2 \times \mathbb{Z}_2$
& $(n^{\,}_{2c},\, m^{\,}_{2c},\, \nu^{\,}_{2c})$ 
\\
$4d$ & 2D & $C_1$ 
& $\mathbb{Z}^{\,}_1$         
& $\mathbb{Z}^{\,} \times \mathbb{Z}^{\,}_2$ 
& $\mathbb{Z}^{\,} \times \mathbb{Z}^{\,}_2$ 
& $(n^{\,}_{4d},\, x^{\,}_{4d})$ 
\\ \hline
\end{tabular*}
\caption{No.\ 7, $p2mg$.}
\label{tab:p2mg}
\end{table*}
    Equivalence relations: Type (i)
    \begin{align}
        n^{\,}_{2a}-2k,\, n^{\,}_{2b}+2k\\ 
        n^{\,}_{2a}-2k,\, n^{\,}_{2c}+2k\\ 
        n^{\,}_{2a}-2k,\, n^{\,}_{4b}+k.
    \end{align}
    
    Reduced indices:
    \begin{equation}
        \begin{split}
        (n^{\,}_{2a},\, n^{\,}_{2b},\, n^{\,}_{2c},\, n^{\,}_{4d}) \in \mathbb{Z}^4_{\,} \rightarrow (n^{\,}_{2a},\, n^{\,}_{2b},\, n^{\,}_{2c}) \in (\mathbb{Z}^{\,}_2)^3
        \end{split}
    \end{equation} 
    
    \item ($p2gg$)
\begin{table*}[h!]
\centering
\begin{tabular*}{\textwidth}{@{\extracolsep{\fill}}c c c c c c c}
\hline
WP & dim & site symm. & $G_{\mathrm{spa}}$ & Classification & Gappable decorations & Indices \\
\hline
$2a$ & 0D & $C_2$ 
& $\mathbb{Z}^{\mathrm{F}}_4$   
& $\mathbb{Z}^{\,} \times \mathbb{Z}^{\,}_2$ 
& $\mathbb{Z}^{\,} \times \mathbb{Z}^{\,}_2$ 
& $(n^{\,}_{2a},\, r^{\,}_{2a})$ 
\\
$2b$ & 0D & $C_2$ 
& $\mathbb{Z}^{\mathrm{F}}_4$   
& $\mathbb{Z}^{\,} \times \mathbb{Z}^{\,}_2$ 
& $\mathbb{Z}^{\,} \times \mathbb{Z}^{\,}_2$ 
& $(n^{\,}_{2b},\, r^{\,}_{2b})$ 
\\
$4c$ & 2D & $C_1$ 
& $\mathbb{Z}^{\,}_1$         
& $\mathbb{Z}^{\,} \times \mathbb{Z}^{\,}_2$ 
& $\mathbb{Z}^{\,} \times \mathbb{Z}^{\,}_2$ 
& $(n^{\,}_{4c},\, x^{\,}_{4c})$ \\
\hline
\end{tabular*}
\caption{No.\ 8, $p2gg$.}
\label{tab:p2gg}
\end{table*}

    Equivalence relations: Type (i)
    \begin{align}
        n^{\,}_{2a}-2k,\, n^{\,}_{2b}+2k\\
        n^{\,}_{2a}-2k,\, n^{\,}_{4c}+k.
    \end{align}
    
    Reduced indices:
    \begin{equation}
        \begin{split}
        (n^{\,}_{2a},\, n^{\,}_{2b},\, n^{\,}_{4c}) \in \mathbb{Z}^3_{\,} \rightarrow
            (n^{\,}_{2a},\, n^{\,}_{2b}) \in (\mathbb{Z}^{\,}_2)^2
        \end{split}
    \end{equation} 

    \clearpage
    \item ($c2mm$)
    \begin{table*}[h!]
\centering
\begin{tabular*}{\textwidth}{@{\extracolsep{\fill}}c c c c c c c}
\hline
WP & dim & site symm. & $G_{\mathrm{spa}}$ & Classification & Gappable decorations & Indices \\
\hline
$2a$ & 0D & $C_{2v}$
& $\mathbb{Z}^{\mathrm{F}}_4 \rtimes \mathbb{Z}^{\mathrm{F}}_4 / \mathbb{Z}^{\mathrm{F}}_2$ 
& $\mathbb{Z}^{\,} \times \mathbb{Z}^{\,}_2 \times \mathbb{Z}^{\,}_2$ 
& $\mathbb{Z}^{\,} \times \mathbb{Z}^{\,}_2 \times \mathbb{Z}^{\,}_2$ 
& $(n^{\,}_{2a},\, r^{\,}_{2a},\, m_{2a})$
\\
$2b$ & 0D & $C_{2v}$
& $\mathbb{Z}^{\mathrm{F}}_4 \rtimes \mathbb{Z}^{\mathrm{F}}_4 / \mathbb{Z}^{\mathrm{F}}_2$
& $\mathbb{Z}^{\,} \times \mathbb{Z}^{\,}_2 \times \mathbb{Z}^{\,}_2$ 
& $\mathbb{Z}^{\,} \times \mathbb{Z}^{\,}_2 \times \mathbb{Z}^{\,}_2$ 
& $(n^{\,}_{2b},\, r^{\,}_{2b},\, m_{2b})$
\\
$4c$ & 0D & $C_{2}$
& $\mathbb{Z}^{\mathrm{F}}_4$
& $\mathbb{Z}^{\,} \times \mathbb{Z}^{\,}_2$ 
& $\mathbb{Z}^{\,} \times \mathbb{Z}^{\,}_2$ 
& $(n^{\,}_{4c},\, r^{\,}_{4c})$
\\
$4d$ & 1D & $C_{s}$
& $\mathbb{Z}^{\mathrm{F}}_4$
& $\mathbb{Z}^{\,} \times \mathbb{Z}^{\,}_2 \times \mathbb{Z}^{\,}_2$ 
& $\mathbb{Z}^{\,} \times \mathbb{Z}^{\,}_2 \times \mathbb{Z}^{\,}_2$ 
& $(n^{\,}_{4d},\, m^{\,}_{4d},\, \nu^{\,}_{4d})$
\\
$4e$ & 1D & $C_{s}$
& $\mathbb{Z}^{\mathrm{F}}_4$ 
& $\mathbb{Z}^{\,} \times \mathbb{Z}^{\,}_2 \times \mathbb{Z}^{\,}_2$ 
& $\mathbb{Z}^{\,} \times \mathbb{Z}^{\,}_2 \times \mathbb{Z}^{\,}_2$ 
& $(n^{\,}_{4e},\, m^{\,}_{4e},\, \nu^{\,}_{4e})$
\\
$8f$ & 2D & $C_{1}$
& $\mathbb{Z}^{\,}_1$         
& $\mathbb{Z}^{\,} \times \mathbb{Z}^{\,}_2$ 
& $\mathbb{Z}^{\,} \times \mathbb{Z}^{\,}_2$ 
& $(n^{\,}_{8f},\, x^{\,}_{8f})$ 
\\ \hline
\end{tabular*}
\caption{No.\ 9, $c2mm$.}
\label{tab:c2mm}
\end{table*}

    Equivalence relations: Type (i)
    \begin{align}
        n^{\,}_{2a}-2k,\, n^{\,}_{2b}+2k\\
        n^{\,}_{2a}-4k,\, n^{\,}_{4c}+2k\\
        n^{\,}_{2a}-2k,\, n^{\,}_{4d}+k\\
        n^{\,}_{2a}-2k,\, n^{\,}_{4e}+k\\
        n^{\,}_{2a}-4k,\, n^{\,}_{8f}+k.
    \end{align}
    Type (ii)
    \begin{align}
        -m^{\,}_{4e}\\
        -m^{\,}_{4d}.
    \end{align}
    
    Reduced indices:
    \begin{equation}
        \begin{split}
        (n^{\,}_{2a},\, n^{\,}_{2b},\, n^{\,}_{4c},\, n^{\,}_{4d},\, n^{\,}_{4e},\, n^{\,}_{8f},\, m^{\,}_{4e}, \, m^{\,}_{4d}) \in \mathbb{Z}^6_{\,} \times (\mathbb{Z}^{\,}_2)^2\rightarrow
(n^{\,}_{2a},\, n^{\,}_{2b}\, n^{\,}_{4c}) \in (\mathbb{Z}^{\,}_2)^3
        \end{split}
    \end{equation} 
    
    \item ($p4$)

    \begin{table*}[h!]
\centering
\begin{tabular*}{\textwidth}{@{\extracolsep{\fill}}c c c c c c c}
\hline
WP & dim & site symm. & $G_{\mathrm{spa}}$ & Classification & Gappable decorations & Indices \\
\hline
$1a$ & 0D & $C_4$ 
& $\mathbb{Z}^{\mathrm{F}}_8$ 
& $\mathbb{Z} \times \mathbb{Z}^{\,}_2$ & $\mathbb{Z} \times \mathbb{Z}^{\,}_2$ 
& $(n^{\,}_{1a},\, r^{\,}_{1a})$
\\
$1b$ & 0D & $C_4$ 
& $\mathbb{Z}^{\mathrm{F}}_8$ 
& $\mathbb{Z} \times \mathbb{Z}^{\,}_2$ & $\mathbb{Z} \times \mathbb{Z}^{\,}_2$ 
& $(n^{\,}_{1b},\, r^{\,}_{1b})$
\\
$2c$ & 0D & $C_2$ 
& $\mathbb{Z}^{\mathrm{F}}_4$ 
& $\mathbb{Z} \times \mathbb{Z}^{\,}_2$ & $\mathbb{Z} \times \mathbb{Z}^{\,}_2$ 
& $(n^{\,}_{2c},\, r^{\,}_{2c})$
\\
$4d$ & 2D & $C_1$ 
& $\mathbb{Z}_1$ 
& $\mathbb{Z} \times \mathbb{Z}^{\,}_2$ & $\mathbb{Z} \times \mathbb{Z}^{\,}_2$ 
& $(n^{\,}_{4d},\, x^{\,}_{4d})$
\\ \hline
\end{tabular*}
\caption{No.\ 10, $p4$.}
\label{tab:p4}
\end{table*}
    Equivalence relations: Type (i)
    \begin{align}
            n^{\,}_{1a}-4k,\, n^{\,}_{1b}+4k\\
    n^{\,}_{1a}-4k,\, n^{\,}_{2c}+ 2k\\ 
    n^{\,}_{1a}-4k,\, n^{\,}_{4d}+ k.
    \end{align}
    
    Reduced indices:
    \begin{equation}
        \begin{split}
        (n^{\,}_{1a},\, n^{\,}_{1b},\, n^{\,}_{2c},\, n^{\,}_{4d}) \in \mathbb{Z}^4_{\,} \rightarrow
(n^{\,}_{1a},\, n^{\,}_{1b},\, n^{\,}_{2c}) \in (\mathbb{Z}^{\,}_4)^2 \times \mathbb{Z}^{\,}_2
        \end{split}
    \end{equation} 
    \clearpage
    \item ($p4mm$)

    \begin{table*}[h!]
\centering
\begin{tabular*}{\textwidth}{@{\extracolsep{\fill}}c c c c c c c}
\hline
WP & dim & site symm. & $G_{\mathrm{spa}}$ & Classification & Gappable decorations & Indices \\
\hline
$1a$ & 0D & $C_{4v}$ 
& $\mathbb{Z}^{\mathrm{F}}_{8} \rtimes \mathbb{Z}^{\mathrm{F}}_4/ \mathbb{Z}^{\mathrm{F}}_2$ 
& $\mathbb{Z} \times \mathbb{Z}^{\,}_2 \times \mathbb{Z}^{\,}_2$ 
& $\mathbb{Z} \times \mathbb{Z}^{\,}_2 \times \mathbb{Z}^{\,}_2$ 
& $(n^{\,}_{1a},\, r^{\,}_{1a},\, m^{\,}_{1a})$
\\
$1b$ & 0D & $C_{4v}$ 
& $\mathbb{Z}^{\mathrm{F}}_{8} \rtimes \mathbb{Z}^{\mathrm{F}}_4/ \mathbb{Z}^{\mathrm{F}}_2$ 
& $\mathbb{Z} \times \mathbb{Z}^{\,}_2 \times \mathbb{Z}^{\,}_2$ 
& $\mathbb{Z} \times \mathbb{Z}^{\,}_2 \times \mathbb{Z}^{\,}_2$ 
& $(n^{\,}_{1b},\, r^{\,}_{1b},\, m^{\,}_{1b})$
\\
$2c$ & 0D & $C_{2v}$ 
& $\mathbb{Z}^{\mathrm{F}}_{4} \rtimes \mathbb{Z}^{\mathrm{F}}_4/ \mathbb{Z}^{\mathrm{F}}_2$
& $\mathbb{Z} \times \mathbb{Z}^{\,}_2 \times \mathbb{Z}^{\,}_2$ 
& $\mathbb{Z} \times \mathbb{Z}^{\,}_2 \times \mathbb{Z}^{\,}_2$ 
& $(n^{\,}_{2c},\, r^{\,}_{2c},\, m^{\,}_{2c})$
\\
$4d$ & 1D & $C_{s}$ 
& $\mathbb{Z}^{\mathrm{F}}_4$ 
& $\mathbb{Z} \times \mathbb{Z}^{\,}_2 \times \mathbb{Z}^{\,}_2$ 
& $\mathbb{Z} \times \mathbb{Z}^{\,}_2 \times \mathbb{Z}^{\,}_2$ 
& $(n^{\,}_{4d},\, m^{\,}_{4d},\, \nu^{\,}_{4d})$
\\
$4e$ & 1D & $C_{s}$ 
& $\mathbb{Z}^{\mathrm{F}}_4$ 
& $\mathbb{Z} \times \mathbb{Z}^{\,}_2 \times \mathbb{Z}^{\,}_2$ 
& $\mathbb{Z} \times \mathbb{Z}^{\,}_2 \times \mathbb{Z}^{\,}_2$ 
& $(n^{\,}_{4e},\, m^{\,}_{4e},\, \nu^{\,}_{4e})$
\\
$4f$ & 1D & $C_{s}$ 
& $\mathbb{Z}^{\mathrm{F}}_4$ 
& $\mathbb{Z} \times \mathbb{Z}^{\,}_2 \times \mathbb{Z}^{\,}_2$ 
& $\mathbb{Z} \times \mathbb{Z}^{\,}_2 \times \mathbb{Z}^{\,}_2$ 
& $(n^{\,}_{4f},\, m^{\,}_{4f},\, \nu^{\,}_{4f})$
\\
$8g$ & 2D & $C_{1}$ 
& $\mathbb{Z}^{\,}_1$ 
& $\mathbb{Z} \times \mathbb{Z}^{\,}_2$ 
& $\mathbb{Z} \times \mathbb{Z}^{\,}_2$ 
& $(n^{\,}_{8g},\, x^{\,}_{8g})$
\\ \hline
\end{tabular*}
\caption{No.\ 11, $p4mm$.}
\label{tab:p4mm}
\end{table*}
    Equivalence relations: Type (i)
    \begin{align}       
        n^{\,}_{1a}-4k,\, n^{\,}_{1b}+4k\\ 
        n^{\,}_{1a}-4k,\, n^{\,}_{2c}+ 2k\\ 
        n^{\,}_{1a}-4k,\, n^{\,}_{4d}+ k \\
        n^{\,}_{1a}-4k,\, n^{\,}_{4e}+ k \\
        n^{\,}_{1a}-4k,\, n^{\,}_{4f}+ k \\
        n^{\,}_{1a}-4k,\, n^{\,}_{8g}+ k.
    \end{align}
    Type (ii)
     \begin{align}  
        -m^{\,}_{4d}\\
        -m^{\,}_{4e}\\
        -m^{\,}_{4f}.
    \end{align}
    
    Reduced indices:
    \begin{equation}
        \begin{split}
        (n^{\,}_{1a},\, n^{\,}_{1b},\, n^{\,}_{2c},\, n^{\,}_{4d},\, n^{\,}_{4e}, \, n^{\,}_{4f}, \, n^{\,}_{8g},\, m^{\,}_{4d},\, m^{\,}_{4e},\, m^{\,}_{4f}) \in \mathbb{Z}^7_{\,} \times (\mathbb{Z}^{\,}_2)^3 \rightarrow
(n^{\,}_{1a},\, n^{\,}_{1b},\, n^{\,}_{2c}) \in (\mathbb{Z}^{\,}_4)^2 \times \mathbb{Z}^{\,}_2
        \end{split}
    \end{equation} 
    
    \item ($p4gm$)

    \begin{table*}[h!]
\centering
\begin{tabular*}{\textwidth}{@{\extracolsep{\fill}}c c c c c c c}
\hline
WP & dim & site symm. & $G_{\mathrm{spa}}$ & Classification & Gappable decorations & Indices \\
\hline
$2a$ & 0D & $C_4$
& $\mathbb{Z}^{\mathrm{F}}_8$ 
& $\mathbb{Z} \times \mathbb{Z}^{\,}_2$ 
& $\mathbb{Z} \times \mathbb{Z}^{\,}_2$ 
& $(n^{\,}_{2a},\, r^{\,}_{2a})$
\\
$2b$ & 0D & $C_{2v}$
& $\mathbb{Z}^{\mathrm{F}}_{4} \rtimes \mathbb{Z}^{\mathrm{F}}_4/ \mathbb{Z}^{\mathrm{F}}_2$
& $\mathbb{Z} \times \mathbb{Z}^{\,}_2 \times \mathbb{Z}^{\,}_2$ 
& $\mathbb{Z} \times \mathbb{Z}^{\,}_2 \times \mathbb{Z}^{\,}_2$ 
& $(n^{\,}_{2b},\, r^{\,}_{2b},\, m^{\,}_{2b})$
\\
$4c$ & 1D & $C_s$
& $\mathbb{Z}^{\mathrm{F}}_4$ 
& $\mathbb{Z} \times \mathbb{Z}^{\,}_2 \times \mathbb{Z}^{\,}_2$ 
& $\mathbb{Z} \times \mathbb{Z}^{\,}_2 \times \mathbb{Z}^{\,}_2$ 
& $(n^{\,}_{4c},\, m^{\,}_{4c},\, \nu^{\,}_{4c})$
\\
$8d$ & 2D & $C_1$
& $\mathbb{Z}^{\,}_1$ 
& $\mathbb{Z} \times \mathbb{Z}^{\,}_2$ 
& $\mathbb{Z} \times \mathbb{Z}^{\,}_2$ 
& $(n^{\,}_{8d},\, x^{\,}_{8d})$
\\ \hline
\end{tabular*}
\caption{No.\ 12, $p4gm$.}
\label{tab:p4gm}
\end{table*}
    Equivalence relations: Type (i)
    \begin{align}
        n^{\,}_{2a}-4k,\, n^{\,}_{2b}+4k\\ 
        n^{\,}_{2a}-4k,\, n^{\,}_{4c}+ 2k\\ 
        n^{\,}_{2a}-4k,\, n^{\,}_{8d}+ k.
    \end{align}
    Type (ii)
    \begin{align}
        -m^{\,}_{4c}.
    \end{align}

    Reduced indices:
    \begin{equation}
        \begin{split}
        (n^{\,}_{2a},\, n^{\,}_{2b},\, n^{\,}_{4c},\, n^{\,}_{8d},\, m^{\,}_{4c}) \in \mathbb{Z}^4_{\,} \times \mathbb{Z}^{\,}_2\rightarrow
(n^{\,}_{2a},\, n^{\,}_{2b},\, n^{\,}_{4c}) \in (\mathbb{Z}^{\,}_4)^2 \times \mathbb{Z}^{\,}_2
        \end{split}
    \end{equation} 
    
    \item ($p3$)

    \begin{table*}[h!]
\centering
\begin{tabular*}{\textwidth}{@{\extracolsep{\fill}}c c c c c c c}
\hline
WP & dim & site symm. & $G_{\mathrm{spa}}$ & Classification & Gappable decorations & Indices \\
\hline
$1a$ & 0D & $C_3$ 
& $\mathbb{Z}^{\mathrm{F}}_6$ 
& $\mathbb{Z}$ 
& $\mathbb{Z}$ 
& $n^{\,}_{1a}$
\\
$1b$ & 0D & $C_3$ 
& $\mathbb{Z}^{\mathrm{F}}_6$ 
& $\mathbb{Z}$ 
& $\mathbb{Z}$ 
& $n^{\,}_{1b}$
\\
$1c$ & 0D & $C_3$ 
& $\mathbb{Z}^{\mathrm{F}}_6$ 
& $\mathbb{Z}$ 
& $\mathbb{Z}$ 
& $n^{\,}_{1c}$
\\
$3d$ & 2D & $C_1$ 
& $\mathbb{Z}^{\,}_1$ 
& $\mathbb{Z} \times \mathbb{Z}^{\,}_2 $ 
& $\mathbb{Z} \times \mathbb{Z}^{\,}_2 $ 
& $(n^{\,}_{3d},\, x^{\,}_{3d})$ 
\\ \hline
\end{tabular*}
\caption{No.\ 13, $p3$.}
\label{tab:p3}
\end{table*}
    Equivalence relations: Type (i)
    \begin{align}
            n^{\,}_{1a}-3k,\, n^{\,}_{1b}+3k\\ 
            n^{\,}_{1a}-3k,\, n^{\,}_{1c}+ 3k\\
            n^{\,}_{1a}-3k,\, n^{\,}_{3d}+ k.
    \end{align}
    
    Reduced indices:
    \begin{equation}
        \begin{split}
        (n^{\,}_{1a},\, n^{\,}_{1b},\, n^{\,}_{1c},\, n^{\,}_{3d}) \in (\mathbb{Z})^4 \rightarrow
    (n^{\,}_{1a},\, n^{\,}_{1b},\, n^{\,}_{1c}) \in (\mathbb{Z}_3)^3
        \end{split}
    \end{equation} 
    
    \item ($p3m1$)

\begin{table*}[h!]
\centering
\begin{tabular*}{\textwidth}{@{\extracolsep{\fill}}c c c c c c c}
\hline
WP & dim & site symm. & $G_{\mathrm{spa}}$ & Classification & Gappable decorations & Indices \\
\hline
$1a$ & 0D & $C_{3v}$ 
& $\mathbb{Z}^{\mathrm{F}}_6 \rtimes \mathbb{Z}^{\mathrm{F}}_4/\mathbb{Z}^{\mathrm{F}}_2$
& $\mathbb{Z} \times \mathbb{Z}^{\,}_2$
& $\mathbb{Z} \times \mathbb{Z}^{\,}_2$
& $(n^{\,}_{1a},\, m^{\,}_{1a})$
\\
$1b$ & 0D & $C_{3v}$ 
& $\mathbb{Z}^{\mathrm{F}}_6 \rtimes \mathbb{Z}^{\mathrm{F}}_4/\mathbb{Z}^{\mathrm{F}}_2$
& $\mathbb{Z} \times \mathbb{Z}^{\,}_2$
& $\mathbb{Z} \times \mathbb{Z}^{\,}_2$
& $(n^{\,}_{1b},\, m^{\,}_{1b})$
\\
$1c$ & 0D & $C_{3v}$ 
& $\mathbb{Z}^{\mathrm{F}}_6 \rtimes \mathbb{Z}^{\mathrm{F}}_4/\mathbb{Z}^{\mathrm{F}}_2$
& $\mathbb{Z} \times \mathbb{Z}^{\,}_2$
& $\mathbb{Z} \times \mathbb{Z}^{\,}_2$
& $(n^{\,}_{1c},\, m^{\,}_{1c})$
\\
$3d$ & 1D & $C_{s}$ 
& $\mathbb{Z}^{\mathrm{F}}_4$
& $\mathbb{Z} \times \mathbb{Z}^{\,}_2 \times \mathbb{Z}^{\,}_2$
& $\mathbb{Z} \times \mathbb{Z}^{\,}_2 \times \mathbb{Z}^{\,}_2$
& $(n^{\,}_{3d},\, m^{\,}_{3d},\, \nu^{\,}_{3d})$
\\
$6e$ & 2D & $C_{1}$ 
& $\mathbb{Z}^{\,}_1$ 
& $\mathbb{Z} \times \mathbb{Z}^{\,}_2$ 
& $\mathbb{Z} \times \mathbb{Z}^{\,}_2$
& $(n^{\,}_{6e},\, x^{\,}_{6e})$
\\ \hline
\end{tabular*}
\caption{No.\ 14, $p3m1$.}
\label{tab:p3m1}
\end{table*}
    Equivalence relations: Type (i)
    \begin{align}
        n^{\,}_{1a}-3k,\, n^{\,}_{1b}+3k\\ 
        n^{\,}_{1a}-3k,\, n^{\,}_{1c}+ 3k\\
        n^{\,}_{1a}-3k,\, n^{\,}_{3d} + k\\
        n^{\,}_{1a}-6k,\, n^{\,}_{6e} + k.
    \end{align}
    Type (ii)
    \begin{align}
        -m^{\,}_{1a},\,-m^{\,}_{3d}\\
        -m^{\,}_{1b},\,-m^{\,}_{3d}\\
        -m^{\,}_{1c},\,-m^{\,}_{3d}.
    \end{align}
    
    Reduced indices:
    \begin{equation}
        \begin{split}
        (n^{\,}_{1a},\, n^{\,}_{1b},\, n^{\,}_{1c},\, n^{\,}_{3d}, \, n^{\,}_{6e},\, m^{\,}_{1a},\, m^{\,}_{1b},\, m^{\,}_{1c}\, m^{\,}_{3d}) \in (\mathbb{Z})^5 \times (\mathbb{Z}_2)^4 \\ \rightarrow(n^{\,}_{1a},\, n^{\,}_{1b},\, n^{\,}_{1c}, \, m \equiv m^{\,}_{1a} m^{\,}_{1b} m^{\,}_{1c} m^{\,}_{3d}) \in (\mathbb{Z}^{\,}_3)^3 \times \mathbb{Z}^{\,}_2
        \end{split}
    \end{equation} 
    
    \item ($p31m$)

\begin{table*}[h!]
\centering
\begin{tabular*}{\textwidth}{@{\extracolsep{\fill}}c c c c c c c}
\hline
WP & dim & site symm. & $G_{\mathrm{spa}}$ & Classification & Gappable decorations & Indices \\
\hline
$1a$ & 0D & $C_{3v}$ 
& $\mathbb{Z}^{\mathrm{F}}_6 \rtimes \mathbb{Z}^{\mathrm{F}}_4/\mathbb{Z}^{\mathrm{F}}_2$ 
& $\mathbb{Z} \times \mathbb{Z}^{\,}_2$ 
& $\mathbb{Z} \times \mathbb{Z}^{\,}_2$
& $(n^{\,}_{1a},\, m^{\,}_{1a})$
\\
$2b$ & 0D & $C_{3}$ 
& $\mathbb{Z}^{\mathrm{F}}_6$ 
& $\mathbb{Z}$ 
& $\mathbb{Z}$ 
& $n^{\,}_{2b}$
\\
$3c$ & 1D & $C_{s}$ 
& $\mathbb{Z}^{\mathrm{F}}_2$ 
& $\mathbb{Z} \times \mathbb{Z}^{\,}_2 \times \mathbb{Z}^{\,}_2$ 
& $\mathbb{Z}\times \mathbb{Z}^{\,}_2 \times \mathbb{Z}^{\,}_2$ 
& $(n^{\,}_{3c},\ m^{\,}_{3c},\ \nu_{3c})$
\\
$6d$ & 2D & $C_{1}$ 
& $\mathbb{Z}^{\,}_1$ 
& $\mathbb{Z} \times \mathbb{Z}^{\,}_2$ 
& $\mathbb{Z} \times \mathbb{Z}^{\,}_2$ 
& $(n^{\,}_{6d}, x^{\,}_{6d})$
\\ \hline
\end{tabular*}
\caption{No.\ 15, $p31m$.}
\label{tab:p31m}
\end{table*}
    
    Equivalence relations: Type (i)
    \begin{align}
        n^{\,}_{1a}-6k,\, n^{\,}_{2b} + 3k\\ 
        n^{\,}_{1a}-3k,\, n^{\,}_{3c} + k\\
        n^{\,}_{1a}-6k,\, n^{\,}_{6d} +k.
    \end{align}
    Type (ii)
    \begin{align}
       -m^{\,}_{1a}, \, -m^{\,}_{3c}.
    \end{align}

    Reduced indices:
    \begin{equation}
        \begin{split}
        (n^{\,}_{1a},\, n^{\,}_{2b},\, n^{\,}_{3c},\, n^{\,}_{6d},\, m^{\,}_{1a},\, m^{\,}_{3c}) \in (\mathbb{Z})^4 \times (\mathbb{Z}_2)^2  \rightarrow
(n^{\,}_{1a},\, n^{\,}_{2b}, \, m\equiv m^{\,}_{1a}m^{\,}_{3c}) \in (\mathbb{Z}_3)^2 \times \mathbb{Z}_2
        \end{split}
    \end{equation} 
    
    \item ($p6$)

\begin{table*}[h!]
\centering
\begin{tabular*}{\textwidth}{@{\extracolsep{\fill}}c c c c c c c}
\hline
WP & dim & site symm. & $G_{\mathrm{spa}}$ & Classification & Gappable decorations & Indices \\
\hline
$1a$ & 0D & $C_6$ 
& $\mathbb{Z}^{\mathrm{F}}_{12}$ 
& $\mathbb{Z} \times \mathbb{Z}^{\,}_2$ 
& $\mathbb{Z} \times \mathbb{Z}^{\,}_2$ 
& $(n^{\,}_{1a},\, r^{\,}_{1a})$
\\
$2b$ & 0D & $C_3$ 
& $\mathbb{Z}^{\mathrm{F}}_6$ 
& $\mathbb{Z}$ 
& $\mathbb{Z}$ 
& $n^{\,}_{2b}$
\\
$3c$ & 0D & $C_2$ 
& $\mathbb{Z}^{\mathrm{F}}_4$
& $\mathbb{Z} \times \mathbb{Z}^{\,}_2$ 
& $\mathbb{Z} \times \mathbb{Z}^{\,}_2$
& $(n^{\,}_{3c},\, r^{\,}_{3c})$
\\
$6d$ & 2D & $C_{1}$ 
& $\mathbb{Z}^{\,}_1$ 
& $\mathbb{Z} \times \mathbb{Z}^{\,}_2$ 
& $\mathbb{Z} \times \mathbb{Z}^{\,}_2$ 
& $(n^{\,}_{6d},\, x^{\,}_{6d})$
\\ \hline
\end{tabular*}
\caption{No.\ 16, $p6$.}
\label{tab:p6}
\end{table*}
    Equivalence relations: Type (i)
    \begin{align}
        n^{\,}_{1a}-6k,\, n^{\,}_{2b} + 3k\\ 
        n^{\,}_{1a}-6k,\, n^{\,}_{3c} + 2k\\
        n^{\,}_{1a}-6k,\, n^{\,}_{6d} + k.
    \end{align}
    
    Reduced indices:
    \begin{equation}
        \begin{split}
       (n^{\,}_{1a},\, n^{\,}_{2b},\, n^{\,}_{3c},\, n^{\,}_{6d}) \in  (\mathbb{Z})^4 \rightarrow(n^{\,}_{1a},\, n^{\,}_{2b},\, n^{\,}_{3c}) \in \mathbb{Z}^{\,}_6 \times \mathbb{Z}^{\,}_3 \times \mathbb{Z}_2
        \end{split}
    \end{equation} 
    \clearpage
    \item ($p6mm$)
\begin{table*}[h!]
\centering
\begin{tabular*}{\textwidth}{@{\extracolsep{\fill}}c c c c c c c}
\hline
WP & dim & site symm. & $G_{\mathrm{spa}}$ & Classification & Gappable decorations & Indices \\
\hline
$1a$ & 0D & $C_{6v}$ 
& $\mathbb{Z}^{\mathrm{F}}_{12} \rtimes \mathbb{Z}^{\mathrm{F}}_4 \ \mathbb{Z}^{\mathrm{F}}_2$ 
& $\mathbb{Z} \times \mathbb{Z}_2 \times \mathbb{Z}_2$
& $\mathbb{Z} \times \mathbb{Z}_2 \times \mathbb{Z}_2$
& $(n_{1a},\, r_{1a},\, m_{1a})$
\\
$2b$ & 0D & $C_{3v}$ 
& $\mathbb{Z}^{\mathrm{F}}_{6} \rtimes \mathbb{Z}^{\mathrm{F}}_4 \ \mathbb{Z}^{\mathrm{F}}_2$ 
& $\mathbb{Z} \times \mathbb{Z}_2$
& $\mathbb{Z} \times \mathbb{Z}_2$
& $(n_{2b},\, m_{2b})$
\\
$3c$ & 0D & $C_{2v}$ 
& $\mathbb{Z}^{\mathrm{F}}_{4} \rtimes \mathbb{Z}^{\mathrm{F}}_4 \ \mathbb{Z}^{\mathrm{F}}_2$ 
& $\mathbb{Z} \times \mathbb{Z}_2 \times \mathbb{Z}_2$
& $\mathbb{Z} \times \mathbb{Z}_2 \times \mathbb{Z}_2$
& $(n_{3c},\, r_{3c},\, m_{3c})$
\\
$6d$ & 1D & $C_{s}$ 
& $\mathbb{Z}^{\mathrm{F}}_{4}$ 
& $\mathbb{Z} \times \mathbb{Z}_2 \times \mathbb{Z}_2$
& $\mathbb{Z} \times \mathbb{Z}_2 \times \mathbb{Z}_2$
& $(n_{6d},\, m_{6d},\, \nu_{6d})$
\\
$6e$ & 1D & $C_{s}$ 
& $\mathbb{Z}^{\mathrm{F}}_{4}$ 
& $\mathbb{Z} \times \mathbb{Z}_2 \times \mathbb{Z}_2$
& $\mathbb{Z} \times \mathbb{Z}_2 \times \mathbb{Z}_2$
& $(n_{6e},\, m_{6e},\, \nu_{6e})$
\\
$12f$ & 2D & $C_{1}$ 
& $\mathbb{Z}^{\,}_{1}$ 
& $\mathbb{Z} \times \mathbb{Z}_2$
& $\mathbb{Z} \times \mathbb{Z}_2$
& $(n_{12f},\, x_{12f})$
\\ \hline
\end{tabular*}
\caption{No.\ 17, $p6mm$.}
\label{tab:p6mm}
\end{table*}
    
    Equivalence relations: Type (i)
    \begin{align}
        n^{\,}_{1a}-6k,\, n^{\,}_{2b}+3k, \\
        n^{\,}_{1a}-6k,\, n^{\,}_{3c}+2k, \\
        n^{\,}_{1a}-6k,\, n^{\,}_{6d}+k, \\
        n^{\,}_{1a}-6k,\, n^{\,}_{6e}+k, \\
        n^{\,}_{1a}-12k,\, n^{\,}_{12f}+k.
    \end{align}
    Type (ii)
    \begin{align}
        -m^{\,}_{2b},\, -m^{\,}_{6e}\\
        -m^{\,}_{6e}\\
        -m^{\,}_{6d}.
    \end{align}
    While $m^{\,}_{2b}$'s sign can be flipped without altering any other index, $m^{\,}_{6e}$ and $m^{\,}_{6d}$ can be changed individually. Overall, this leads to trivialization of all these three indices.

    Reduced indices:
    \begin{equation}
        \begin{split}
        (n^{\,}_{1a}, \,n^{\,}_{2b}, \, n^{\,}_{3c}, \,n^{\,}_{6d},\, n^{\,}_{6e}, \,n^{\,}_{12f},\, m^{\,}_{2b},\, m^{\,}_{6e},\, m^{\,}_{6d}) \in \mathbb{Z}^6_{\,} \times (\mathbb{Z}^{\,}_2)^3  \rightarrow  (n^{\,}_{1a}, \, n^{\,}_{2b}, \, n^{\,}_{3c}) \in \mathbb{Z}^{\,}_6 \times \mathbb{Z}^{\,}_3 \times \mathbb{Z}^{\,}_2
        \end{split}
    \end{equation} 
\end{enumerate}

\clearpage
\subsection*{Classification of cFSPTs in wallpaper groups with corresponding indices}\label{App:full classif table}
Finally, Table~\ref{tab:final classification appendix} shows the classification for cFSPTs, for each of the 17 wallpaper groups in two dimensions.
\begin{table*}[h!]
\centering
\begin{tabular*}{\textwidth}{@{\extracolsep{\fill}}l l c c c}
\hline \hline
No.\ & Wallpaper &  Classification & Noninteracting classification & Indices\\
\hline
1 &  $p1$   
&  $\mathbb{Z} \times \magenta{\mathbb{Z}^{\,}_2}$ 
&  $\mathbb{Z} \times \magenta{\mathbb{Z}^{\,}_2}$ 
& $(n^{\,}_{1a}, \, \magenta{x^{\,}_{1a}})$
\\ \hline
2 & $p2$ 
& $\mathbb{Z}\times (\mathbb{Z}_{2})^4 \times \cyan{(\mathbb{Z}_{2})^4} \times \magenta{\mathbb{Z}^{\,}_2}$ 
& $\mathbb{Z} \times (\mathbb{Z}^{\,}_2)^4 \times \magenta{\mathbb{Z}^{\,}_2}$ 
& \makecell{$(n,\, n^{\,}_{1a}, n^{\,}_{1b}, n^{\,}_{1c},\, n^{\,}_{1d},$\\
$ \cyan{r^{\,}_{1a}}, \cyan{r^{\,}_{1b}}, \cyan{r^{\,}_{1c}}, \cyan{r^{\,}_{1d}}, \magenta{x^{\,}_{2e}})$}
\\ \hline
3 & $p1m1$ 
& $\mathbb{Z} \times (\mathbb{Z}^{\,}_2)^2 \times \cyan{(\mathbb{Z}^{\,}_2)^4} \times \magenta{\mathbb{Z}^{\,}_2}$ 
&  $\mathbb{Z} \times (\mathbb{Z}^{\,}_2)^2 \times \magenta{\mathbb{Z}^{\,}_2}$ 
& \makecell{$(n,\, n^{\,}_{1a}, \,n^{\,}_{1b}, \, \cyan{m^{\,}_{1a}}, \, \cyan{\cyan{m^{\,}_{1b}}},$\\ $\cyan{\nu^{\,}_{1a}}, \, \cyan{\nu^{\,}_{1b}}, \, \magenta{x^{\,}_{2c}})$} 
\\ \hline
4 & $p1g1$ 
& $\mathbb{Z} \times \magenta{\mathbb{Z}^{\,}_2}$ 
&  $\mathbb{Z} \times \magenta{\mathbb{Z}^{\,}_2}$ 
& $(n^{\,}_{2a}, \, x^{\,}_{2a})$ 
\\ \hline
5 & $c1m1$  
&  $\mathbb{Z} \times \mathbb{Z}^{\,}_2 \times \cyan{(\mathbb{Z}^{\,}_2)^2} \times \magenta{\mathbb{Z}^{\,}_2}$ 
& $\mathbb{Z} \times \mathbb{Z}^{\,}_2 \times \magenta{\mathbb{Z}^{\,}_2}$ 
        & $(n,\, n^{\,}_{2a},\, \cyan{m^{\,}_{2a}},\, \cyan{\nu^{\,}_{2a}},\, \magenta{x^{\,}_{4b}})$ 
\\ \hline
6 &   $p2mm$ 
& $\mathbb{Z} \times (\mathbb{Z}^{\,}_2)^{4} \times \cyan{(\mathbb{Z}^{\,}_2)^{12}} \times \magenta{\mathbb{Z}^{\,}_2}$ 
&  $\mathbb{Z} \times (\mathbb{Z}^{\,}_2)^4 \times \magenta{\mathbb{Z}^{\,}_2}$ 
& \makecell{$(n, n^{\,}_{1a}, n^{\,}_{1b},\, n^{\,}_{1c}, \, n^{\,}_{1d},\, \cyan{r^{\,}_{1a}},\, \cyan{m^{\,}_{1a}},$ \\ 
$\cyan{r^{\,}_{1b}},\, \cyan{m^{\,}_{1b}},\, \cyan{r^{\,}_{1c}},\, \cyan{m^{\,}_{1c}},\, \cyan{r^{\,}_{1d}},\, \cyan{m^{\,}_{1d}},$\\
$\cyan{\nu^{\,}_{2e}},\, \cyan{\nu^{\,}_{2f}},\, \cyan{\nu^{\,}_{2g}},\, \cyan{\nu^{\,}_{2h}}, \magenta{x^{\,}_{4i}})$ }
\\ \hline
7 & $p2mg$ 
& $\mathbb{Z} \times (\mathbb{Z}^{\,}_2)^3 \times \cyan{(\mathbb{Z}^{\,}_2)^4} \times \magenta{\mathbb{Z}^{\,}_2}$ 
& $\mathbb{Z} \times (\mathbb{Z}^{\,}_2)^3 \times \magenta{\mathbb{Z}^{\,}_2}$ 
& \makecell{$(n,\, n^{\,}_{2a},\, n^{\,}_{2b},\, n^{\,}_{2c},\, \cyan{r^{\,}_{2a}},\, \cyan{r^{\,}_{2b}},$\\
$\cyan{m^{\,}_{2c}},\, \cyan{\nu^{\,}_{2c}},\, \magenta{x^{\,}_{4d}})$}
\\ \hline
8 & $p2gg$ 
& $\mathbb{Z} \times (\mathbb{Z}^{\,}_2)^2 \times \cyan{(\mathbb{Z}^{\,}_2)^2} \times \magenta{\mathbb{Z}^{\,}_2}$ 
& $\mathbb{Z} \times (\mathbb{Z}^{\,}_2)^2 \times \magenta{\mathbb{Z}^{\,}_2}$ 
& \makecell{$(n,\, n^{\,}_{2a},\, n^{\,}_{2b},\, \cyan{r^{\,}_{2a}},\, \cyan{r^{\,}_{2b}},\, \magenta{x^{\,}_{4c}})$}
\\ \hline
9 & $c2mm$ 
& $\mathbb{Z} \times (\mathbb{Z}^{\,}_2)^{3} \times  \cyan{(\mathbb{Z}^{\,}_2)^{7}}  \times \magenta{\mathbb{Z}^{\,}_2}$ 
& $\mathbb{Z} \times (\mathbb{Z}^{\,}_2)^3 \times \magenta{\mathbb{Z}^{\,}_2}$ 
& \makecell{$(n,\, n^{\,}_{2a},\, n^{\,}_{2b},\,n^{\,}_{4c},$\\ $\cyan{r^{\,}_{2a}},\,\cyan{m^{\,}_{2a}},\, \cyan{r^{\,}_{2b}},\, \cyan{m^{\,}_{2b}},\, \cyan{r^{\,}_{4c}},$\\
$\cyan{\nu^{\,}_{4d}},\, \cyan{\nu^{\,}_{4e}},\, \magenta{x^{\,}_{8f}})$}
\\ \hline
10 & $p4$ 
& $\mathbb{Z} \times (\mathbb{Z}^{\,}_4)^2 \times \mathbb{Z}^{\,}_2 \times \cyan{(\mathbb{Z}^{\,}_2)^3} \times \magenta{\mathbb{Z}^{\,}_2}$
& $\mathbb{Z} \times (\mathbb{Z}^{\,}_4)^2 \times \mathbb{Z}^{\,}_2 \times \magenta{\mathbb{Z}^{\,}_2}$
& \makecell{
$(n,\, n^{\,}_{1a},\, n^{\,}_{1b},\, n^{\,}_{2c},$\\
$\cyan{r^{\,}_{1a}},\, \cyan{r^{\,}_{1b}},\, \cyan{r^{\,}_{2c}},\, \magenta{x^{\,}_{4d}})$} 
\\ \hline
11 & $p4mm$ 
& $\mathbb{Z} \times (\mathbb{Z}^{\,}_4)^2 \times \mathbb{Z}^{\,}_2 \times \cyan{(\mathbb{Z}^{\,}_2)^{9}}\times \magenta{\mathbb{Z}^{\,}_2}$
& $\mathbb{Z} \times (\mathbb{Z}^{\,}_4)^2 \times \mathbb{Z}^{\,}_2  \times \magenta{\mathbb{Z}^{\,}_2}$ 
&  \makecell{$(n,\, n^{\,}_{1a},\, n^{\,}_{1b},\, n^{\,}_{2c},$\\
$\cyan{r^{\,}_{1a}},\, \cyan{m^{\,}_{1a}},\, \cyan{r^{\,}_{1b}},\, \cyan{m^{\,}_{1b}},\, \cyan{r^{\,}_{2c}},\, \cyan{m^{\,}_{2c}},$ \\
$\cyan{\nu^{\,}_{4d}},\,\cyan{\nu^{\,}_{4e}},\, \cyan{\nu^{\,}_{4f}},\, \magenta{x^{\,}_{8g}})$ }
\\ \hline
12 & $p4gm$  
& $\mathbb{Z} \times (\mathbb{Z}^{\,}_4)^2 \times \mathbb{Z}^{\,}_2 \times \cyan{(\mathbb{Z}^{\,}_2)^{4}} \times \magenta{\mathbb{Z}^{\,}_2}$
&  $\mathbb{Z} \times (\mathbb{Z}^{\,}_4)^2 \times \mathbb{Z}^{\,}_2 \times \magenta{\mathbb{Z}^{\,}_2}$ 
&  \makecell{$(n,\, n^{\,}_{2a},\, n^{\,}_{2b},\, n^{\,}_{4c},$\\
$\cyan{r^{\,}_{2a}},\, \cyan{r^{\,}_{2b}},\, \cyan{m^{\,}_{2b}},\, \cyan{\nu^{\,}_{4c}},\, \magenta{x^{\,}_{8d}})$ }
\\ \hline
13 & $p3$ 
& $\mathbb{Z} \times (\mathbb{Z}^{\,}_3)^3 \times \magenta{\mathbb{Z}^{\,}_2}$ 
& $\mathbb{Z} \times (\mathbb{Z}^{\,}_3)^3 \times \magenta{\mathbb{Z}^{\,}_2}$ 
& $(n,\, n^{\,}_{1a},\, n^{\,}_{1b},\, n^{\,}_{1c},\, \magenta{x^{\,}_{3d}})$ 
\\ \hline
14 & $p3m1$  
& $\mathbb{Z} \times (\mathbb{Z}^{\,}_3)^3 \times \cyan{(\mathbb{Z}^{\,}_2)^2} \times \magenta{\mathbb{Z}^{\,}_2}$
& $\mathbb{Z} \times (\mathbb{Z}^{\,}_3)^3 \times \magenta{\mathbb{Z}^{\,}_2}$
& \makecell{$(n,\, n^{\,}_{1a},\, n^{\,}_{1b},\, n^{\,}_{1c},$ \\
$\cyan{m \equiv m^{\,}_{1a} m^{\,}_{1b} m^{\,}_{1c} m^{\,}_{3d}}, \,\cyan{\nu^{\,}_{3d}},\, \magenta{x^{\,}_{6e}})$}
\\ \hline
15 & $p31m$ 
& $\mathbb{Z} \times (\mathbb{Z}^{\,}_3)^2 \times \cyan{(\mathbb{Z}^{\,}_2)^2} \times \magenta{\mathbb{Z}^{\,}_2}$
& $\mathbb{Z} \times (\mathbb{Z}^{\,}_3)^2 \times \magenta{\mathbb{Z}^{\,}_2}$
& $(n,\, n^{\,}_{1a},\, n^{\,}_{2b},\, \cyan{m\equiv m^{\,}_{1a} m^{\,}_{3c}}, \cyan{\nu_{3c}},\, \magenta{x^{\,}_{6d}})$
\\ \hline
16 & $p6$  
& $\mathbb{Z} \times \mathbb{Z}^{\,}_6 \times \mathbb{Z}^{\,}_3 \times \mathbb{Z}^{\,}_2 \times \cyan{(\mathbb{Z}^{\,}_2)^2} \times \magenta{\mathbb{Z}^{\,}_2}$ 
& $\mathbb{Z} \times \mathbb{Z}^{\,}_6 \times \mathbb{Z}^{\,}_3 \times \mathbb{Z}^{\,}_2 \times \magenta{\mathbb{Z}^{\,}_2}$ 
& $(n,\, n^{\,}_{1a},\, n^{\,}_{2b},\, n_{3c},\, \cyan{r_{1a}},\, \cyan{r_{3c}},\, \magenta{x^{\,}_{6d}})$
\\ \hline
17 & $p6mm$  &  $\mathbb{Z}\times \mathbb{Z}^{\,}_6 \times \mathbb{Z}^{\,}_3 \times \mathbb{Z}^{\,}_2 \times \cyan{(\mathbb{Z}^{\,}_2)^6} \times \magenta{\mathbb{Z}^{\,}_2}$ 
&  $\mathbb{Z} \times \mathbb{Z}^{\,}_6 \times \mathbb{Z}^{\,}_3\times \mathbb{Z}^{\,}_2 \times \magenta{\mathbb{Z}^{\,}_2}$ 
& \makecell{$(n,\, n^{\,}_{1a}, n^{\,}_{2b},\, n^{\,}_{3c},\, \cyan{r^{\,}_{1a}},\, \cyan{m^{\,}_{1a}},\, \cyan{r^{\,}_{3c}},$\\
$\cyan{m^{\,}_{3c}},\, \cyan{\nu^{\,}_{6d}},\, \cyan{\nu^{\,}_{6e}},\, \magenta{x^{\,}_{12f}})$}
\\ \hline \hline
\end{tabular*}
\caption{Classification of cFSPTs in wallpaper groups. Every wallpaper group has a $\mathbb{Z}$ factor in its classification due to the total number of Kramers' pairs in the unit-cell (with index $n$), and a $\mathbb{Z}_2$ factor (with index $x_{wp}$, with $wp$ the general Wyckoff position label) deriving from the intrinsic 2D SPT phase that does not require crystalline symmetries to be realized (colored in \magenta{magenta}). Terms in the classification and indices colored in black correspond to phases that exist both in the absence or presence of interactions. The terms corresponding to intrinsically interacting phases in the classification and their corresponding indices are highlighted in \cyan{cyan}. Note that the ``Indices" column refers to the ``Classification" column, and the ordering is consistent between the two.}
\label{tab:final classification appendix}
\end{table*}

\end{appendix}

\twocolumngrid
\bibliography{references}

\begin{thebibliography}{137}%
\makeatletter
\providecommand \@ifxundefined [1]{%
 \@ifx{#1\undefined}
}%
\providecommand \@ifnum [1]{%
 \ifnum #1\expandafter \@firstoftwo
 \else \expandafter \@secondoftwo
 \fi
}%
\providecommand \@ifx [1]{%
 \ifx #1\expandafter \@firstoftwo
 \else \expandafter \@secondoftwo
 \fi
}%
\providecommand \natexlab [1]{#1}%
\providecommand \enquote  [1]{``#1''}%
\providecommand \bibnamefont  [1]{#1}%
\providecommand \bibfnamefont [1]{#1}%
\providecommand \citenamefont [1]{#1}%
\providecommand \href@noop [0]{\@secondoftwo}%
\providecommand \href [0]{\begingroup \@sanitize@url \@href}%
\providecommand \@href[1]{\@@startlink{#1}\@@href}%
\providecommand \@@href[1]{\endgroup#1\@@endlink}%
\providecommand \@sanitize@url [0]{\catcode `\\12\catcode `\$12\catcode
  `\&12\catcode `\#12\catcode `\^12\catcode `\_12\catcode `\%12\relax}%
\providecommand \@@startlink[1]{}%
\providecommand \@@endlink[0]{}%
\providecommand \url  [0]{\begingroup\@sanitize@url \@url }%
\providecommand \@url [1]{\endgroup\@href {#1}{\urlprefix }}%
\providecommand \urlprefix  [0]{URL }%
\providecommand \Eprint [0]{\href }%
\providecommand \doibase [0]{https://doi.org/}%
\providecommand \selectlanguage [0]{\@gobble}%
\providecommand \bibinfo  [0]{\@secondoftwo}%
\providecommand \bibfield  [0]{\@secondoftwo}%
\providecommand \translation [1]{[#1]}%
\providecommand \BibitemOpen [0]{}%
\providecommand \bibitemStop [0]{}%
\providecommand \bibitemNoStop [0]{.\EOS\space}%
\providecommand \EOS [0]{\spacefactor3000\relax}%
\providecommand \BibitemShut  [1]{\csname bibitem#1\endcsname}%
\let\auto@bib@innerbib\@empty
\bibitem [{\citenamefont {Schnyder}\ \emph {et~al.}(2009)\citenamefont
  {Schnyder}, \citenamefont {Ryu}, \citenamefont {Furusaki},\ and\
  \citenamefont {Ludwig}}]{Schnyder2009}%
  \BibitemOpen
  \bibfield  {author} {\bibinfo {author} {\bibfnamefont {A.~P.}\ \bibnamefont
  {Schnyder}}, \bibinfo {author} {\bibfnamefont {S.}~\bibnamefont {Ryu}},
  \bibinfo {author} {\bibfnamefont {A.}~\bibnamefont {Furusaki}},\ and\
  \bibinfo {author} {\bibfnamefont {A.~W.~W.}\ \bibnamefont {Ludwig}},\
  }\bibfield  {title} {\bibinfo {title} {{Classification of Topological
  Insulators and Superconductors}},\ }\href {https://doi.org/10.1063/1.3149481}
  {\bibfield  {journal} {\bibinfo  {journal} {AIP Conference Proceedings}\
  }\textbf {\bibinfo {volume} {1134}},\ \bibinfo {pages} {10} (\bibinfo {year}
  {2009})},\ \Eprint
  {https://arxiv.org/abs/https://pubs.aip.org/aip/acp/article-pdf/1134/1/10/11584263/10\_1\_online.pdf}
  {https://pubs.aip.org/aip/acp/article-pdf/1134/1/10/11584263/10\_1\_online.pdf}
  \BibitemShut {NoStop}%
\bibitem [{\citenamefont {Kitaev}(2009)}]{Kitaev2009}%
  \BibitemOpen
  \bibfield  {author} {\bibinfo {author} {\bibfnamefont {A.}~\bibnamefont
  {Kitaev}},\ }\bibfield  {title} {\bibinfo {title} {{Periodic table for
  topological insulators and superconductors}},\ }\href
  {https://doi.org/10.1063/1.3149495} {\bibfield  {journal} {\bibinfo
  {journal} {AIP Conference Proceedings}\ }\textbf {\bibinfo {volume} {1134}},\
  \bibinfo {pages} {22} (\bibinfo {year} {2009})},\ \Eprint
  {https://arxiv.org/abs/https://pubs.aip.org/aip/acp/article-pdf/1134/1/22/11584243/22\_1\_online.pdf}
  {https://pubs.aip.org/aip/acp/article-pdf/1134/1/22/11584243/22\_1\_online.pdf}
  \BibitemShut {NoStop}%
\bibitem [{\citenamefont {{Ryu}}\ \emph {et~al.}(2010)\citenamefont {{Ryu}},
  \citenamefont {{Schnyder}}, \citenamefont {{Furusaki}},\ and\ \citenamefont
  {{Ludwig}}}]{Ryu2010}%
  \BibitemOpen
  \bibfield  {author} {\bibinfo {author} {\bibfnamefont {S.}~\bibnamefont
  {{Ryu}}}, \bibinfo {author} {\bibfnamefont {A.~P.}\ \bibnamefont
  {{Schnyder}}}, \bibinfo {author} {\bibfnamefont {A.}~\bibnamefont
  {{Furusaki}}},\ and\ \bibinfo {author} {\bibfnamefont {A.~W.~W.}\
  \bibnamefont {{Ludwig}}},\ }\bibfield  {title} {\bibinfo {title}
  {{Topological insulators and superconductors: tenfold way and dimensional
  hierarchy}},\ }\href {https://doi.org/10.1088/1367-2630/12/6/065010}
  {\bibfield  {journal} {\bibinfo  {journal} {New Journal of Physics}\ }\textbf
  {\bibinfo {volume} {12}},\ \bibinfo {eid} {065010} (\bibinfo {year}
  {2010})},\ \Eprint {https://arxiv.org/abs/0912.2157} {arXiv:0912.2157
  [cond-mat.mes-hall]} \BibitemShut {NoStop}%
\bibitem [{\citenamefont {v.Klitzing}\ \emph {et~al.}(1980)\citenamefont
  {v.Klitzing}, \citenamefont {Dorda},\ and\ \citenamefont
  {Pepper}}]{Klitzing1980}%
  \BibitemOpen
  \bibfield  {author} {\bibinfo {author} {\bibfnamefont {K.}~\bibnamefont
  {v.Klitzing}}, \bibinfo {author} {\bibfnamefont {G.}~\bibnamefont {Dorda}},\
  and\ \bibinfo {author} {\bibfnamefont {M.}~\bibnamefont {Pepper}},\
  }\bibfield  {title} {\bibinfo {title} {{New Method for High-Accuracy
  Determination of the Fine-Structure Constant Based on Quantized Hall
  Resistance}},\ }\href {https://doi.org/10.1103/PhysRevLett.45.494} {\bibfield
   {journal} {\bibinfo  {journal} {Phys. Rev. Lett.}\ }\textbf {\bibinfo
  {volume} {45}},\ \bibinfo {pages} {494} (\bibinfo {year} {1980})}\BibitemShut
  {NoStop}%
\bibitem [{\citenamefont {Chen}\ \emph {et~al.}(2010)\citenamefont {Chen},
  \citenamefont {Gu},\ and\ \citenamefont {Wen}}]{Chen2010}%
  \BibitemOpen
  \bibfield  {author} {\bibinfo {author} {\bibfnamefont {X.}~\bibnamefont
  {Chen}}, \bibinfo {author} {\bibfnamefont {Z.-C.}\ \bibnamefont {Gu}},\ and\
  \bibinfo {author} {\bibfnamefont {X.-G.}\ \bibnamefont {Wen}},\ }\bibfield
  {title} {\bibinfo {title} {Local unitary transformation, long-range quantum
  entanglement, wave function renormalization, and topological order},\ }\href
  {https://doi.org/10.1103/PhysRevB.82.155138} {\bibfield  {journal} {\bibinfo
  {journal} {Phys. Rev. B}\ }\textbf {\bibinfo {volume} {82}},\ \bibinfo
  {pages} {155138} (\bibinfo {year} {2010})}\BibitemShut {NoStop}%
\bibitem [{\citenamefont {Freed}\ and\ \citenamefont
  {Hopkins}(2021)}]{Freed2021}%
  \BibitemOpen
  \bibfield  {author} {\bibinfo {author} {\bibfnamefont {D.~S.}\ \bibnamefont
  {Freed}}\ and\ \bibinfo {author} {\bibfnamefont {M.~J.}\ \bibnamefont
  {Hopkins}},\ }\bibfield  {title} {\bibinfo {title} {Reflection positivity and
  invertible topological phases},\ }\href@noop {} {\bibfield  {journal}
  {\bibinfo  {journal} {Geometry \& Topology}\ }\textbf {\bibinfo {volume}
  {25}},\ \bibinfo {pages} {1165} (\bibinfo {year} {2021})}\BibitemShut
  {NoStop}%
\bibitem [{\citenamefont {Bernevig}\ \emph {et~al.}(2006)\citenamefont
  {Bernevig}, \citenamefont {Hughes},\ and\ \citenamefont
  {Zhang}}]{Bernevig2006}%
  \BibitemOpen
  \bibfield  {author} {\bibinfo {author} {\bibfnamefont {B.~A.}\ \bibnamefont
  {Bernevig}}, \bibinfo {author} {\bibfnamefont {T.~L.}\ \bibnamefont
  {Hughes}},\ and\ \bibinfo {author} {\bibfnamefont {S.-C.}\ \bibnamefont
  {Zhang}},\ }\bibfield  {title} {\bibinfo {title} {{Quantum Spin Hall Effect
  and Topological Phase Transition in HgTe Quantum Wells}},\ }\href
  {https://doi.org/10.1126/science.1133734} {\bibfield  {journal} {\bibinfo
  {journal} {Science}\ }\textbf {\bibinfo {volume} {314}},\ \bibinfo {pages}
  {1757–1761} (\bibinfo {year} {2006})}\BibitemShut {NoStop}%
\bibitem [{\citenamefont {König}\ \emph {et~al.}(2007)\citenamefont {König},
  \citenamefont {Wiedmann}, \citenamefont {Brüne}, \citenamefont {Roth},
  \citenamefont {Buhmann}, \citenamefont {Molenkamp}, \citenamefont {Qi},\ and\
  \citenamefont {Zhang}}]{Konig2007}%
  \BibitemOpen
  \bibfield  {author} {\bibinfo {author} {\bibfnamefont {M.}~\bibnamefont
  {König}}, \bibinfo {author} {\bibfnamefont {S.}~\bibnamefont {Wiedmann}},
  \bibinfo {author} {\bibfnamefont {C.}~\bibnamefont {Brüne}}, \bibinfo
  {author} {\bibfnamefont {A.}~\bibnamefont {Roth}}, \bibinfo {author}
  {\bibfnamefont {H.}~\bibnamefont {Buhmann}}, \bibinfo {author} {\bibfnamefont
  {L.~W.}\ \bibnamefont {Molenkamp}}, \bibinfo {author} {\bibfnamefont {X.-L.}\
  \bibnamefont {Qi}},\ and\ \bibinfo {author} {\bibfnamefont {S.-C.}\
  \bibnamefont {Zhang}},\ }\bibfield  {title} {\bibinfo {title} {{Quantum Spin
  Hall Insulator State in HgTe Quantum Wells}},\ }\href
  {https://doi.org/10.1126/science.1148047} {\bibfield  {journal} {\bibinfo
  {journal} {Science}\ }\textbf {\bibinfo {volume} {318}},\ \bibinfo {pages}
  {766–770} (\bibinfo {year} {2007})}\BibitemShut {NoStop}%
\bibitem [{\citenamefont {Roth}\ \emph {et~al.}(2009)\citenamefont {Roth},
  \citenamefont {Brüne}, \citenamefont {Buhmann}, \citenamefont {Molenkamp},
  \citenamefont {Maciejko}, \citenamefont {Qi},\ and\ \citenamefont
  {Zhang}}]{Roth2009}%
  \BibitemOpen
  \bibfield  {author} {\bibinfo {author} {\bibfnamefont {A.}~\bibnamefont
  {Roth}}, \bibinfo {author} {\bibfnamefont {C.}~\bibnamefont {Brüne}},
  \bibinfo {author} {\bibfnamefont {H.}~\bibnamefont {Buhmann}}, \bibinfo
  {author} {\bibfnamefont {L.~W.}\ \bibnamefont {Molenkamp}}, \bibinfo {author}
  {\bibfnamefont {J.}~\bibnamefont {Maciejko}}, \bibinfo {author}
  {\bibfnamefont {X.-L.}\ \bibnamefont {Qi}},\ and\ \bibinfo {author}
  {\bibfnamefont {S.-C.}\ \bibnamefont {Zhang}},\ }\bibfield  {title} {\bibinfo
  {title} {{Nonlocal Transport in the Quantum Spin Hall State}},\ }\href
  {https://doi.org/10.1126/science.1174736} {\bibfield  {journal} {\bibinfo
  {journal} {Science}\ }\textbf {\bibinfo {volume} {325}},\ \bibinfo {pages}
  {294–297} (\bibinfo {year} {2009})}\BibitemShut {NoStop}%
\bibitem [{\citenamefont {Hsieh}\ \emph {et~al.}(2009)\citenamefont {Hsieh},
  \citenamefont {Qian}, \citenamefont {Wray}, \citenamefont {Xia},
  \citenamefont {Hor}, \citenamefont {Cava},\ and\ \citenamefont
  {Hasan}}]{hsieh2009}%
  \BibitemOpen
  \bibfield  {author} {\bibinfo {author} {\bibfnamefont {D.}~\bibnamefont
  {Hsieh}}, \bibinfo {author} {\bibfnamefont {D.}~\bibnamefont {Qian}},
  \bibinfo {author} {\bibfnamefont {L.}~\bibnamefont {Wray}}, \bibinfo {author}
  {\bibfnamefont {Y.}~\bibnamefont {Xia}}, \bibinfo {author} {\bibfnamefont
  {Y.~S.}\ \bibnamefont {Hor}}, \bibinfo {author} {\bibfnamefont {R.~J.}\
  \bibnamefont {Cava}},\ and\ \bibinfo {author} {\bibfnamefont {M.~Z.}\
  \bibnamefont {Hasan}},\ }\href@noop {} {\bibinfo {title} {{A topological
  Dirac insulator in a quantum spin Hall phase (experimental realization of a
  3D Topological Insulator)}}} (\bibinfo {year} {2009}),\ \Eprint
  {https://arxiv.org/abs/0910.2420} {arXiv:0910.2420 [cond-mat.mes-hall]}
  \BibitemShut {NoStop}%
\bibitem [{\citenamefont {{Xia}}\ \emph {et~al.}(2009)\citenamefont {{Xia}},
  \citenamefont {{Qian}}, \citenamefont {{Hsieh}}, \citenamefont {{Wray}},
  \citenamefont {{Pal}}, \citenamefont {{Lin}}, \citenamefont {{Bansil}},
  \citenamefont {{Grauer}}, \citenamefont {{Hor}}, \citenamefont {{Cava}},\
  and\ \citenamefont {{Hasan}}}]{Xia2009}%
  \BibitemOpen
  \bibfield  {author} {\bibinfo {author} {\bibfnamefont {Y.}~\bibnamefont
  {{Xia}}}, \bibinfo {author} {\bibfnamefont {D.}~\bibnamefont {{Qian}}},
  \bibinfo {author} {\bibfnamefont {D.}~\bibnamefont {{Hsieh}}}, \bibinfo
  {author} {\bibfnamefont {L.}~\bibnamefont {{Wray}}}, \bibinfo {author}
  {\bibfnamefont {A.}~\bibnamefont {{Pal}}}, \bibinfo {author} {\bibfnamefont
  {H.}~\bibnamefont {{Lin}}}, \bibinfo {author} {\bibfnamefont
  {A.}~\bibnamefont {{Bansil}}}, \bibinfo {author} {\bibfnamefont
  {D.}~\bibnamefont {{Grauer}}}, \bibinfo {author} {\bibfnamefont {Y.~S.}\
  \bibnamefont {{Hor}}}, \bibinfo {author} {\bibfnamefont {R.~J.}\ \bibnamefont
  {{Cava}}},\ and\ \bibinfo {author} {\bibfnamefont {M.~Z.}\ \bibnamefont
  {{Hasan}}},\ }\bibfield  {title} {\bibinfo {title} {{Observation of a
  large-gap topological-insulator class with a single Dirac cone on the
  surface}},\ }\href {https://doi.org/10.1038/nphys1274} {\bibfield  {journal}
  {\bibinfo  {journal} {Nature Physics}\ }\textbf {\bibinfo {volume} {5}},\
  \bibinfo {pages} {398} (\bibinfo {year} {2009})},\ \Eprint
  {https://arxiv.org/abs/0908.3513} {arXiv:0908.3513} \BibitemShut {NoStop}%
\bibitem [{\citenamefont {Fu}\ and\ \citenamefont {Kane}(2007)}]{Fu2007}%
  \BibitemOpen
  \bibfield  {author} {\bibinfo {author} {\bibfnamefont {L.}~\bibnamefont
  {Fu}}\ and\ \bibinfo {author} {\bibfnamefont {C.~L.}\ \bibnamefont {Kane}},\
  }\bibfield  {title} {\bibinfo {title} {Topological insulators with inversion
  symmetry},\ }\href {https://doi.org/10.1103/PhysRevB.76.045302} {\bibfield
  {journal} {\bibinfo  {journal} {Phys. Rev. B}\ }\textbf {\bibinfo {volume}
  {76}},\ \bibinfo {pages} {045302} (\bibinfo {year} {2007})}\BibitemShut
  {NoStop}%
\bibitem [{\citenamefont {Fu}(2011)}]{Fu2011}%
  \BibitemOpen
  \bibfield  {author} {\bibinfo {author} {\bibfnamefont {L.}~\bibnamefont
  {Fu}},\ }\bibfield  {title} {\bibinfo {title} {Topological crystalline
  insulators},\ }\href {https://doi.org/10.1103/PhysRevLett.106.106802}
  {\bibfield  {journal} {\bibinfo  {journal} {Phys. Rev. Lett.}\ }\textbf
  {\bibinfo {volume} {106}},\ \bibinfo {pages} {106802} (\bibinfo {year}
  {2011})}\BibitemShut {NoStop}%
\bibitem [{\citenamefont {Slager}\ \emph {et~al.}(2013)\citenamefont {Slager},
  \citenamefont {Mesaros}, \citenamefont {Juricic},\ and\ \citenamefont
  {Zaanen}}]{Slager2013}%
  \BibitemOpen
  \bibfield  {author} {\bibinfo {author} {\bibfnamefont {R.-J.}\ \bibnamefont
  {Slager}}, \bibinfo {author} {\bibfnamefont {A.}~\bibnamefont {Mesaros}},
  \bibinfo {author} {\bibfnamefont {V.}~\bibnamefont {Juricic}},\ and\ \bibinfo
  {author} {\bibfnamefont {J.}~\bibnamefont {Zaanen}},\ }\bibfield  {title}
  {\bibinfo {title} {The space group classification of topological
  band-insulators},\ }\href {https://doi.org/10.1038/nphys2513} {\bibfield
  {journal} {\bibinfo  {journal} {Nature Physics}\ }\textbf {\bibinfo {volume}
  {9}},\ \bibinfo {pages} {98} (\bibinfo {year} {2013})}\BibitemShut {NoStop}%
\bibitem [{\citenamefont {Zhang}\ \emph {et~al.}(2013)\citenamefont {Zhang},
  \citenamefont {Kane},\ and\ \citenamefont {Mele}}]{Zhang2013}%
  \BibitemOpen
  \bibfield  {author} {\bibinfo {author} {\bibfnamefont {F.}~\bibnamefont
  {Zhang}}, \bibinfo {author} {\bibfnamefont {C.~L.}\ \bibnamefont {Kane}},\
  and\ \bibinfo {author} {\bibfnamefont {E.~J.}\ \bibnamefont {Mele}},\
  }\bibfield  {title} {\bibinfo {title} {Topological mirror
  superconductivity},\ }\href {https://doi.org/10.1103/PhysRevLett.111.056403}
  {\bibfield  {journal} {\bibinfo  {journal} {Phys. Rev. Lett.}\ }\textbf
  {\bibinfo {volume} {111}},\ \bibinfo {pages} {056403} (\bibinfo {year}
  {2013})}\BibitemShut {NoStop}%
\bibitem [{\citenamefont {Chiu}\ \emph {et~al.}(2013)\citenamefont {Chiu},
  \citenamefont {Yao},\ and\ \citenamefont {Ryu}}]{Chiu2013}%
  \BibitemOpen
  \bibfield  {author} {\bibinfo {author} {\bibfnamefont {C.-K.}\ \bibnamefont
  {Chiu}}, \bibinfo {author} {\bibfnamefont {H.}~\bibnamefont {Yao}},\ and\
  \bibinfo {author} {\bibfnamefont {S.}~\bibnamefont {Ryu}},\ }\bibfield
  {title} {\bibinfo {title} {Classification of topological insulators and
  superconductors in the presence of reflection symmetry},\ }\href
  {https://doi.org/10.1103/PhysRevB.88.075142} {\bibfield  {journal} {\bibinfo
  {journal} {Phys. Rev. B}\ }\textbf {\bibinfo {volume} {88}},\ \bibinfo
  {pages} {075142} (\bibinfo {year} {2013})}\BibitemShut {NoStop}%
\bibitem [{\citenamefont {Morimoto}\ and\ \citenamefont
  {Furusaki}(2013)}]{Morimoto2013}%
  \BibitemOpen
  \bibfield  {author} {\bibinfo {author} {\bibfnamefont {T.}~\bibnamefont
  {Morimoto}}\ and\ \bibinfo {author} {\bibfnamefont {A.}~\bibnamefont
  {Furusaki}},\ }\bibfield  {title} {\bibinfo {title} {Topological
  classification with additional symmetries from clifford algebras},\ }\href
  {https://doi.org/10.1103/PhysRevB.88.125129} {\bibfield  {journal} {\bibinfo
  {journal} {Phys. Rev. B}\ }\textbf {\bibinfo {volume} {88}},\ \bibinfo
  {pages} {125129} (\bibinfo {year} {2013})}\BibitemShut {NoStop}%
\bibitem [{\citenamefont {Fang}\ \emph {et~al.}(2013)\citenamefont {Fang},
  \citenamefont {Gilbert},\ and\ \citenamefont {Bernevig}}]{Fang2013}%
  \BibitemOpen
  \bibfield  {author} {\bibinfo {author} {\bibfnamefont {C.}~\bibnamefont
  {Fang}}, \bibinfo {author} {\bibfnamefont {M.~J.}\ \bibnamefont {Gilbert}},\
  and\ \bibinfo {author} {\bibfnamefont {B.~A.}\ \bibnamefont {Bernevig}},\
  }\bibfield  {title} {\bibinfo {title} {Entanglement spectrum classification
  of ${C}_{n}$-invariant noninteracting topological insulators in two
  dimensions},\ }\href {https://doi.org/10.1103/PhysRevB.87.035119} {\bibfield
  {journal} {\bibinfo  {journal} {Phys. Rev. B}\ }\textbf {\bibinfo {volume}
  {87}},\ \bibinfo {pages} {035119} (\bibinfo {year} {2013})}\BibitemShut
  {NoStop}%
\bibitem [{\citenamefont {Jadaun}\ \emph {et~al.}(2013)\citenamefont {Jadaun},
  \citenamefont {Xiao}, \citenamefont {Niu},\ and\ \citenamefont
  {Banerjee}}]{Jadaun2013}%
  \BibitemOpen
  \bibfield  {author} {\bibinfo {author} {\bibfnamefont {P.}~\bibnamefont
  {Jadaun}}, \bibinfo {author} {\bibfnamefont {D.}~\bibnamefont {Xiao}},
  \bibinfo {author} {\bibfnamefont {Q.}~\bibnamefont {Niu}},\ and\ \bibinfo
  {author} {\bibfnamefont {S.~K.}\ \bibnamefont {Banerjee}},\ }\bibfield
  {title} {\bibinfo {title} {Topological classification of crystalline
  insulators with space group symmetry},\ }\href
  {https://doi.org/10.1103/PhysRevB.88.085110} {\bibfield  {journal} {\bibinfo
  {journal} {Phys. Rev. B}\ }\textbf {\bibinfo {volume} {88}},\ \bibinfo
  {pages} {085110} (\bibinfo {year} {2013})}\BibitemShut {NoStop}%
\bibitem [{\citenamefont {Koshino}\ \emph {et~al.}(2014)\citenamefont
  {Koshino}, \citenamefont {Morimoto},\ and\ \citenamefont
  {Sato}}]{Koshino2014}%
  \BibitemOpen
  \bibfield  {author} {\bibinfo {author} {\bibfnamefont {M.}~\bibnamefont
  {Koshino}}, \bibinfo {author} {\bibfnamefont {T.}~\bibnamefont {Morimoto}},\
  and\ \bibinfo {author} {\bibfnamefont {M.}~\bibnamefont {Sato}},\ }\bibfield
  {title} {\bibinfo {title} {Topological zero modes and dirac points protected
  by spatial symmetry and chiral symmetry},\ }\href
  {https://doi.org/10.1103/PhysRevB.90.115207} {\bibfield  {journal} {\bibinfo
  {journal} {Phys. Rev. B}\ }\textbf {\bibinfo {volume} {90}},\ \bibinfo
  {pages} {115207} (\bibinfo {year} {2014})}\BibitemShut {NoStop}%
\bibitem [{\citenamefont {Liu}\ \emph {et~al.}(2014)\citenamefont {Liu},
  \citenamefont {Zhang},\ and\ \citenamefont {VanLeeuwen}}]{Liu2014}%
  \BibitemOpen
  \bibfield  {author} {\bibinfo {author} {\bibfnamefont {C.-X.}\ \bibnamefont
  {Liu}}, \bibinfo {author} {\bibfnamefont {R.-X.}\ \bibnamefont {Zhang}},\
  and\ \bibinfo {author} {\bibfnamefont {B.~K.}\ \bibnamefont {VanLeeuwen}},\
  }\bibfield  {title} {\bibinfo {title} {Topological nonsymmorphic crystalline
  insulators},\ }\href {https://doi.org/10.1103/PhysRevB.90.085304} {\bibfield
  {journal} {\bibinfo  {journal} {Phys. Rev. B}\ }\textbf {\bibinfo {volume}
  {90}},\ \bibinfo {pages} {085304} (\bibinfo {year} {2014})}\BibitemShut
  {NoStop}%
\bibitem [{\citenamefont {Shiozaki}\ and\ \citenamefont
  {Sato}(2014)}]{Shiozaki2014}%
  \BibitemOpen
  \bibfield  {author} {\bibinfo {author} {\bibfnamefont {K.}~\bibnamefont
  {Shiozaki}}\ and\ \bibinfo {author} {\bibfnamefont {M.}~\bibnamefont
  {Sato}},\ }\bibfield  {title} {\bibinfo {title} {Topology of crystalline
  insulators and superconductors},\ }\href
  {https://doi.org/10.1103/PhysRevB.90.165114} {\bibfield  {journal} {\bibinfo
  {journal} {Phys. Rev. B}\ }\textbf {\bibinfo {volume} {90}},\ \bibinfo
  {pages} {165114} (\bibinfo {year} {2014})}\BibitemShut {NoStop}%
\bibitem [{\citenamefont {Ando}\ and\ \citenamefont {Fu}(2015)}]{Ando2015}%
  \BibitemOpen
  \bibfield  {author} {\bibinfo {author} {\bibfnamefont {Y.}~\bibnamefont
  {Ando}}\ and\ \bibinfo {author} {\bibfnamefont {L.}~\bibnamefont {Fu}},\
  }\bibfield  {title} {\bibinfo {title} {Topological crystalline insulators and
  topological superconductors: From concepts to materials},\ }\href
  {https://doi.org/10.1146/annurev-conmatphys-031214-014501} {\bibfield
  {journal} {\bibinfo  {journal} {Annual Review of Condensed Matter Physics}\
  }\textbf {\bibinfo {volume} {6}},\ \bibinfo {pages} {361} (\bibinfo {year}
  {2015})}\BibitemShut {NoStop}%
\bibitem [{\citenamefont {Trifunovic}\ and\ \citenamefont
  {Brouwer}(2017)}]{Trifunovic2017}%
  \BibitemOpen
  \bibfield  {author} {\bibinfo {author} {\bibfnamefont {L.}~\bibnamefont
  {Trifunovic}}\ and\ \bibinfo {author} {\bibfnamefont {P.}~\bibnamefont
  {Brouwer}},\ }\bibfield  {title} {\bibinfo {title} {Bott periodicity for the
  topological classification of gapped states of matter with reflection
  symmetry},\ }\href {https://doi.org/10.1103/PhysRevB.96.195109} {\bibfield
  {journal} {\bibinfo  {journal} {Phys. Rev. B}\ }\textbf {\bibinfo {volume}
  {96}},\ \bibinfo {pages} {195109} (\bibinfo {year} {2017})}\BibitemShut
  {NoStop}%
\bibitem [{\citenamefont {Benalcazar}\ \emph {et~al.}(2017)\citenamefont
  {Benalcazar}, \citenamefont {Bernevig},\ and\ \citenamefont
  {Hughes}}]{Benalcazar2017}%
  \BibitemOpen
  \bibfield  {author} {\bibinfo {author} {\bibfnamefont {W.~A.}\ \bibnamefont
  {Benalcazar}}, \bibinfo {author} {\bibfnamefont {B.~A.}\ \bibnamefont
  {Bernevig}},\ and\ \bibinfo {author} {\bibfnamefont {T.~L.}\ \bibnamefont
  {Hughes}},\ }\bibfield  {title} {\bibinfo {title} {{Electric multipole
  moments, topological multipole moment pumping, and chiral hinge states in
  crystalline insulators}},\ }\href
  {https://doi.org/10.1103/PhysRevB.96.245115} {\bibfield  {journal} {\bibinfo
  {journal} {Phys. Rev. B}\ }\textbf {\bibinfo {volume} {96}},\ \bibinfo
  {pages} {245115} (\bibinfo {year} {2017})}\BibitemShut {NoStop}%
\bibitem [{\citenamefont {Schindler}\ \emph
  {et~al.}(2018{\natexlab{a}})\citenamefont {Schindler}, \citenamefont {Cook},
  \citenamefont {Vergniory}, \citenamefont {Wang}, \citenamefont {Parkin},
  \citenamefont {Bernevig},\ and\ \citenamefont {Neupert}}]{Schindler2018}%
  \BibitemOpen
  \bibfield  {author} {\bibinfo {author} {\bibfnamefont {F.}~\bibnamefont
  {Schindler}}, \bibinfo {author} {\bibfnamefont {A.~M.}\ \bibnamefont {Cook}},
  \bibinfo {author} {\bibfnamefont {M.~G.}\ \bibnamefont {Vergniory}}, \bibinfo
  {author} {\bibfnamefont {Z.}~\bibnamefont {Wang}}, \bibinfo {author}
  {\bibfnamefont {S.~S.~P.}\ \bibnamefont {Parkin}}, \bibinfo {author}
  {\bibfnamefont {B.~A.}\ \bibnamefont {Bernevig}},\ and\ \bibinfo {author}
  {\bibfnamefont {T.}~\bibnamefont {Neupert}},\ }\bibfield  {title} {\bibinfo
  {title} {Higher-order topological insulators},\ }\bibfield  {journal}
  {\bibinfo  {journal} {Science Advances}\ }\textbf {\bibinfo {volume} {4}},\
  \href {https://doi.org/10.1126/sciadv.aat0346} {10.1126/sciadv.aat0346}
  (\bibinfo {year} {2018}{\natexlab{a}})\BibitemShut {NoStop}%
\bibitem [{\citenamefont {Trifunovic}\ and\ \citenamefont
  {Brouwer}(2019)}]{Trifunovic2019}%
  \BibitemOpen
  \bibfield  {author} {\bibinfo {author} {\bibfnamefont {L.}~\bibnamefont
  {Trifunovic}}\ and\ \bibinfo {author} {\bibfnamefont {P.~W.}\ \bibnamefont
  {Brouwer}},\ }\bibfield  {title} {\bibinfo {title} {{Higher-Order
  Bulk-Boundary Correspondence for Topological Crystalline Phases}},\ }\href
  {https://doi.org/10.1103/PhysRevX.9.011012} {\bibfield  {journal} {\bibinfo
  {journal} {Phys. Rev. X}\ }\textbf {\bibinfo {volume} {9}},\ \bibinfo {pages}
  {011012} (\bibinfo {year} {2019})}\BibitemShut {NoStop}%
\bibitem [{\citenamefont {Fang}\ and\ \citenamefont {Fu}(2019)}]{Fang2019}%
  \BibitemOpen
  \bibfield  {author} {\bibinfo {author} {\bibfnamefont {C.}~\bibnamefont
  {Fang}}\ and\ \bibinfo {author} {\bibfnamefont {L.}~\bibnamefont {Fu}},\
  }\bibfield  {title} {\bibinfo {title} {{New classes of topological
  crystalline insulators having surface rotation anomaly}},\ }\bibfield
  {journal} {\bibinfo  {journal} {Science Advances}\ }\textbf {\bibinfo
  {volume} {5}},\ \href {https://doi.org/10.1126/sciadv.aat2374}
  {10.1126/sciadv.aat2374} (\bibinfo {year} {2019})\BibitemShut {NoStop}%
\bibitem [{\citenamefont {Song}\ \emph
  {et~al.}(2017{\natexlab{a}})\citenamefont {Song}, \citenamefont {Fang},\ and\
  \citenamefont {Fang}}]{Song2017hoti}%
  \BibitemOpen
  \bibfield  {author} {\bibinfo {author} {\bibfnamefont {Z.}~\bibnamefont
  {Song}}, \bibinfo {author} {\bibfnamefont {Z.}~\bibnamefont {Fang}},\ and\
  \bibinfo {author} {\bibfnamefont {C.}~\bibnamefont {Fang}},\ }\bibfield
  {title} {\bibinfo {title} {{$(d\ensuremath{-}2)$-Dimensional Edge States of
  Rotation Symmetry Protected Topological States}},\ }\href
  {https://doi.org/10.1103/PhysRevLett.119.246402} {\bibfield  {journal}
  {\bibinfo  {journal} {Phys. Rev. Lett.}\ }\textbf {\bibinfo {volume} {119}},\
  \bibinfo {pages} {246402} (\bibinfo {year} {2017}{\natexlab{a}})}\BibitemShut
  {NoStop}%
\bibitem [{\citenamefont {Po}\ \emph {et~al.}(2017)\citenamefont {Po},
  \citenamefont {Vishwanath},\ and\ \citenamefont {Watanabe}}]{Po2017}%
  \BibitemOpen
  \bibfield  {author} {\bibinfo {author} {\bibfnamefont {H.~C.}\ \bibnamefont
  {Po}}, \bibinfo {author} {\bibfnamefont {A.}~\bibnamefont {Vishwanath}},\
  and\ \bibinfo {author} {\bibfnamefont {H.}~\bibnamefont {Watanabe}},\
  }\bibfield  {title} {\bibinfo {title} {{Symmetry-based indicators of band
  topology in the 230 space groups}},\ }\bibfield  {journal} {\bibinfo
  {journal} {Nature Communications}\ }\textbf {\bibinfo {volume} {8}},\ \href
  {https://doi.org/10.1038/s41467-017-00133-2} {10.1038/s41467-017-00133-2}
  (\bibinfo {year} {2017})\BibitemShut {NoStop}%
\bibitem [{\citenamefont {Po}(2020)}]{Po2020}%
  \BibitemOpen
  \bibfield  {author} {\bibinfo {author} {\bibfnamefont {H.~C.}\ \bibnamefont
  {Po}},\ }\bibfield  {title} {\bibinfo {title} {{Symmetry indicators of band
  topology}},\ }\href {https://doi.org/10.1088/1361-648x/ab7adb} {\bibfield
  {journal} {\bibinfo  {journal} {Journal of Physics: Condensed Matter}\
  }\textbf {\bibinfo {volume} {32}},\ \bibinfo {pages} {263001} (\bibinfo
  {year} {2020})}\BibitemShut {NoStop}%
\bibitem [{\citenamefont {Bradlyn}\ \emph {et~al.}(2017)\citenamefont
  {Bradlyn}, \citenamefont {Elcoro}, \citenamefont {Cano}, \citenamefont
  {Vergniory}, \citenamefont {Wang}, \citenamefont {Felser}, \citenamefont
  {Aroyo},\ and\ \citenamefont {Bernevig}}]{Bradlyn2017}%
  \BibitemOpen
  \bibfield  {author} {\bibinfo {author} {\bibfnamefont {B.}~\bibnamefont
  {Bradlyn}}, \bibinfo {author} {\bibfnamefont {L.}~\bibnamefont {Elcoro}},
  \bibinfo {author} {\bibfnamefont {J.}~\bibnamefont {Cano}}, \bibinfo {author}
  {\bibfnamefont {M.~G.}\ \bibnamefont {Vergniory}}, \bibinfo {author}
  {\bibfnamefont {Z.}~\bibnamefont {Wang}}, \bibinfo {author} {\bibfnamefont
  {C.}~\bibnamefont {Felser}}, \bibinfo {author} {\bibfnamefont {M.~I.}\
  \bibnamefont {Aroyo}},\ and\ \bibinfo {author} {\bibfnamefont {B.~A.}\
  \bibnamefont {Bernevig}},\ }\bibfield  {title} {\bibinfo {title} {Topological
  quantum chemistry},\ }\href {https://doi.org/10.1038/nature23268} {\bibfield
  {journal} {\bibinfo  {journal} {Nature}\ }\textbf {\bibinfo {volume} {547}},\
  \bibinfo {pages} {298} (\bibinfo {year} {2017})}\BibitemShut {NoStop}%
\bibitem [{\citenamefont {Kruthoff}\ \emph {et~al.}(2017)\citenamefont
  {Kruthoff}, \citenamefont {de~Boer}, \citenamefont {van Wezel}, \citenamefont
  {Kane},\ and\ \citenamefont {Slager}}]{Kruthoff2017}%
  \BibitemOpen
  \bibfield  {author} {\bibinfo {author} {\bibfnamefont {J.}~\bibnamefont
  {Kruthoff}}, \bibinfo {author} {\bibfnamefont {J.}~\bibnamefont {de~Boer}},
  \bibinfo {author} {\bibfnamefont {J.}~\bibnamefont {van Wezel}}, \bibinfo
  {author} {\bibfnamefont {C.~L.}\ \bibnamefont {Kane}},\ and\ \bibinfo
  {author} {\bibfnamefont {R.-J.}\ \bibnamefont {Slager}},\ }\bibfield  {title}
  {\bibinfo {title} {{Topological Classification of Crystalline Insulators
  through Band Structure Combinatorics}},\ }\href
  {https://doi.org/10.1103/PhysRevX.7.041069} {\bibfield  {journal} {\bibinfo
  {journal} {Phys. Rev. X}\ }\textbf {\bibinfo {volume} {7}},\ \bibinfo {pages}
  {041069} (\bibinfo {year} {2017})}\BibitemShut {NoStop}%
\bibitem [{\citenamefont {Vergniory}\ \emph {et~al.}(2019)\citenamefont
  {Vergniory}, \citenamefont {Elcoro}, \citenamefont {Felser} \emph
  {et~al.}}]{Vergniory2019}%
  \BibitemOpen
  \bibfield  {author} {\bibinfo {author} {\bibfnamefont {M.}~\bibnamefont
  {Vergniory}}, \bibinfo {author} {\bibfnamefont {L.}~\bibnamefont {Elcoro}},
  \bibinfo {author} {\bibfnamefont {C.}~\bibnamefont {Felser}}, \emph
  {et~al.},\ }\bibfield  {title} {\bibinfo {title} {{A complete catalogue of
  high-quality topological materials}},\ }\href
  {https://doi.org/https://doi.org/10.1038/s41586-019-0954-4} {\bibfield
  {journal} {\bibinfo  {journal} {Nature}\ }\textbf {\bibinfo {volume} {566}},\
  \bibinfo {pages} {480–485} (\bibinfo {year} {2019})}\BibitemShut {NoStop}%
\bibitem [{\citenamefont {Cano}\ and\ \citenamefont
  {Bradlyn}(2021)}]{Cano2021}%
  \BibitemOpen
  \bibfield  {author} {\bibinfo {author} {\bibfnamefont {J.}~\bibnamefont
  {Cano}}\ and\ \bibinfo {author} {\bibfnamefont {B.}~\bibnamefont {Bradlyn}},\
  }\bibfield  {title} {\bibinfo {title} {{Band Representations and Topological
  Quantum Chemistry}},\ }\href
  {https://doi.org/10.1146/annurev-conmatphys-041720-124134} {\bibfield
  {journal} {\bibinfo  {journal} {Annual Review of Condensed Matter Physics}\
  }\textbf {\bibinfo {volume} {12}},\ \bibinfo {pages} {225} (\bibinfo {year}
  {2021})},\ \Eprint
  {https://arxiv.org/abs/https://doi.org/10.1146/annurev-conmatphys-041720-124134}
  {https://doi.org/10.1146/annurev-conmatphys-041720-124134} \BibitemShut
  {NoStop}%
\bibitem [{\citenamefont {{Elcoro}}\ \emph {et~al.}(2021)\citenamefont
  {{Elcoro}}, \citenamefont {{Wieder}}, \citenamefont {{Song}}, \citenamefont
  {{Xu}}, \citenamefont {{Bradlyn}},\ and\ \citenamefont
  {{Bernevig}}}]{Elcoro2021}%
  \BibitemOpen
  \bibfield  {author} {\bibinfo {author} {\bibfnamefont {L.}~\bibnamefont
  {{Elcoro}}}, \bibinfo {author} {\bibfnamefont {B.~J.}\ \bibnamefont
  {{Wieder}}}, \bibinfo {author} {\bibfnamefont {Z.}~\bibnamefont {{Song}}},
  \bibinfo {author} {\bibfnamefont {Y.}~\bibnamefont {{Xu}}}, \bibinfo {author}
  {\bibfnamefont {B.}~\bibnamefont {{Bradlyn}}},\ and\ \bibinfo {author}
  {\bibfnamefont {B.~A.}\ \bibnamefont {{Bernevig}}},\ }\bibfield  {title}
  {\bibinfo {title} {{Magnetic topological quantum chemistry}},\ }\href
  {https://doi.org/10.1038/s41467-021-26241-8} {\bibfield  {journal} {\bibinfo
  {journal} {Nature Communications}\ }\textbf {\bibinfo {volume} {12}},\
  \bibinfo {eid} {5965} (\bibinfo {year} {2021})},\ \Eprint
  {https://arxiv.org/abs/2010.00598} {arXiv:2010.00598} \BibitemShut {NoStop}%
\bibitem [{\citenamefont {Schindler}\ \emph
  {et~al.}(2018{\natexlab{b}})\citenamefont {Schindler}, \citenamefont {Wang},
  \citenamefont {Vergniory}, \citenamefont {Cook}, \citenamefont {Murani},
  \citenamefont {Sengupta}, \citenamefont {Kasumov}, \citenamefont {Deblock},
  \citenamefont {Jeon}, \citenamefont {Drozdov}, \citenamefont {Bouchiat},
  \citenamefont {Guéron}, \citenamefont {Yazdani}, \citenamefont {Bernevig},\
  and\ \citenamefont {Neupert}}]{Schindler2018Bismuth}%
  \BibitemOpen
  \bibfield  {author} {\bibinfo {author} {\bibfnamefont {F.}~\bibnamefont
  {Schindler}}, \bibinfo {author} {\bibfnamefont {Z.}~\bibnamefont {Wang}},
  \bibinfo {author} {\bibfnamefont {M.~G.}\ \bibnamefont {Vergniory}}, \bibinfo
  {author} {\bibfnamefont {A.~M.}\ \bibnamefont {Cook}}, \bibinfo {author}
  {\bibfnamefont {A.}~\bibnamefont {Murani}}, \bibinfo {author} {\bibfnamefont
  {S.}~\bibnamefont {Sengupta}}, \bibinfo {author} {\bibfnamefont {A.~Y.}\
  \bibnamefont {Kasumov}}, \bibinfo {author} {\bibfnamefont {R.}~\bibnamefont
  {Deblock}}, \bibinfo {author} {\bibfnamefont {S.}~\bibnamefont {Jeon}},
  \bibinfo {author} {\bibfnamefont {I.}~\bibnamefont {Drozdov}}, \bibinfo
  {author} {\bibfnamefont {H.}~\bibnamefont {Bouchiat}}, \bibinfo {author}
  {\bibfnamefont {S.}~\bibnamefont {Guéron}}, \bibinfo {author} {\bibfnamefont
  {A.}~\bibnamefont {Yazdani}}, \bibinfo {author} {\bibfnamefont {B.~A.}\
  \bibnamefont {Bernevig}},\ and\ \bibinfo {author} {\bibfnamefont
  {T.}~\bibnamefont {Neupert}},\ }\bibfield  {title} {\bibinfo {title}
  {{Higher-order topology in bismuth}},\ }\href
  {https://doi.org/10.1038/s41567-018-0224-7} {\bibfield  {journal} {\bibinfo
  {journal} {Nature Physics}\ }\textbf {\bibinfo {volume} {14}},\ \bibinfo
  {pages} {918–924} (\bibinfo {year} {2018}{\natexlab{b}})}\BibitemShut
  {NoStop}%
\bibitem [{\citenamefont {{Chuang-Han Hsu and Xiaoting Zhou and Qiong Ma and
  Nuh Gedik and Arun Bansil and Vitor M Pereira and Hsin Lin and Liang Fu and
  Su-Yang Xu and Tay-Rong Chang}}(2019)}]{Hsu2019}%
  \BibitemOpen
  \bibfield  {author} {\bibinfo {author} {\bibfnamefont {t.~.~P.}\ \bibnamefont
  {{Chuang-Han Hsu and Xiaoting Zhou and Qiong Ma and Nuh Gedik and Arun Bansil
  and Vitor M Pereira and Hsin Lin and Liang Fu and Su-Yang Xu and Tay-Rong
  Chang}}},\ }\href {https://doi.org/10.1088/2053-1583/ab1607} {\bibfield
  {journal} {\bibinfo  {journal} {2D Materials}\ }\textbf {\bibinfo {volume}
  {6}},\ \bibinfo {pages} {031004} (\bibinfo {year} {2019})}\BibitemShut
  {NoStop}%
\bibitem [{\citenamefont {Yoon}\ \emph {et~al.}(2020)\citenamefont {Yoon},
  \citenamefont {Liu}, \citenamefont {Min},\ and\ \citenamefont
  {Zhang}}]{Yoon2020}%
  \BibitemOpen
  \bibfield  {author} {\bibinfo {author} {\bibfnamefont {C.}~\bibnamefont
  {Yoon}}, \bibinfo {author} {\bibfnamefont {C.-C.}\ \bibnamefont {Liu}},
  \bibinfo {author} {\bibfnamefont {H.}~\bibnamefont {Min}},\ and\ \bibinfo
  {author} {\bibfnamefont {F.}~\bibnamefont {Zhang}},\ }\href@noop {} {\bibinfo
  {title} {{Quasi-One-Dimensional Higher-Order Topological Insulators}}}
  (\bibinfo {year} {2020}),\ \Eprint {https://arxiv.org/abs/2005.14710}
  {arXiv:2005.14710 [cond-mat.mes-hall]} \BibitemShut {NoStop}%
\bibitem [{\citenamefont {Aggarwal}\ \emph {et~al.}(2021)\citenamefont
  {Aggarwal}, \citenamefont {Zhu}, \citenamefont {Hughes},\ and\ \citenamefont
  {Madhavan}}]{Aggarwal2021}%
  \BibitemOpen
  \bibfield  {author} {\bibinfo {author} {\bibfnamefont {L.}~\bibnamefont
  {Aggarwal}}, \bibinfo {author} {\bibfnamefont {P.}~\bibnamefont {Zhu}},
  \bibinfo {author} {\bibfnamefont {T.~L.}\ \bibnamefont {Hughes}},\ and\
  \bibinfo {author} {\bibfnamefont {V.}~\bibnamefont {Madhavan}},\ }\bibfield
  {title} {\bibinfo {title} {{Evidence for higher order topology in Bi and
  Bi0.92Sb0.08}},\ }\bibfield  {journal} {\bibinfo  {journal} {Nature
  Communications}\ }\textbf {\bibinfo {volume} {12}},\ \href
  {https://doi.org/10.1038/s41467-021-24683-8} {10.1038/s41467-021-24683-8}
  (\bibinfo {year} {2021})\BibitemShut {NoStop}%
\bibitem [{\citenamefont {{Shumiya}}\ \emph {et~al.}(2022)\citenamefont
  {{Shumiya}}, \citenamefont {{Hossain}}, \citenamefont {{Yin}}, \citenamefont
  {{Wang}}, \citenamefont {{Litskevich}}, \citenamefont {{Yoon}}, \citenamefont
  {{Li}}, \citenamefont {{Yang}}, \citenamefont {{Jiang}}, \citenamefont
  {{Cheng}}, \citenamefont {{Lin}}, \citenamefont {{Zhang}}, \citenamefont
  {{Cheng}}, \citenamefont {{Cochran}}, \citenamefont {{Multer}}, \citenamefont
  {{Yang}}, \citenamefont {{Casas}}, \citenamefont {{Chang}}, \citenamefont
  {{Neupert}}, \citenamefont {{Yuan}}, \citenamefont {{Jia}}, \citenamefont
  {{Lin}}, \citenamefont {{Yao}}, \citenamefont {{Balicas}}, \citenamefont
  {{Zhang}}, \citenamefont {{Yao}},\ and\ \citenamefont
  {{Hasan}}}]{Shumiya2022}%
  \BibitemOpen
  \bibfield  {author} {\bibinfo {author} {\bibfnamefont {N.}~\bibnamefont
  {{Shumiya}}}, \bibinfo {author} {\bibfnamefont {M.~S.}\ \bibnamefont
  {{Hossain}}}, \bibinfo {author} {\bibfnamefont {J.-X.}\ \bibnamefont
  {{Yin}}}, \bibinfo {author} {\bibfnamefont {Z.}~\bibnamefont {{Wang}}},
  \bibinfo {author} {\bibfnamefont {M.}~\bibnamefont {{Litskevich}}}, \bibinfo
  {author} {\bibfnamefont {C.}~\bibnamefont {{Yoon}}}, \bibinfo {author}
  {\bibfnamefont {Y.}~\bibnamefont {{Li}}}, \bibinfo {author} {\bibfnamefont
  {Y.}~\bibnamefont {{Yang}}}, \bibinfo {author} {\bibfnamefont {Y.-X.}\
  \bibnamefont {{Jiang}}}, \bibinfo {author} {\bibfnamefont {G.}~\bibnamefont
  {{Cheng}}}, \bibinfo {author} {\bibfnamefont {Y.-C.}\ \bibnamefont {{Lin}}},
  \bibinfo {author} {\bibfnamefont {Q.}~\bibnamefont {{Zhang}}}, \bibinfo
  {author} {\bibfnamefont {Z.-J.}\ \bibnamefont {{Cheng}}}, \bibinfo {author}
  {\bibfnamefont {T.~A.}\ \bibnamefont {{Cochran}}}, \bibinfo {author}
  {\bibfnamefont {D.}~\bibnamefont {{Multer}}}, \bibinfo {author}
  {\bibfnamefont {X.~P.}\ \bibnamefont {{Yang}}}, \bibinfo {author}
  {\bibfnamefont {B.}~\bibnamefont {{Casas}}}, \bibinfo {author} {\bibfnamefont
  {T.-R.}\ \bibnamefont {{Chang}}}, \bibinfo {author} {\bibfnamefont
  {T.}~\bibnamefont {{Neupert}}}, \bibinfo {author} {\bibfnamefont
  {Z.}~\bibnamefont {{Yuan}}}, \bibinfo {author} {\bibfnamefont
  {S.}~\bibnamefont {{Jia}}}, \bibinfo {author} {\bibfnamefont
  {H.}~\bibnamefont {{Lin}}}, \bibinfo {author} {\bibfnamefont
  {N.}~\bibnamefont {{Yao}}}, \bibinfo {author} {\bibfnamefont
  {L.}~\bibnamefont {{Balicas}}}, \bibinfo {author} {\bibfnamefont
  {F.}~\bibnamefont {{Zhang}}}, \bibinfo {author} {\bibfnamefont
  {Y.}~\bibnamefont {{Yao}}},\ and\ \bibinfo {author} {\bibfnamefont {M.~Z.}\
  \bibnamefont {{Hasan}}},\ }\bibfield  {title} {\bibinfo {title} {{Evidence of
  a room-temperature quantum spin Hall edge state in a higher-order topological
  insulator}},\ }\href {https://doi.org/10.1038/s41563-022-01304-3} {\bibfield
  {journal} {\bibinfo  {journal} {Nature Materials}\ }\textbf {\bibinfo
  {volume} {21}},\ \bibinfo {pages} {1111} (\bibinfo {year} {2022})},\ \Eprint
  {https://arxiv.org/abs/2110.05718} {arXiv:2110.05718 [cond-mat.mes-hall]}
  \BibitemShut {NoStop}%
\bibitem [{\citenamefont {{Serra-Garcia}}\ \emph {et~al.}(2018)\citenamefont
  {{Serra-Garcia}}, \citenamefont {{Peri}}, \citenamefont {{S{\"u}sstrunk}},
  \citenamefont {{Bilal}}, \citenamefont {{Larsen}}, \citenamefont
  {{Villanueva}},\ and\ \citenamefont {{Huber}}}]{Serra-Garcia2018}%
  \BibitemOpen
  \bibfield  {author} {\bibinfo {author} {\bibfnamefont {M.}~\bibnamefont
  {{Serra-Garcia}}}, \bibinfo {author} {\bibfnamefont {V.}~\bibnamefont
  {{Peri}}}, \bibinfo {author} {\bibfnamefont {R.}~\bibnamefont
  {{S{\"u}sstrunk}}}, \bibinfo {author} {\bibfnamefont {O.~R.}\ \bibnamefont
  {{Bilal}}}, \bibinfo {author} {\bibfnamefont {T.}~\bibnamefont {{Larsen}}},
  \bibinfo {author} {\bibfnamefont {L.~G.}\ \bibnamefont {{Villanueva}}},\ and\
  \bibinfo {author} {\bibfnamefont {S.~D.}\ \bibnamefont {{Huber}}},\
  }\bibfield  {title} {\bibinfo {title} {{Observation of a phononic quadrupole
  topological insulator}},\ }\href {https://doi.org/10.1038/nature25156}
  {\bibfield  {journal} {\bibinfo  {journal} {\nat}\ }\textbf {\bibinfo
  {volume} {555}},\ \bibinfo {pages} {342} (\bibinfo {year} {2018})},\ \Eprint
  {https://arxiv.org/abs/1708.05015} {arXiv:1708.05015} \BibitemShut {NoStop}%
\bibitem [{\citenamefont {Soldini}\ \emph
  {et~al.}(2023{\natexlab{a}})\citenamefont {Soldini}, \citenamefont {Küster},
  \citenamefont {Wagner}, \citenamefont {Das}, \citenamefont {Aldarawsheh},
  \citenamefont {Thomale}, \citenamefont {Lounis}, \citenamefont {Parkin},
  \citenamefont {Sessi},\ and\ \citenamefont {Neupert}}]{Soldini2023Shiba}%
  \BibitemOpen
  \bibfield  {author} {\bibinfo {author} {\bibfnamefont {M.~O.}\ \bibnamefont
  {Soldini}}, \bibinfo {author} {\bibfnamefont {F.}~\bibnamefont {Küster}},
  \bibinfo {author} {\bibfnamefont {G.}~\bibnamefont {Wagner}}, \bibinfo
  {author} {\bibfnamefont {S.}~\bibnamefont {Das}}, \bibinfo {author}
  {\bibfnamefont {A.}~\bibnamefont {Aldarawsheh}}, \bibinfo {author}
  {\bibfnamefont {R.}~\bibnamefont {Thomale}}, \bibinfo {author} {\bibfnamefont
  {S.}~\bibnamefont {Lounis}}, \bibinfo {author} {\bibfnamefont {S.~S.~P.}\
  \bibnamefont {Parkin}}, \bibinfo {author} {\bibfnamefont {P.}~\bibnamefont
  {Sessi}},\ and\ \bibinfo {author} {\bibfnamefont {T.}~\bibnamefont
  {Neupert}},\ }\bibfield  {title} {\bibinfo {title} {{Two-dimensional Shiba
  lattices as a possible platform for crystalline topological
  superconductivity}},\ }\href {https://doi.org/10.1038/s41567-023-02104-5}
  {\bibfield  {journal} {\bibinfo  {journal} {Nature Physics}\ }\textbf
  {\bibinfo {volume} {19}},\ \bibinfo {pages} {1848–1854} (\bibinfo {year}
  {2023}{\natexlab{a}})}\BibitemShut {NoStop}%
\bibitem [{\citenamefont {Wang}\ \emph {et~al.}(2024)\citenamefont {Wang},
  \citenamefont {Fan}, \citenamefont {Chen}, \citenamefont {Jiang},
  \citenamefont {Gao}, \citenamefont {Lado},\ and\ \citenamefont
  {Yang}}]{wang2024}%
  \BibitemOpen
  \bibfield  {author} {\bibinfo {author} {\bibfnamefont {H.}~\bibnamefont
  {Wang}}, \bibinfo {author} {\bibfnamefont {P.}~\bibnamefont {Fan}}, \bibinfo
  {author} {\bibfnamefont {J.}~\bibnamefont {Chen}}, \bibinfo {author}
  {\bibfnamefont {L.}~\bibnamefont {Jiang}}, \bibinfo {author} {\bibfnamefont
  {H.-J.}\ \bibnamefont {Gao}}, \bibinfo {author} {\bibfnamefont {J.~L.}\
  \bibnamefont {Lado}},\ and\ \bibinfo {author} {\bibfnamefont
  {K.}~\bibnamefont {Yang}},\ }\href@noop {} {\bibinfo {title} {Realizing
  topological quantum magnets with atomic spins on surfaces}} (\bibinfo {year}
  {2024}),\ \Eprint {https://arxiv.org/abs/2403.14145} {arXiv:2403.14145
  [cond-mat.mes-hall]} \BibitemShut {NoStop}%
\bibitem [{\citenamefont {Fidkowski}\ and\ \citenamefont
  {Kitaev}(2010)}]{Fidkowski2010}%
  \BibitemOpen
  \bibfield  {author} {\bibinfo {author} {\bibfnamefont {L.}~\bibnamefont
  {Fidkowski}}\ and\ \bibinfo {author} {\bibfnamefont {A.}~\bibnamefont
  {Kitaev}},\ }\bibfield  {title} {\bibinfo {title} {Effects of interactions on
  the topological classification of free fermion systems},\ }\href
  {https://doi.org/10.1103/PhysRevB.81.134509} {\bibfield  {journal} {\bibinfo
  {journal} {Phys. Rev. B}\ }\textbf {\bibinfo {volume} {81}},\ \bibinfo
  {pages} {134509} (\bibinfo {year} {2010})}\BibitemShut {NoStop}%
\bibitem [{\citenamefont {Fidkowski}\ and\ \citenamefont
  {Kitaev}(2011)}]{Fidkowski2011}%
  \BibitemOpen
  \bibfield  {author} {\bibinfo {author} {\bibfnamefont {L.}~\bibnamefont
  {Fidkowski}}\ and\ \bibinfo {author} {\bibfnamefont {A.}~\bibnamefont
  {Kitaev}},\ }\bibfield  {title} {\bibinfo {title} {Topological phases of
  fermions in one dimension},\ }\href
  {https://doi.org/10.1103/PhysRevB.83.075103} {\bibfield  {journal} {\bibinfo
  {journal} {Phys. Rev. B}\ }\textbf {\bibinfo {volume} {83}},\ \bibinfo
  {pages} {075103} (\bibinfo {year} {2011})}\BibitemShut {NoStop}%
\bibitem [{\citenamefont {Ryu}\ and\ \citenamefont {Zhang}(2012)}]{Ryu2012}%
  \BibitemOpen
  \bibfield  {author} {\bibinfo {author} {\bibfnamefont {S.}~\bibnamefont
  {Ryu}}\ and\ \bibinfo {author} {\bibfnamefont {S.-C.}\ \bibnamefont
  {Zhang}},\ }\bibfield  {title} {\bibinfo {title} {Interacting topological
  phases and modular invariance},\ }\href
  {https://doi.org/10.1103/PhysRevB.85.245132} {\bibfield  {journal} {\bibinfo
  {journal} {Phys. Rev. B}\ }\textbf {\bibinfo {volume} {85}},\ \bibinfo
  {pages} {245132} (\bibinfo {year} {2012})}\BibitemShut {NoStop}%
\bibitem [{\citenamefont {Qi}(2013)}]{Qi2013}%
  \BibitemOpen
  \bibfield  {author} {\bibinfo {author} {\bibfnamefont {X.-L.}\ \bibnamefont
  {Qi}},\ }\bibfield  {title} {\bibinfo {title} {A new class of (2 +
  1)-dimensional topological superconductors with $\mathbb{Z}_8$ topological
  classification},\ }\href {https://doi.org/10.1088/1367-2630/15/6/065002}
  {\bibfield  {journal} {\bibinfo  {journal} {New Journal of Physics}\ }\textbf
  {\bibinfo {volume} {15}},\ \bibinfo {pages} {065002} (\bibinfo {year}
  {2013})}\BibitemShut {NoStop}%
\bibitem [{\citenamefont {Fidkowski}\ \emph {et~al.}(2013)\citenamefont
  {Fidkowski}, \citenamefont {Chen},\ and\ \citenamefont
  {Vishwanath}}]{Fidkowski2013}%
  \BibitemOpen
  \bibfield  {author} {\bibinfo {author} {\bibfnamefont {L.}~\bibnamefont
  {Fidkowski}}, \bibinfo {author} {\bibfnamefont {X.}~\bibnamefont {Chen}},\
  and\ \bibinfo {author} {\bibfnamefont {A.}~\bibnamefont {Vishwanath}},\
  }\bibfield  {title} {\bibinfo {title} {Non-abelian topological order on the
  surface of a 3d topological superconductor from an exactly solved model},\
  }\href {https://doi.org/10.1103/PhysRevX.3.041016} {\bibfield  {journal}
  {\bibinfo  {journal} {Phys. Rev. X}\ }\textbf {\bibinfo {volume} {3}},\
  \bibinfo {pages} {041016} (\bibinfo {year} {2013})}\BibitemShut {NoStop}%
\bibitem [{\citenamefont {Yao}\ and\ \citenamefont {Ryu}(2013)}]{Yao2013}%
  \BibitemOpen
  \bibfield  {author} {\bibinfo {author} {\bibfnamefont {H.}~\bibnamefont
  {Yao}}\ and\ \bibinfo {author} {\bibfnamefont {S.}~\bibnamefont {Ryu}},\
  }\bibfield  {title} {\bibinfo {title} {Interaction effect on topological
  classification of superconductors in two dimensions},\ }\href
  {https://doi.org/10.1103/PhysRevB.88.064507} {\bibfield  {journal} {\bibinfo
  {journal} {Phys. Rev. B}\ }\textbf {\bibinfo {volume} {88}},\ \bibinfo
  {pages} {064507} (\bibinfo {year} {2013})}\BibitemShut {NoStop}%
\bibitem [{\citenamefont {Metlitski}\ \emph {et~al.}(2014)\citenamefont
  {Metlitski}, \citenamefont {Fidkowski}, \citenamefont {Chen},\ and\
  \citenamefont {Vishwanath}}]{Metlitski2014}%
  \BibitemOpen
  \bibfield  {author} {\bibinfo {author} {\bibfnamefont {M.~A.}\ \bibnamefont
  {Metlitski}}, \bibinfo {author} {\bibfnamefont {L.}~\bibnamefont
  {Fidkowski}}, \bibinfo {author} {\bibfnamefont {X.}~\bibnamefont {Chen}},\
  and\ \bibinfo {author} {\bibfnamefont {A.}~\bibnamefont {Vishwanath}},\
  }\bibfield  {title} {\bibinfo {title} {Interaction effects on 3d topological
  superconductors: surface topological order from vortex condensation, the 16
  fold way and fermionic kramers doublets},\ }\href
  {https://arxiv.org/abs/1406.3032} {\bibfield  {journal} {\bibinfo  {journal}
  {arXiv preprint arXiv:1406.3032}\ } (\bibinfo {year} {2014})}\BibitemShut
  {NoStop}%
\bibitem [{\citenamefont {Wang}\ and\ \citenamefont
  {Senthil}(2014)}]{Wang2014}%
  \BibitemOpen
  \bibfield  {author} {\bibinfo {author} {\bibfnamefont {C.}~\bibnamefont
  {Wang}}\ and\ \bibinfo {author} {\bibfnamefont {T.}~\bibnamefont {Senthil}},\
  }\bibfield  {title} {\bibinfo {title} {Interacting fermionic topological
  insulators/superconductors in three dimensions},\ }\href
  {https://doi.org/10.1103/PhysRevB.89.195124} {\bibfield  {journal} {\bibinfo
  {journal} {Phys. Rev. B}\ }\textbf {\bibinfo {volume} {89}},\ \bibinfo
  {pages} {195124} (\bibinfo {year} {2014})}\BibitemShut {NoStop}%
\bibitem [{\citenamefont {You}\ and\ \citenamefont {Xu}(2014)}]{You2014}%
  \BibitemOpen
  \bibfield  {author} {\bibinfo {author} {\bibfnamefont {Y.-Z.}\ \bibnamefont
  {You}}\ and\ \bibinfo {author} {\bibfnamefont {C.}~\bibnamefont {Xu}},\
  }\bibfield  {title} {\bibinfo {title} {Symmetry-protected topological states
  of interacting fermions and bosons},\ }\href
  {https://doi.org/10.1103/PhysRevB.90.245120} {\bibfield  {journal} {\bibinfo
  {journal} {Phys. Rev. B}\ }\textbf {\bibinfo {volume} {90}},\ \bibinfo
  {pages} {245120} (\bibinfo {year} {2014})}\BibitemShut {NoStop}%
\bibitem [{\citenamefont {Morimoto}\ \emph {et~al.}(2015)\citenamefont
  {Morimoto}, \citenamefont {Furusaki},\ and\ \citenamefont
  {Mudry}}]{Morimoto2015}%
  \BibitemOpen
  \bibfield  {author} {\bibinfo {author} {\bibfnamefont {T.}~\bibnamefont
  {Morimoto}}, \bibinfo {author} {\bibfnamefont {A.}~\bibnamefont {Furusaki}},\
  and\ \bibinfo {author} {\bibfnamefont {C.}~\bibnamefont {Mudry}},\ }\bibfield
   {title} {\bibinfo {title} {Breakdown of the topological classification
  $\mathbb{Z}$ for gapped phases of noninteracting fermions by quartic
  interactions},\ }\href {https://doi.org/10.1103/PhysRevB.92.125104}
  {\bibfield  {journal} {\bibinfo  {journal} {Phys. Rev. B}\ }\textbf {\bibinfo
  {volume} {92}},\ \bibinfo {pages} {125104} (\bibinfo {year}
  {2015})}\BibitemShut {NoStop}%
\bibitem [{\citenamefont {Yoshida}\ and\ \citenamefont
  {Furusaki}(2015)}]{Yoshida2015}%
  \BibitemOpen
  \bibfield  {author} {\bibinfo {author} {\bibfnamefont {T.}~\bibnamefont
  {Yoshida}}\ and\ \bibinfo {author} {\bibfnamefont {A.}~\bibnamefont
  {Furusaki}},\ }\bibfield  {title} {\bibinfo {title} {Correlation effects on
  topological crystalline insulators},\ }\href
  {https://doi.org/10.1103/PhysRevB.92.085114} {\bibfield  {journal} {\bibinfo
  {journal} {Phys. Rev. B}\ }\textbf {\bibinfo {volume} {92}},\ \bibinfo
  {pages} {085114} (\bibinfo {year} {2015})}\BibitemShut {NoStop}%
\bibitem [{\citenamefont {Song}\ and\ \citenamefont
  {Schnyder}(2017)}]{SongXY2017}%
  \BibitemOpen
  \bibfield  {author} {\bibinfo {author} {\bibfnamefont {X.-Y.}\ \bibnamefont
  {Song}}\ and\ \bibinfo {author} {\bibfnamefont {A.~P.}\ \bibnamefont
  {Schnyder}},\ }\bibfield  {title} {\bibinfo {title} {Interaction effects on
  the classification of crystalline topological insulators and
  superconductors},\ }\href {https://doi.org/10.1103/PhysRevB.95.195108}
  {\bibfield  {journal} {\bibinfo  {journal} {Phys. Rev. B}\ }\textbf {\bibinfo
  {volume} {95}},\ \bibinfo {pages} {195108} (\bibinfo {year}
  {2017})}\BibitemShut {NoStop}%
\bibitem [{\citenamefont {Aksoy}\ \emph
  {et~al.}(2021{\natexlab{a}})\citenamefont {Aksoy}, \citenamefont {Chen},
  \citenamefont {Ryu}, \citenamefont {Furusaki},\ and\ \citenamefont
  {Mudry}}]{Aksoy2021a}%
  \BibitemOpen
  \bibfield  {author} {\bibinfo {author} {\bibfnamefont {{\"O}.~M.}\
  \bibnamefont {Aksoy}}, \bibinfo {author} {\bibfnamefont {J.-H.}\ \bibnamefont
  {Chen}}, \bibinfo {author} {\bibfnamefont {S.}~\bibnamefont {Ryu}}, \bibinfo
  {author} {\bibfnamefont {A.}~\bibnamefont {Furusaki}},\ and\ \bibinfo
  {author} {\bibfnamefont {C.}~\bibnamefont {Mudry}},\ }\bibfield  {title}
  {\bibinfo {title} {Stability against contact interactions of a topological
  superconductor in two-dimensional space protected by time-reversal and
  reflection symmetries},\ }\href {https://doi.org/10.1103/PhysRevB.103.205121}
  {\bibfield  {journal} {\bibinfo  {journal} {Phys. Rev. B}\ }\textbf {\bibinfo
  {volume} {103}},\ \bibinfo {pages} {205121} (\bibinfo {year}
  {2021}{\natexlab{a}})}\BibitemShut {NoStop}%
\bibitem [{\citenamefont {Chen}\ \emph
  {et~al.}(2011{\natexlab{a}})\citenamefont {Chen}, \citenamefont {Gu},\ and\
  \citenamefont {Wen}}]{Chen2011a}%
  \BibitemOpen
  \bibfield  {author} {\bibinfo {author} {\bibfnamefont {X.}~\bibnamefont
  {Chen}}, \bibinfo {author} {\bibfnamefont {Z.-C.}\ \bibnamefont {Gu}},\ and\
  \bibinfo {author} {\bibfnamefont {X.-G.}\ \bibnamefont {Wen}},\ }\bibfield
  {title} {\bibinfo {title} {Classification of gapped symmetric phases in
  one-dimensional spin systems},\ }\href
  {https://doi.org/10.1103/PhysRevB.83.035107} {\bibfield  {journal} {\bibinfo
  {journal} {Phys. Rev. B}\ }\textbf {\bibinfo {volume} {83}},\ \bibinfo
  {pages} {035107} (\bibinfo {year} {2011}{\natexlab{a}})}\BibitemShut
  {NoStop}%
\bibitem [{\citenamefont {Chen}\ \emph
  {et~al.}(2011{\natexlab{b}})\citenamefont {Chen}, \citenamefont {Gu},\ and\
  \citenamefont {Wen}}]{Chen2011b}%
  \BibitemOpen
  \bibfield  {author} {\bibinfo {author} {\bibfnamefont {X.}~\bibnamefont
  {Chen}}, \bibinfo {author} {\bibfnamefont {Z.-C.}\ \bibnamefont {Gu}},\ and\
  \bibinfo {author} {\bibfnamefont {X.-G.}\ \bibnamefont {Wen}},\ }\bibfield
  {title} {\bibinfo {title} {Complete classification of one-dimensional gapped
  quantum phases in interacting spin systems},\ }\href
  {https://doi.org/10.1103/PhysRevB.84.235128} {\bibfield  {journal} {\bibinfo
  {journal} {Phys. Rev. B}\ }\textbf {\bibinfo {volume} {84}},\ \bibinfo
  {pages} {235128} (\bibinfo {year} {2011}{\natexlab{b}})}\BibitemShut
  {NoStop}%
\bibitem [{\citenamefont {Chen}\ \emph
  {et~al.}(2011{\natexlab{c}})\citenamefont {Chen}, \citenamefont {Liu},\ and\
  \citenamefont {Wen}}]{Chen2011c}%
  \BibitemOpen
  \bibfield  {author} {\bibinfo {author} {\bibfnamefont {X.}~\bibnamefont
  {Chen}}, \bibinfo {author} {\bibfnamefont {Z.-X.}\ \bibnamefont {Liu}},\ and\
  \bibinfo {author} {\bibfnamefont {X.-G.}\ \bibnamefont {Wen}},\ }\bibfield
  {title} {\bibinfo {title} {Two-dimensional symmetry-protected topological
  orders and their protected gapless edge excitations},\ }\href
  {https://doi.org/10.1103/PhysRevB.84.235141} {\bibfield  {journal} {\bibinfo
  {journal} {Phys. Rev. B}\ }\textbf {\bibinfo {volume} {84}},\ \bibinfo
  {pages} {235141} (\bibinfo {year} {2011}{\natexlab{c}})}\BibitemShut
  {NoStop}%
\bibitem [{\citenamefont {Chen}\ \emph {et~al.}(2013)\citenamefont {Chen},
  \citenamefont {Gu}, \citenamefont {Liu},\ and\ \citenamefont
  {Wen}}]{Chen2013}%
  \BibitemOpen
  \bibfield  {author} {\bibinfo {author} {\bibfnamefont {X.}~\bibnamefont
  {Chen}}, \bibinfo {author} {\bibfnamefont {Z.-C.}\ \bibnamefont {Gu}},
  \bibinfo {author} {\bibfnamefont {Z.-X.}\ \bibnamefont {Liu}},\ and\ \bibinfo
  {author} {\bibfnamefont {X.-G.}\ \bibnamefont {Wen}},\ }\bibfield  {title}
  {\bibinfo {title} {Symmetry protected topological orders and the group
  cohomology of their symmetry group},\ }\href
  {https://doi.org/10.1103/PhysRevB.87.155114} {\bibfield  {journal} {\bibinfo
  {journal} {Phys. Rev. B}\ }\textbf {\bibinfo {volume} {87}},\ \bibinfo
  {pages} {155114} (\bibinfo {year} {2013})}\BibitemShut {NoStop}%
\bibitem [{\citenamefont {Chen}\ \emph {et~al.}(2014)\citenamefont {Chen},
  \citenamefont {Lu},\ and\ \citenamefont {Vishwanath}}]{Chen2014}%
  \BibitemOpen
  \bibfield  {author} {\bibinfo {author} {\bibfnamefont {X.}~\bibnamefont
  {Chen}}, \bibinfo {author} {\bibfnamefont {Y.-M.}\ \bibnamefont {Lu}},\ and\
  \bibinfo {author} {\bibfnamefont {A.}~\bibnamefont {Vishwanath}},\ }\bibfield
   {title} {\bibinfo {title} {{Symmetry-protected topological phases from
  decorated domain walls}},\ }\bibfield  {journal} {\bibinfo  {journal} {Nature
  Communications}\ }\textbf {\bibinfo {volume} {5}},\ \href
  {https://doi.org/10.1038/ncomms4507} {10.1038/ncomms4507} (\bibinfo {year}
  {2014})\BibitemShut {NoStop}%
\bibitem [{\citenamefont {Kapustin}(2014{\natexlab{a}})}]{Kapustin2014a}%
  \BibitemOpen
  \bibfield  {author} {\bibinfo {author} {\bibfnamefont {A.}~\bibnamefont
  {Kapustin}},\ }\href@noop {} {\bibinfo {title} {Symmetry protected
  topological phases, anomalies, and cobordisms: Beyond group cohomology}}
  (\bibinfo {year} {2014}{\natexlab{a}}),\ \Eprint
  {https://arxiv.org/abs/1403.1467} {arXiv:1403.1467 [cond-mat.str-el]}
  \BibitemShut {NoStop}%
\bibitem [{\citenamefont {Kapustin}(2014{\natexlab{b}})}]{Kapustin2014c}%
  \BibitemOpen
  \bibfield  {author} {\bibinfo {author} {\bibfnamefont {A.}~\bibnamefont
  {Kapustin}},\ }\href@noop {} {\bibinfo {title} {Bosonic topological
  insulators and paramagnets: a view from cobordisms}} (\bibinfo {year}
  {2014}{\natexlab{b}}),\ \Eprint {https://arxiv.org/abs/1404.6659}
  {arXiv:1404.6659 [cond-mat.str-el]} \BibitemShut {NoStop}%
\bibitem [{\citenamefont {Gu}\ and\ \citenamefont {Wen}(2014)}]{Gu2014}%
  \BibitemOpen
  \bibfield  {author} {\bibinfo {author} {\bibfnamefont {Z.-C.}\ \bibnamefont
  {Gu}}\ and\ \bibinfo {author} {\bibfnamefont {X.-G.}\ \bibnamefont {Wen}},\
  }\bibfield  {title} {\bibinfo {title} {{Symmetry-protected topological orders
  for interacting fermions: Fermionic topological nonlinear
  $\ensuremath{\sigma}$ models and a special group supercohomology theory}},\
  }\href {https://doi.org/10.1103/PhysRevB.90.115141} {\bibfield  {journal}
  {\bibinfo  {journal} {Phys. Rev. B}\ }\textbf {\bibinfo {volume} {90}},\
  \bibinfo {pages} {115141} (\bibinfo {year} {2014})}\BibitemShut {NoStop}%
\bibitem [{\citenamefont {Kapustin}\ \emph {et~al.}(2015)\citenamefont
  {Kapustin}, \citenamefont {Thorngren}, \citenamefont {Turzillo},\ and\
  \citenamefont {Wang}}]{Kapustin2015}%
  \BibitemOpen
  \bibfield  {author} {\bibinfo {author} {\bibfnamefont {A.}~\bibnamefont
  {Kapustin}}, \bibinfo {author} {\bibfnamefont {R.}~\bibnamefont {Thorngren}},
  \bibinfo {author} {\bibfnamefont {A.}~\bibnamefont {Turzillo}},\ and\
  \bibinfo {author} {\bibfnamefont {Z.}~\bibnamefont {Wang}},\ }\bibfield
  {title} {\bibinfo {title} {Fermionic symmetry protected topological phases
  and cobordisms},\ }\href {https://doi.org/10.1007/JHEP12(2015)052} {\bibfield
   {journal} {\bibinfo  {journal} {Journal of High Energy Physics}\ }\textbf
  {\bibinfo {volume} {2015}},\ \bibinfo {pages} {1} (\bibinfo {year}
  {2015})}\BibitemShut {NoStop}%
\bibitem [{\citenamefont {Wang}\ and\ \citenamefont {Gu}(2020)}]{Wang2020}%
  \BibitemOpen
  \bibfield  {author} {\bibinfo {author} {\bibfnamefont {Q.-R.}\ \bibnamefont
  {Wang}}\ and\ \bibinfo {author} {\bibfnamefont {Z.-C.}\ \bibnamefont {Gu}},\
  }\bibfield  {title} {\bibinfo {title} {Construction and classification of
  symmetry-protected topological phases in interacting fermion systems},\
  }\href {https://doi.org/10.1103/PhysRevX.10.031055} {\bibfield  {journal}
  {\bibinfo  {journal} {Phys. Rev. X}\ }\textbf {\bibinfo {volume} {10}},\
  \bibinfo {pages} {031055} (\bibinfo {year} {2020})}\BibitemShut {NoStop}%
\bibitem [{\citenamefont {Barkeshli}\ \emph {et~al.}(2022)\citenamefont
  {Barkeshli}, \citenamefont {Chen}, \citenamefont {Hsin},\ and\ \citenamefont
  {Manjunath}}]{Barkeshli2022}%
  \BibitemOpen
  \bibfield  {author} {\bibinfo {author} {\bibfnamefont {M.}~\bibnamefont
  {Barkeshli}}, \bibinfo {author} {\bibfnamefont {Y.-A.}\ \bibnamefont {Chen}},
  \bibinfo {author} {\bibfnamefont {P.-S.}\ \bibnamefont {Hsin}},\ and\
  \bibinfo {author} {\bibfnamefont {N.}~\bibnamefont {Manjunath}},\ }\bibfield
  {title} {\bibinfo {title} {Classification of $(2+1)$d invertible fermionic
  topological phases with symmetry},\ }\href
  {https://doi.org/10.1103/PhysRevB.105.235143} {\bibfield  {journal} {\bibinfo
   {journal} {Phys. Rev. B}\ }\textbf {\bibinfo {volume} {105}},\ \bibinfo
  {pages} {235143} (\bibinfo {year} {2022})}\BibitemShut {NoStop}%
\bibitem [{\citenamefont {Wang}\ \emph {et~al.}(2017)\citenamefont {Wang},
  \citenamefont {Lin},\ and\ \citenamefont {Gu}}]{Wang2017}%
  \BibitemOpen
  \bibfield  {author} {\bibinfo {author} {\bibfnamefont {C.}~\bibnamefont
  {Wang}}, \bibinfo {author} {\bibfnamefont {C.-H.}\ \bibnamefont {Lin}},\ and\
  \bibinfo {author} {\bibfnamefont {Z.-C.}\ \bibnamefont {Gu}},\ }\bibfield
  {title} {\bibinfo {title} {Interacting fermionic symmetry-protected
  topological phases in two dimensions},\ }\href
  {https://doi.org/10.1103/PhysRevB.95.195147} {\bibfield  {journal} {\bibinfo
  {journal} {Phys. Rev. B}\ }\textbf {\bibinfo {volume} {95}},\ \bibinfo
  {pages} {195147} (\bibinfo {year} {2017})}\BibitemShut {NoStop}%
\bibitem [{\citenamefont {Cheng}\ \emph {et~al.}(2018)\citenamefont {Cheng},
  \citenamefont {Tantivasadakarn},\ and\ \citenamefont {Wang}}]{Cheng2018}%
  \BibitemOpen
  \bibfield  {author} {\bibinfo {author} {\bibfnamefont {M.}~\bibnamefont
  {Cheng}}, \bibinfo {author} {\bibfnamefont {N.}~\bibnamefont
  {Tantivasadakarn}},\ and\ \bibinfo {author} {\bibfnamefont {C.}~\bibnamefont
  {Wang}},\ }\bibfield  {title} {\bibinfo {title} {Loop braiding statistics and
  interacting fermionic symmetry-protected topological phases in three
  dimensions},\ }\href {https://doi.org/10.1103/PhysRevX.8.011054} {\bibfield
  {journal} {\bibinfo  {journal} {Phys. Rev. X}\ }\textbf {\bibinfo {volume}
  {8}},\ \bibinfo {pages} {011054} (\bibinfo {year} {2018})}\BibitemShut
  {NoStop}%
\bibitem [{\citenamefont {Tantivasadakarn}\ and\ \citenamefont
  {Vishwanath}(2018)}]{Tantivasadakarn2018}%
  \BibitemOpen
  \bibfield  {author} {\bibinfo {author} {\bibfnamefont {N.}~\bibnamefont
  {Tantivasadakarn}}\ and\ \bibinfo {author} {\bibfnamefont {A.}~\bibnamefont
  {Vishwanath}},\ }\bibfield  {title} {\bibinfo {title} {{Full commuting
  projector Hamiltonians of interacting symmetry-protected topological phases
  of fermions}},\ }\href {https://doi.org/10.1103/PhysRevB.98.165104}
  {\bibfield  {journal} {\bibinfo  {journal} {Phys. Rev. B}\ }\textbf {\bibinfo
  {volume} {98}},\ \bibinfo {pages} {165104} (\bibinfo {year}
  {2018})}\BibitemShut {NoStop}%
\bibitem [{\citenamefont {Sullivan}\ and\ \citenamefont
  {Cheng}(2020)}]{Sullivan2020}%
  \BibitemOpen
  \bibfield  {author} {\bibinfo {author} {\bibfnamefont {J.}~\bibnamefont
  {Sullivan}}\ and\ \bibinfo {author} {\bibfnamefont {M.}~\bibnamefont
  {Cheng}},\ }\bibfield  {title} {\bibinfo {title} {{Interacting edge states of
  fermionic symmetry-protected topological phases in two dimensions}},\ }\href
  {https://doi.org/10.21468/SciPostPhys.9.2.016} {\bibfield  {journal}
  {\bibinfo  {journal} {SciPost Phys.}\ }\textbf {\bibinfo {volume} {9}},\
  \bibinfo {pages} {016} (\bibinfo {year} {2020})}\BibitemShut {NoStop}%
\bibitem [{\citenamefont {Cheng}\ and\ \citenamefont {Wang}(2022)}]{Cheng2022}%
  \BibitemOpen
  \bibfield  {author} {\bibinfo {author} {\bibfnamefont {M.}~\bibnamefont
  {Cheng}}\ and\ \bibinfo {author} {\bibfnamefont {C.}~\bibnamefont {Wang}},\
  }\bibfield  {title} {\bibinfo {title} {Rotation symmetry-protected
  topological phases of fermions},\ }\href
  {https://doi.org/10.1103/PhysRevB.105.195154} {\bibfield  {journal} {\bibinfo
   {journal} {Phys. Rev. B}\ }\textbf {\bibinfo {volume} {105}},\ \bibinfo
  {pages} {195154} (\bibinfo {year} {2022})}\BibitemShut {NoStop}%
\bibitem [{\citenamefont {Thorngren}\ and\ \citenamefont
  {Else}(2018)}]{Thorngren2018}%
  \BibitemOpen
  \bibfield  {author} {\bibinfo {author} {\bibfnamefont {R.}~\bibnamefont
  {Thorngren}}\ and\ \bibinfo {author} {\bibfnamefont {D.~V.}\ \bibnamefont
  {Else}},\ }\bibfield  {title} {\bibinfo {title} {Gauging spatial symmetries
  and the classification of topological crystalline phases},\ }\href
  {https://doi.org/10.1103/PhysRevX.8.011040} {\bibfield  {journal} {\bibinfo
  {journal} {Phys. Rev. X}\ }\textbf {\bibinfo {volume} {8}},\ \bibinfo {pages}
  {011040} (\bibinfo {year} {2018})}\BibitemShut {NoStop}%
\bibitem [{\citenamefont {Else}\ and\ \citenamefont
  {Thorngren}(2019)}]{Else2019a}%
  \BibitemOpen
  \bibfield  {author} {\bibinfo {author} {\bibfnamefont {D.~V.}\ \bibnamefont
  {Else}}\ and\ \bibinfo {author} {\bibfnamefont {R.}~\bibnamefont
  {Thorngren}},\ }\bibfield  {title} {\bibinfo {title} {Crystalline topological
  phases as defect networks},\ }\href
  {https://doi.org/10.1103/PhysRevB.99.115116} {\bibfield  {journal} {\bibinfo
  {journal} {Phys. Rev. B}\ }\textbf {\bibinfo {volume} {99}},\ \bibinfo
  {pages} {115116} (\bibinfo {year} {2019})}\BibitemShut {NoStop}%
\bibitem [{\citenamefont {Freed}\ and\ \citenamefont
  {Hopkins}(2019)}]{Freed2019}%
  \BibitemOpen
  \bibfield  {author} {\bibinfo {author} {\bibfnamefont {D.~S.}\ \bibnamefont
  {Freed}}\ and\ \bibinfo {author} {\bibfnamefont {M.~J.}\ \bibnamefont
  {Hopkins}},\ }\href@noop {} {\bibinfo {title} {Invertible phases of matter
  with spatial symmetry}} (\bibinfo {year} {2019}),\ \Eprint
  {https://arxiv.org/abs/1901.06419} {arXiv:1901.06419 [math-ph]} \BibitemShut
  {NoStop}%
\bibitem [{\citenamefont {Debray}(2021)}]{Debray2021}%
  \BibitemOpen
  \bibfield  {author} {\bibinfo {author} {\bibfnamefont {A.}~\bibnamefont
  {Debray}},\ }\href@noop {} {\bibinfo {title} {Invertible phases for mixed
  spatial symmetries and the fermionic crystalline equivalence principle}}
  (\bibinfo {year} {2021}),\ \Eprint {https://arxiv.org/abs/2102.02941}
  {arXiv:2102.02941 [math-ph]} \BibitemShut {NoStop}%
\bibitem [{\citenamefont {Song}\ \emph
  {et~al.}(2017{\natexlab{b}})\citenamefont {Song}, \citenamefont {Huang},
  \citenamefont {Fu},\ and\ \citenamefont {Hermele}}]{Song2017}%
  \BibitemOpen
  \bibfield  {author} {\bibinfo {author} {\bibfnamefont {H.}~\bibnamefont
  {Song}}, \bibinfo {author} {\bibfnamefont {S.-J.}\ \bibnamefont {Huang}},
  \bibinfo {author} {\bibfnamefont {L.}~\bibnamefont {Fu}},\ and\ \bibinfo
  {author} {\bibfnamefont {M.}~\bibnamefont {Hermele}},\ }\bibfield  {title}
  {\bibinfo {title}
  {\href{https://link.aps.org/doi/10.1103/PhysRevX.7.011020}{Topological Phases
  Protected by Point Group Symmetry}},\ }\href
  {https://doi.org/10.1103/PhysRevX.7.011020} {\bibfield  {journal} {\bibinfo
  {journal} {Phys. Rev. X}\ }\textbf {\bibinfo {volume} {7}},\ \bibinfo {pages}
  {011020} (\bibinfo {year} {2017}{\natexlab{b}})}\BibitemShut {NoStop}%
\bibitem [{\citenamefont {Huang}\ \emph {et~al.}(2017)\citenamefont {Huang},
  \citenamefont {Song}, \citenamefont {Huang},\ and\ \citenamefont
  {Hermele}}]{Huang2017}%
  \BibitemOpen
  \bibfield  {author} {\bibinfo {author} {\bibfnamefont {S.-J.}\ \bibnamefont
  {Huang}}, \bibinfo {author} {\bibfnamefont {H.}~\bibnamefont {Song}},
  \bibinfo {author} {\bibfnamefont {Y.-P.}\ \bibnamefont {Huang}},\ and\
  \bibinfo {author} {\bibfnamefont {M.}~\bibnamefont {Hermele}},\ }\bibfield
  {title} {\bibinfo {title}
  {\href{https://link.aps.org/doi/10.1103/PhysRevB.96.205106}{Building
  crystalline topological phases from lower-dimensional states}},\ }\href
  {https://doi.org/10.1103/PhysRevB.96.205106} {\bibfield  {journal} {\bibinfo
  {journal} {Phys. Rev. B}\ }\textbf {\bibinfo {volume} {96}},\ \bibinfo
  {pages} {205106} (\bibinfo {year} {2017})}\BibitemShut {NoStop}%
\bibitem [{\citenamefont {Zhang}\ \emph {et~al.}(2020)\citenamefont {Zhang},
  \citenamefont {Wang}, \citenamefont {Yang}, \citenamefont {Qi},\ and\
  \citenamefont {Gu}}]{Zhang2020}%
  \BibitemOpen
  \bibfield  {author} {\bibinfo {author} {\bibfnamefont {J.-H.}\ \bibnamefont
  {Zhang}}, \bibinfo {author} {\bibfnamefont {Q.-R.}\ \bibnamefont {Wang}},
  \bibinfo {author} {\bibfnamefont {S.}~\bibnamefont {Yang}}, \bibinfo {author}
  {\bibfnamefont {Y.}~\bibnamefont {Qi}},\ and\ \bibinfo {author}
  {\bibfnamefont {Z.-C.}\ \bibnamefont {Gu}},\ }\bibfield  {title} {\bibinfo
  {title} {Construction and classification of point-group symmetry-protected
  topological phases in two-dimensional interacting fermionic systems},\ }\href
  {https://doi.org/10.1103/PhysRevB.101.100501} {\bibfield  {journal} {\bibinfo
   {journal} {Phys. Rev. B}\ }\textbf {\bibinfo {volume} {101}},\ \bibinfo
  {pages} {100501(R)} (\bibinfo {year} {2020})}\BibitemShut {NoStop}%
\bibitem [{\citenamefont {{Zhang}}\ \emph {et~al.}(2022)\citenamefont
  {{Zhang}}, \citenamefont {{Qi}},\ and\ \citenamefont {{Gu}}}]{Zhang2022a}%
  \BibitemOpen
  \bibfield  {author} {\bibinfo {author} {\bibfnamefont {J.-H.}\ \bibnamefont
  {{Zhang}}}, \bibinfo {author} {\bibfnamefont {Y.}~\bibnamefont {{Qi}}},\ and\
  \bibinfo {author} {\bibfnamefont {Z.-C.}\ \bibnamefont {{Gu}}},\ }\bibfield
  {title} {\bibinfo {title} {{Construction and classification of crystalline
  topological superconductor and insulators in three-dimensional interacting
  fermion systems}},\ }\href {https://doi.org/10.48550/arXiv.2204.13558}
  {\bibfield  {journal} {\bibinfo  {journal} {arXiv e-prints}\ ,\ \bibinfo
  {eid} {arXiv:2204.13558}} (\bibinfo {year} {2022})},\ \Eprint
  {https://arxiv.org/abs/2204.13558} {arXiv:2204.13558 [cond-mat.str-el]}
  \BibitemShut {NoStop}%
\bibitem [{\citenamefont {Zhang}\ \emph {et~al.}(2022)\citenamefont {Zhang},
  \citenamefont {Yang}, \citenamefont {Qi},\ and\ \citenamefont
  {Gu}}]{Zhang2022b}%
  \BibitemOpen
  \bibfield  {author} {\bibinfo {author} {\bibfnamefont {J.-H.}\ \bibnamefont
  {Zhang}}, \bibinfo {author} {\bibfnamefont {S.}~\bibnamefont {Yang}},
  \bibinfo {author} {\bibfnamefont {Y.}~\bibnamefont {Qi}},\ and\ \bibinfo
  {author} {\bibfnamefont {Z.-C.}\ \bibnamefont {Gu}},\ }\bibfield  {title}
  {\bibinfo {title}
  {\href{https://link.aps.org/doi/10.1103/PhysRevResearch.4.033081}{Real-space
  construction of crystalline topological superconductors and insulators in 2D
  interacting fermionic systems}},\ }\href
  {https://doi.org/10.1103/PhysRevResearch.4.033081} {\bibfield  {journal}
  {\bibinfo  {journal} {Phys. Rev. Res.}\ }\textbf {\bibinfo {volume} {4}},\
  \bibinfo {pages} {033081} (\bibinfo {year} {2022})}\BibitemShut {NoStop}%
\bibitem [{\citenamefont {Rasmussen}\ and\ \citenamefont
  {Lu}(2018)}]{Rasmussen2018}%
  \BibitemOpen
  \bibfield  {author} {\bibinfo {author} {\bibfnamefont {A.}~\bibnamefont
  {Rasmussen}}\ and\ \bibinfo {author} {\bibfnamefont {Y.-M.}\ \bibnamefont
  {Lu}},\ }\href@noop {} {\bibinfo {title} {Intrinsically interacting
  topological crystalline insulators and superconductors}} (\bibinfo {year}
  {2018}),\ \Eprint {https://arxiv.org/abs/1810.12317} {arXiv:1810.12317
  [cond-mat.str-el]} \BibitemShut {NoStop}%
\bibitem [{\citenamefont {Rasmussen}\ and\ \citenamefont
  {Lu}(2020)}]{Rasmussen2020}%
  \BibitemOpen
  \bibfield  {author} {\bibinfo {author} {\bibfnamefont {A.}~\bibnamefont
  {Rasmussen}}\ and\ \bibinfo {author} {\bibfnamefont {Y.-M.}\ \bibnamefont
  {Lu}},\ }\bibfield  {title} {\bibinfo {title} {Classification and
  construction of higher-order symmetry-protected topological phases of
  interacting bosons},\ }\href {https://doi.org/10.1103/PhysRevB.101.085137}
  {\bibfield  {journal} {\bibinfo  {journal} {Phys. Rev. B}\ }\textbf {\bibinfo
  {volume} {101}},\ \bibinfo {pages} {085137} (\bibinfo {year}
  {2020})}\BibitemShut {NoStop}%
\bibitem [{\citenamefont {Manjunath}\ \emph {et~al.}(2023)\citenamefont
  {Manjunath}, \citenamefont {Calvera},\ and\ \citenamefont
  {Barkeshli}}]{manjunath2023}%
  \BibitemOpen
  \bibfield  {author} {\bibinfo {author} {\bibfnamefont {N.}~\bibnamefont
  {Manjunath}}, \bibinfo {author} {\bibfnamefont {V.}~\bibnamefont {Calvera}},\
  and\ \bibinfo {author} {\bibfnamefont {M.}~\bibnamefont {Barkeshli}},\
  }\href@noop {} {\bibinfo {title} {{Characterization and classification of
  interacting (2+1)D topological crystalline insulators with
  orientation-preserving wallpaper groups}}} (\bibinfo {year} {2023}),\ \Eprint
  {https://arxiv.org/abs/2309.12389} {arXiv:2309.12389 [cond-mat.str-el]}
  \BibitemShut {NoStop}%
\bibitem [{\citenamefont {Potirniche}\ \emph {et~al.}(2017)\citenamefont
  {Potirniche}, \citenamefont {Potter}, \citenamefont {Schleier-Smith},
  \citenamefont {Vishwanath},\ and\ \citenamefont {Yao}}]{Potirniche2017}%
  \BibitemOpen
  \bibfield  {author} {\bibinfo {author} {\bibfnamefont {I.-D.}\ \bibnamefont
  {Potirniche}}, \bibinfo {author} {\bibfnamefont {A.~C.}\ \bibnamefont
  {Potter}}, \bibinfo {author} {\bibfnamefont {M.}~\bibnamefont
  {Schleier-Smith}}, \bibinfo {author} {\bibfnamefont {A.}~\bibnamefont
  {Vishwanath}},\ and\ \bibinfo {author} {\bibfnamefont {N.~Y.}\ \bibnamefont
  {Yao}},\ }\bibfield  {title} {\bibinfo {title} {{Floquet Symmetry-Protected
  Topological Phases in Cold-Atom Systems}},\ }\href
  {https://doi.org/10.1103/PhysRevLett.119.123601} {\bibfield  {journal}
  {\bibinfo  {journal} {Phys. Rev. Lett.}\ }\textbf {\bibinfo {volume} {119}},\
  \bibinfo {pages} {123601} (\bibinfo {year} {2017})}\BibitemShut {NoStop}%
\bibitem [{\citenamefont {{Sompet}}\ \emph {et~al.}(2022)\citenamefont
  {{Sompet}}, \citenamefont {{Hirthe}}, \citenamefont {{Bourgund}},
  \citenamefont {{Chalopin}}, \citenamefont {{Bibo}}, \citenamefont
  {{Koepsell}}, \citenamefont {{Bojovi{\'c}}}, \citenamefont {{Verresen}},
  \citenamefont {{Pollmann}}, \citenamefont {{Salomon}}, \citenamefont
  {{Gross}}, \citenamefont {{Hilker}},\ and\ \citenamefont
  {{Bloch}}}]{Sompet2022}%
  \BibitemOpen
  \bibfield  {author} {\bibinfo {author} {\bibfnamefont {P.}~\bibnamefont
  {{Sompet}}}, \bibinfo {author} {\bibfnamefont {S.}~\bibnamefont {{Hirthe}}},
  \bibinfo {author} {\bibfnamefont {D.}~\bibnamefont {{Bourgund}}}, \bibinfo
  {author} {\bibfnamefont {T.}~\bibnamefont {{Chalopin}}}, \bibinfo {author}
  {\bibfnamefont {J.}~\bibnamefont {{Bibo}}}, \bibinfo {author} {\bibfnamefont
  {J.}~\bibnamefont {{Koepsell}}}, \bibinfo {author} {\bibfnamefont
  {P.}~\bibnamefont {{Bojovi{\'c}}}}, \bibinfo {author} {\bibfnamefont
  {R.}~\bibnamefont {{Verresen}}}, \bibinfo {author} {\bibfnamefont
  {F.}~\bibnamefont {{Pollmann}}}, \bibinfo {author} {\bibfnamefont
  {G.}~\bibnamefont {{Salomon}}}, \bibinfo {author} {\bibfnamefont
  {C.}~\bibnamefont {{Gross}}}, \bibinfo {author} {\bibfnamefont {T.~A.}\
  \bibnamefont {{Hilker}}},\ and\ \bibinfo {author} {\bibfnamefont
  {I.}~\bibnamefont {{Bloch}}},\ }\bibfield  {title} {\bibinfo {title}
  {{Realizing the symmetry-protected Haldane phase in Fermi-Hubbard ladders}},\
  }\href {https://doi.org/10.1038/s41586-022-04688-z} {\bibfield  {journal}
  {\bibinfo  {journal} {\nat}\ }\textbf {\bibinfo {volume} {606}},\ \bibinfo
  {pages} {484} (\bibinfo {year} {2022})},\ \Eprint
  {https://arxiv.org/abs/2103.10421} {arXiv:2103.10421 [cond-mat.quant-gas]}
  \BibitemShut {NoStop}%
\bibitem [{\citenamefont {Kobayashi}\ \emph {et~al.}(2024)\citenamefont
  {Kobayashi}, \citenamefont {Zhang}, \citenamefont {Wang},\ and\ \citenamefont
  {Barkeshli}}]{kobayashi2024}%
  \BibitemOpen
  \bibfield  {author} {\bibinfo {author} {\bibfnamefont {R.}~\bibnamefont
  {Kobayashi}}, \bibinfo {author} {\bibfnamefont {Y.}~\bibnamefont {Zhang}},
  \bibinfo {author} {\bibfnamefont {Y.-Q.}\ \bibnamefont {Wang}},\ and\
  \bibinfo {author} {\bibfnamefont {M.}~\bibnamefont {Barkeshli}},\ }\href@noop
  {} {\bibinfo {title} {{(2+1)D topological phases with RT symmetry: many-body
  invariant, classification, and higher order edge modes}}} (\bibinfo {year}
  {2024}),\ \Eprint {https://arxiv.org/abs/2403.18887} {arXiv:2403.18887
  [cond-mat.str-el]} \BibitemShut {NoStop}%
\bibitem [{\citenamefont {Hahn}\ \emph {et~al.}(2005)\citenamefont {Hahn},
  \citenamefont {Arnold}, \citenamefont {Aroyo}, \citenamefont {Bertraut},
  \citenamefont {Billiet}, \citenamefont {Buerger}, \citenamefont {Burzlaff},
  \citenamefont {Donnay}, \citenamefont {Fischer}, \citenamefont {Fokkema},
  \citenamefont {Gruber}, \citenamefont {Klapper}, \citenamefont {Koch},
  \citenamefont {Konstantinov}, \citenamefont {Langlet}, \citenamefont
  {Looijenga-Vos}, \citenamefont {M\"uller}, \citenamefont {De~Wolff},
  \citenamefont {Wondratschek},\ and\ \citenamefont {Zimmermann}}]{Hahn2005}%
  \BibitemOpen
  \bibfield  {author} {\bibinfo {author} {\bibfnamefont {T.}~\bibnamefont
  {Hahn}}, \bibinfo {author} {\bibfnamefont {H.}~\bibnamefont {Arnold}},
  \bibinfo {author} {\bibfnamefont {M.~I.}\ \bibnamefont {Aroyo}}, \bibinfo
  {author} {\bibfnamefont {E.~F.}\ \bibnamefont {Bertraut}}, \bibinfo {author}
  {\bibfnamefont {Y.}~\bibnamefont {Billiet}}, \bibinfo {author} {\bibfnamefont
  {M.~J.}\ \bibnamefont {Buerger}}, \bibinfo {author} {\bibfnamefont
  {H.}~\bibnamefont {Burzlaff}}, \bibinfo {author} {\bibfnamefont {J.~D.~H.}\
  \bibnamefont {Donnay}}, \bibinfo {author} {\bibfnamefont {W.}~\bibnamefont
  {Fischer}}, \bibinfo {author} {\bibfnamefont {D.~S.}\ \bibnamefont
  {Fokkema}}, \bibinfo {author} {\bibfnamefont {B.}~\bibnamefont {Gruber}},
  \bibinfo {author} {\bibfnamefont {H.}~\bibnamefont {Klapper}}, \bibinfo
  {author} {\bibfnamefont {E.}~\bibnamefont {Koch}}, \bibinfo {author}
  {\bibfnamefont {P.~B.}\ \bibnamefont {Konstantinov}}, \bibinfo {author}
  {\bibfnamefont {G.~A.}\ \bibnamefont {Langlet}}, \bibinfo {author}
  {\bibfnamefont {A.}~\bibnamefont {Looijenga-Vos}}, \bibinfo {author}
  {\bibfnamefont {U.}~\bibnamefont {M\"uller}}, \bibinfo {author}
  {\bibfnamefont {P.~M.}\ \bibnamefont {De~Wolff}}, \bibinfo {author}
  {\bibfnamefont {H.}~\bibnamefont {Wondratschek}},\ and\ \bibinfo {author}
  {\bibfnamefont {H.}~\bibnamefont {Zimmermann}},\ }\href@noop {} {\emph
  {\bibinfo {title} {{International Tables for Crystallography, Vol. A,
  Space-Group Symmetry}}}},\ edited by\ \bibinfo {editor} {\bibfnamefont
  {T.}~\bibnamefont {Hahn}},\ Vol.~\bibinfo {volume} {A}\ (\bibinfo
  {publisher} {Springer},\ \bibinfo {year} {2005})\BibitemShut {NoStop}%
\bibitem [{\citenamefont {Song}\ \emph {et~al.}(2019)\citenamefont {Song},
  \citenamefont {Huang}, \citenamefont {Qi}, \citenamefont {Fang},\ and\
  \citenamefont {Hermele}}]{Song2019}%
  \BibitemOpen
  \bibfield  {author} {\bibinfo {author} {\bibfnamefont {Z.}~\bibnamefont
  {Song}}, \bibinfo {author} {\bibfnamefont {S.-J.}\ \bibnamefont {Huang}},
  \bibinfo {author} {\bibfnamefont {Y.}~\bibnamefont {Qi}}, \bibinfo {author}
  {\bibfnamefont {C.}~\bibnamefont {Fang}},\ and\ \bibinfo {author}
  {\bibfnamefont {M.}~\bibnamefont {Hermele}},\ }\bibfield  {title} {\bibinfo
  {title}
  {\href{https://www.science.org/doi/abs/10.1126/sciadv.aax2007}{Topological
  states from topological crystals}},\ }\bibfield  {journal} {\bibinfo
  {journal} {Science Advances}\ }\textbf {\bibinfo {volume} {5}},\ \href
  {https://doi.org/10.1126/sciadv.aax2007} {10.1126/sciadv.aax2007} (\bibinfo
  {year} {2019})\BibitemShut {NoStop}%
\bibitem [{\citenamefont {Yao}\ and\ \citenamefont {Kivelson}(2010)}]{Yao2010}%
  \BibitemOpen
  \bibfield  {author} {\bibinfo {author} {\bibfnamefont {H.}~\bibnamefont
  {Yao}}\ and\ \bibinfo {author} {\bibfnamefont {S.~A.}\ \bibnamefont
  {Kivelson}},\ }\bibfield  {title} {\bibinfo {title} {{Fragile Mott
  Insulators}},\ }\href {https://doi.org/10.1103/physrevlett.105.166402}
  {\bibfield  {journal} {\bibinfo  {journal} {Physical Review Letters}\
  }\textbf {\bibinfo {volume} {105}},\ \bibinfo {pages} {166402} (\bibinfo
  {year} {2010})}\BibitemShut {NoStop}%
\bibitem [{\citenamefont {Turzillo}\ and\ \citenamefont
  {You}(2019)}]{Turzillo2019}%
  \BibitemOpen
  \bibfield  {author} {\bibinfo {author} {\bibfnamefont {A.}~\bibnamefont
  {Turzillo}}\ and\ \bibinfo {author} {\bibfnamefont {M.}~\bibnamefont {You}},\
  }\bibfield  {title} {\bibinfo {title} {Fermionic matrix product states and
  one-dimensional short-range entangled phases with antiunitary symmetries},\
  }\href {https://doi.org/10.1103/PhysRevB.99.035103} {\bibfield  {journal}
  {\bibinfo  {journal} {Phys. Rev. B}\ }\textbf {\bibinfo {volume} {99}},\
  \bibinfo {pages} {035103} (\bibinfo {year} {2019})}\BibitemShut {NoStop}%
\bibitem [{\citenamefont {Bourne}\ and\ \citenamefont
  {Ogata}(2021)}]{Bourne2021}%
  \BibitemOpen
  \bibfield  {author} {\bibinfo {author} {\bibfnamefont {C.}~\bibnamefont
  {Bourne}}\ and\ \bibinfo {author} {\bibfnamefont {Y.}~\bibnamefont {Ogata}},\
  }\bibfield  {title} {\bibinfo {title} {The classification of symmetry
  protected topological phases of one-dimensional fermion systems},\ }\href
  {https://doi.org/10.1017/fms.2021.19} {\bibfield  {journal} {\bibinfo
  {journal} {Forum of Mathematics, Sigma}\ }\textbf {\bibinfo {volume} {9}},\
  \bibinfo {pages} {e25} (\bibinfo {year} {2021})}\BibitemShut {NoStop}%
\bibitem [{\citenamefont {Aksoy}\ and\ \citenamefont
  {Mudry}(2022)}]{Aksoy2022}%
  \BibitemOpen
  \bibfield  {author} {\bibinfo {author} {\bibfnamefont {{\"O}.~M.}\
  \bibnamefont {Aksoy}}\ and\ \bibinfo {author} {\bibfnamefont
  {C.}~\bibnamefont {Mudry}},\ }\bibfield  {title} {\bibinfo {title}
  {Elementary derivation of the stacking rules of invertible fermionic
  topological phases in one dimension},\ }\href
  {https://doi.org/10.1103/PhysRevB.106.035117} {\bibfield  {journal} {\bibinfo
   {journal} {Phys. Rev. B}\ }\textbf {\bibinfo {volume} {106}},\ \bibinfo
  {pages} {035117} (\bibinfo {year} {2022})}\BibitemShut {NoStop}%
\bibitem [{\citenamefont {Kane}\ and\ \citenamefont {Mele}(2005)}]{Kane_2005}%
  \BibitemOpen
  \bibfield  {author} {\bibinfo {author} {\bibfnamefont {C.~L.}\ \bibnamefont
  {Kane}}\ and\ \bibinfo {author} {\bibfnamefont {E.~J.}\ \bibnamefont
  {Mele}},\ }\bibfield  {title} {\bibinfo {title} {{$\mathbb{Z}_2$ Topological
  Order and the Quantum Spin Hall Effect}},\ }\bibfield  {journal} {\bibinfo
  {journal} {Physical Review Letters}\ }\textbf {\bibinfo {volume} {95}},\
  \href {https://doi.org/10.1103/physrevlett.95.146802}
  {10.1103/physrevlett.95.146802} (\bibinfo {year} {2005})\BibitemShut
  {NoStop}%
\bibitem [{\citenamefont {Scalapino}\ and\ \citenamefont
  {Trugman}(1996)}]{Scalapino1996}%
  \BibitemOpen
  \bibfield  {author} {\bibinfo {author} {\bibfnamefont {D.~J.}\ \bibnamefont
  {Scalapino}}\ and\ \bibinfo {author} {\bibfnamefont {S.~A.}\ \bibnamefont
  {Trugman}},\ }\bibfield  {title} {\bibinfo {title} {{Local antiferromagnetic
  correlations and $d_{x^2-y^2}$ pairing}},\ }\href
  {https://doi.org/10.1080/01418639608240361} {\bibfield  {journal} {\bibinfo
  {journal} {Philosophical Magazine B}\ }\textbf {\bibinfo {volume} {74}},\
  \bibinfo {pages} {607} (\bibinfo {year} {1996})},\ \Eprint
  {https://arxiv.org/abs/https://doi.org/10.1080/01418639608240361}
  {https://doi.org/10.1080/01418639608240361} \BibitemShut {NoStop}%
\bibitem [{\citenamefont {Chakravarty}\ and\ \citenamefont
  {Kivelson}(2001)}]{Chakravarty2001}%
  \BibitemOpen
  \bibfield  {author} {\bibinfo {author} {\bibfnamefont {S.}~\bibnamefont
  {Chakravarty}}\ and\ \bibinfo {author} {\bibfnamefont {S.~A.}\ \bibnamefont
  {Kivelson}},\ }\bibfield  {title} {\bibinfo {title} {Electronic mechanism of
  superconductivity in the cuprates, {$\mathrm{C}_{60},$} and polyacenes},\
  }\href {https://doi.org/10.1103/PhysRevB.64.064511} {\bibfield  {journal}
  {\bibinfo  {journal} {Phys. Rev. B}\ }\textbf {\bibinfo {volume} {64}},\
  \bibinfo {pages} {064511} (\bibinfo {year} {2001})}\BibitemShut {NoStop}%
\bibitem [{\citenamefont {Schumann}(2002)}]{Schumann2002}%
  \BibitemOpen
  \bibfield  {author} {\bibinfo {author} {\bibfnamefont {R.}~\bibnamefont
  {Schumann}},\ }\bibfield  {title} {\bibinfo {title} {Thermodynamics of a
  4-site {H}ubbard model by analytical diagonalization},\ }\href
  {https://doi.org/https://doi.org/10.1002/1521-3889(200201)11:1<49::AID-ANDP49>3.0.CO;2-7}
  {\bibfield  {journal} {\bibinfo  {journal} {Annalen der Physik}\ }\textbf
  {\bibinfo {volume} {11}},\ \bibinfo {pages} {49} (\bibinfo {year}
  {2002})}\BibitemShut {NoStop}%
\bibitem [{\citenamefont {Muechler}\ \emph {et~al.}(2014)\citenamefont
  {Muechler}, \citenamefont {Maciejko}, \citenamefont {Neupert},\ and\
  \citenamefont {Car}}]{Muechler2014}%
  \BibitemOpen
  \bibfield  {author} {\bibinfo {author} {\bibfnamefont {L.}~\bibnamefont
  {Muechler}}, \bibinfo {author} {\bibfnamefont {J.}~\bibnamefont {Maciejko}},
  \bibinfo {author} {\bibfnamefont {T.}~\bibnamefont {Neupert}},\ and\ \bibinfo
  {author} {\bibfnamefont {R.}~\bibnamefont {Car}},\ }\bibfield  {title}
  {\bibinfo {title} {{M\"obius molecules and fragile Mott insulators}},\ }\href
  {https://doi.org/10.1103/PhysRevB.90.245142} {\bibfield  {journal} {\bibinfo
  {journal} {Phys. Rev. B}\ }\textbf {\bibinfo {volume} {90}},\ \bibinfo
  {pages} {245142} (\bibinfo {year} {2014})}\BibitemShut {NoStop}%
\bibitem [{\citenamefont {Soldini}\ \emph
  {et~al.}(2023{\natexlab{b}})\citenamefont {Soldini}, \citenamefont
  {Astrakhantsev}, \citenamefont {Iraola}, \citenamefont {Tiwari},
  \citenamefont {Fischer}, \citenamefont {Valentí}, \citenamefont {Vergniory},
  \citenamefont {Wagner},\ and\ \citenamefont {Neupert}}]{Soldini_2023}%
  \BibitemOpen
  \bibfield  {author} {\bibinfo {author} {\bibfnamefont {M.~O.}\ \bibnamefont
  {Soldini}}, \bibinfo {author} {\bibfnamefont {N.}~\bibnamefont
  {Astrakhantsev}}, \bibinfo {author} {\bibfnamefont {M.}~\bibnamefont
  {Iraola}}, \bibinfo {author} {\bibfnamefont {A.}~\bibnamefont {Tiwari}},
  \bibinfo {author} {\bibfnamefont {M.~H.}\ \bibnamefont {Fischer}}, \bibinfo
  {author} {\bibfnamefont {R.}~\bibnamefont {Valentí}}, \bibinfo {author}
  {\bibfnamefont {M.~G.}\ \bibnamefont {Vergniory}}, \bibinfo {author}
  {\bibfnamefont {G.}~\bibnamefont {Wagner}},\ and\ \bibinfo {author}
  {\bibfnamefont {T.}~\bibnamefont {Neupert}},\ }\bibfield  {title} {\bibinfo
  {title} {{Interacting topological quantum chemistry of Mott atomic limits}},\
  }\href {https://doi.org/10.1103/physrevb.107.245145} {\bibfield  {journal}
  {\bibinfo  {journal} {Physical Review B}\ }\textbf {\bibinfo {volume}
  {107}},\ \bibinfo {pages} {245145} (\bibinfo {year}
  {2023}{\natexlab{b}})}\BibitemShut {NoStop}%
\bibitem [{\citenamefont {{Herzog-Arbeitman}}\ \emph
  {et~al.}(2024)\citenamefont {{Herzog-Arbeitman}}, \citenamefont
  {{Bernevig}},\ and\ \citenamefont {{Song}}}]{Herzog-Arbeitman2023}%
  \BibitemOpen
  \bibfield  {author} {\bibinfo {author} {\bibfnamefont {J.}~\bibnamefont
  {{Herzog-Arbeitman}}}, \bibinfo {author} {\bibfnamefont {A.}~\bibnamefont
  {{Bernevig}}},\ and\ \bibinfo {author} {\bibfnamefont {Z.}~\bibnamefont
  {{Song}}},\ }\bibfield  {title} {\bibinfo {title} {{Interacting topological
  quantum chemistry in 2D with many-body real space invariants}},\ }\href
  {https://doi.org/https://doi.org/10.1038/s41467-024-45395-9} {\bibfield
  {journal} {\bibinfo  {journal} {Nature Communications}\ ,\ \bibinfo {pages}
  {1171}} (\bibinfo {year} {2024})}\BibitemShut {NoStop}%
\bibitem [{\citenamefont {Suzuki}(1971)}]{Suzuki1971}%
  \BibitemOpen
  \bibfield  {author} {\bibinfo {author} {\bibfnamefont {M.}~\bibnamefont
  {Suzuki}},\ }\bibfield  {title} {\bibinfo {title} {{Relationship among
  Exactly Soluble Models of Critical Phenomena. I*): 2D Ising Model, Dimer
  Problem and the Generalized XY-Model}},\ }\href
  {https://doi.org/10.1143/PTP.46.1337} {\bibfield  {journal} {\bibinfo
  {journal} {Progress of Theoretical Physics}\ }\textbf {\bibinfo {volume}
  {46}},\ \bibinfo {pages} {1337} (\bibinfo {year} {1971})},\ \Eprint
  {https://arxiv.org/abs/https://academic.oup.com/ptp/article-pdf/46/5/1337/5268367/46-5-1337.pdf}
  {https://academic.oup.com/ptp/article-pdf/46/5/1337/5268367/46-5-1337.pdf}
  \BibitemShut {NoStop}%
\bibitem [{\citenamefont {Son}\ \emph {et~al.}(2011)\citenamefont {Son},
  \citenamefont {Amico}, \citenamefont {Fazio}, \citenamefont {Hamma},
  \citenamefont {Pascazio},\ and\ \citenamefont {Vedral}}]{Son2011}%
  \BibitemOpen
  \bibfield  {author} {\bibinfo {author} {\bibfnamefont {W.}~\bibnamefont
  {Son}}, \bibinfo {author} {\bibfnamefont {L.}~\bibnamefont {Amico}}, \bibinfo
  {author} {\bibfnamefont {R.}~\bibnamefont {Fazio}}, \bibinfo {author}
  {\bibfnamefont {A.}~\bibnamefont {Hamma}}, \bibinfo {author} {\bibfnamefont
  {S.}~\bibnamefont {Pascazio}},\ and\ \bibinfo {author} {\bibfnamefont
  {V.}~\bibnamefont {Vedral}},\ }\bibfield  {title} {\bibinfo {title} {{Quantum
  phase transition between cluster and antiferromagnetic states}},\ }\href
  {https://doi.org/10.1209/0295-5075/95/50001} {\bibfield  {journal} {\bibinfo
  {journal} {EPL (Europhysics Letters)}\ }\textbf {\bibinfo {volume} {95}},\
  \bibinfo {pages} {50001} (\bibinfo {year} {2011})}\BibitemShut {NoStop}%
\bibitem [{\citenamefont {Verresen}\ \emph {et~al.}(2017)\citenamefont
  {Verresen}, \citenamefont {Moessner},\ and\ \citenamefont
  {Pollmann}}]{Verresen2017}%
  \BibitemOpen
  \bibfield  {author} {\bibinfo {author} {\bibfnamefont {R.}~\bibnamefont
  {Verresen}}, \bibinfo {author} {\bibfnamefont {R.}~\bibnamefont {Moessner}},\
  and\ \bibinfo {author} {\bibfnamefont {F.}~\bibnamefont {Pollmann}},\
  }\bibfield  {title} {\bibinfo {title} {One-dimensional symmetry protected
  topological phases and their transitions},\ }\href
  {https://doi.org/10.1103/PhysRevB.96.165124} {\bibfield  {journal} {\bibinfo
  {journal} {Phys. Rev. B}\ }\textbf {\bibinfo {volume} {96}},\ \bibinfo
  {pages} {165124} (\bibinfo {year} {2017})}\BibitemShut {NoStop}%
\bibitem [{\citenamefont {Iraola}\ \emph {et~al.}(2021)\citenamefont {Iraola},
  \citenamefont {Heinsdorf}, \citenamefont {Tiwari}, \citenamefont {Lessnich},
  \citenamefont {Mertz}, \citenamefont {Ferrari}, \citenamefont {Fischer},
  \citenamefont {Winter}, \citenamefont {Pollmann}, \citenamefont {Neupert},
  \citenamefont {Valent\'{\i}},\ and\ \citenamefont {Vergniory}}]{Iraola2021}%
  \BibitemOpen
  \bibfield  {author} {\bibinfo {author} {\bibfnamefont {M.}~\bibnamefont
  {Iraola}}, \bibinfo {author} {\bibfnamefont {N.}~\bibnamefont {Heinsdorf}},
  \bibinfo {author} {\bibfnamefont {A.}~\bibnamefont {Tiwari}}, \bibinfo
  {author} {\bibfnamefont {D.}~\bibnamefont {Lessnich}}, \bibinfo {author}
  {\bibfnamefont {T.}~\bibnamefont {Mertz}}, \bibinfo {author} {\bibfnamefont
  {F.}~\bibnamefont {Ferrari}}, \bibinfo {author} {\bibfnamefont {M.~H.}\
  \bibnamefont {Fischer}}, \bibinfo {author} {\bibfnamefont {S.~M.}\
  \bibnamefont {Winter}}, \bibinfo {author} {\bibfnamefont {F.}~\bibnamefont
  {Pollmann}}, \bibinfo {author} {\bibfnamefont {T.}~\bibnamefont {Neupert}},
  \bibinfo {author} {\bibfnamefont {R.}~\bibnamefont {Valent\'{\i}}},\ and\
  \bibinfo {author} {\bibfnamefont {M.~G.}\ \bibnamefont {Vergniory}},\
  }\bibfield  {title} {\bibinfo {title} {Towards a topological quantum
  chemistry description of correlated systems: The case of the hubbard diamond
  chain},\ }\href {https://doi.org/10.1103/PhysRevB.104.195125} {\bibfield
  {journal} {\bibinfo  {journal} {Phys. Rev. B}\ }\textbf {\bibinfo {volume}
  {104}},\ \bibinfo {pages} {195125} (\bibinfo {year} {2021})}\BibitemShut
  {NoStop}%
\bibitem [{\citenamefont {Gurarie}(2011)}]{Gurarie2011}%
  \BibitemOpen
  \bibfield  {author} {\bibinfo {author} {\bibfnamefont {V.}~\bibnamefont
  {Gurarie}},\ }\bibfield  {title} {\bibinfo {title} {{Single-particle Green's
  functions and interacting topological insulators}},\ }\href
  {https://doi.org/10.1103/PhysRevB.83.085426} {\bibfield  {journal} {\bibinfo
  {journal} {Phys. Rev. B}\ }\textbf {\bibinfo {volume} {83}},\ \bibinfo
  {pages} {085426} (\bibinfo {year} {2011})}\BibitemShut {NoStop}%
\bibitem [{\citenamefont {Wang}\ \emph {et~al.}(2012)\citenamefont {Wang},
  \citenamefont {Qi},\ and\ \citenamefont {Zhang}}]{Wang2012}%
  \BibitemOpen
  \bibfield  {author} {\bibinfo {author} {\bibfnamefont {Z.}~\bibnamefont
  {Wang}}, \bibinfo {author} {\bibfnamefont {X.-L.}\ \bibnamefont {Qi}},\ and\
  \bibinfo {author} {\bibfnamefont {S.-C.}\ \bibnamefont {Zhang}},\ }\bibfield
  {title} {\bibinfo {title} {{Topological invariants for interacting
  topological insulators with inversion symmetry}},\ }\href
  {https://doi.org/10.1103/PhysRevB.85.165126} {\bibfield  {journal} {\bibinfo
  {journal} {Phys. Rev. B}\ }\textbf {\bibinfo {volume} {85}},\ \bibinfo
  {pages} {165126} (\bibinfo {year} {2012})}\BibitemShut {NoStop}%
\bibitem [{\citenamefont {Wang}\ and\ \citenamefont {Yan}(2013)}]{Wang2013}%
  \BibitemOpen
  \bibfield  {author} {\bibinfo {author} {\bibfnamefont {Z.}~\bibnamefont
  {Wang}}\ and\ \bibinfo {author} {\bibfnamefont {B.}~\bibnamefont {Yan}},\
  }\bibfield  {title} {\bibinfo {title} {{Topological Hamiltonian as an exact
  tool for topological invariants}},\ }\href
  {https://doi.org/10.1088/0953-8984/25/15/155601} {\bibfield  {journal}
  {\bibinfo  {journal} {Journal of Physics: Condensed Matter}\ }\textbf
  {\bibinfo {volume} {25}},\ \bibinfo {pages} {155601} (\bibinfo {year}
  {2013})}\BibitemShut {NoStop}%
\bibitem [{\citenamefont {Lessnich}\ \emph {et~al.}(2021)\citenamefont
  {Lessnich}, \citenamefont {Winter}, \citenamefont {Iraola}, \citenamefont
  {Vergniory},\ and\ \citenamefont {Valent\'{\i}}}]{Lessnich2021}%
  \BibitemOpen
  \bibfield  {author} {\bibinfo {author} {\bibfnamefont {D.}~\bibnamefont
  {Lessnich}}, \bibinfo {author} {\bibfnamefont {S.~M.}\ \bibnamefont
  {Winter}}, \bibinfo {author} {\bibfnamefont {M.}~\bibnamefont {Iraola}},
  \bibinfo {author} {\bibfnamefont {M.~G.}\ \bibnamefont {Vergniory}},\ and\
  \bibinfo {author} {\bibfnamefont {R.}~\bibnamefont {Valent\'{\i}}},\
  }\bibfield  {title} {\bibinfo {title} {{Elementary band representations for
  the single-particle {G}reen’s function of interacting topological
  insulators}},\ }\href {https://doi.org/10.1103/physrevb.104.085116}
  {\bibfield  {journal} {\bibinfo  {journal} {Physical Review B}\ }\textbf
  {\bibinfo {volume} {104}},\ \bibinfo {pages} {085116} (\bibinfo {year}
  {2021})}\BibitemShut {NoStop}%
\bibitem [{\citenamefont {Benalcazar}\ \emph {et~al.}(2019)\citenamefont
  {Benalcazar}, \citenamefont {Li},\ and\ \citenamefont
  {Hughes}}]{Benalcazar2019}%
  \BibitemOpen
  \bibfield  {author} {\bibinfo {author} {\bibfnamefont {W.~A.}\ \bibnamefont
  {Benalcazar}}, \bibinfo {author} {\bibfnamefont {T.}~\bibnamefont {Li}},\
  and\ \bibinfo {author} {\bibfnamefont {T.~L.}\ \bibnamefont {Hughes}},\
  }\bibfield  {title} {\bibinfo {title} {{Quantization of fractional corner
  charge in ${C}_{n}$-symmetric higher-order topological crystalline
  insulators}},\ }\href {https://doi.org/10.1103/PhysRevB.99.245151} {\bibfield
   {journal} {\bibinfo  {journal} {Phys. Rev. B}\ }\textbf {\bibinfo {volume}
  {99}},\ \bibinfo {pages} {245151} (\bibinfo {year} {2019})}\BibitemShut
  {NoStop}%
\bibitem [{\citenamefont {Schindler}\ \emph {et~al.}(2019)\citenamefont
  {Schindler}, \citenamefont {Brzezi\ifmmode~\acute{n}\else \'{n}\fi{}ska},
  \citenamefont {Benalcazar}, \citenamefont {Iraola}, \citenamefont {Bouhon},
  \citenamefont {Tsirkin}, \citenamefont {Vergniory},\ and\ \citenamefont
  {Neupert}}]{Schindler2019}%
  \BibitemOpen
  \bibfield  {author} {\bibinfo {author} {\bibfnamefont {F.}~\bibnamefont
  {Schindler}}, \bibinfo {author} {\bibfnamefont {M.}~\bibnamefont
  {Brzezi\ifmmode~\acute{n}\else \'{n}\fi{}ska}}, \bibinfo {author}
  {\bibfnamefont {W.~A.}\ \bibnamefont {Benalcazar}}, \bibinfo {author}
  {\bibfnamefont {M.}~\bibnamefont {Iraola}}, \bibinfo {author} {\bibfnamefont
  {A.}~\bibnamefont {Bouhon}}, \bibinfo {author} {\bibfnamefont {S.~S.}\
  \bibnamefont {Tsirkin}}, \bibinfo {author} {\bibfnamefont {M.~G.}\
  \bibnamefont {Vergniory}},\ and\ \bibinfo {author} {\bibfnamefont
  {T.}~\bibnamefont {Neupert}},\ }\bibfield  {title} {\bibinfo {title}
  {Fractional corner charges in spin-orbit coupled crystals},\ }\href
  {https://doi.org/10.1103/PhysRevResearch.1.033074} {\bibfield  {journal}
  {\bibinfo  {journal} {Phys. Rev. Res.}\ }\textbf {\bibinfo {volume} {1}},\
  \bibinfo {pages} {033074} (\bibinfo {year} {2019})}\BibitemShut {NoStop}%
\bibitem [{\citenamefont {Fang}\ and\ \citenamefont {Cano}(2021)}]{Fang2021}%
  \BibitemOpen
  \bibfield  {author} {\bibinfo {author} {\bibfnamefont {Y.}~\bibnamefont
  {Fang}}\ and\ \bibinfo {author} {\bibfnamefont {J.}~\bibnamefont {Cano}},\
  }\bibfield  {title} {\bibinfo {title} {{Filling anomaly for general two- and
  three-dimensional ${C}_{4}$ symmetric lattices}},\ }\href
  {https://doi.org/10.1103/PhysRevB.103.165109} {\bibfield  {journal} {\bibinfo
   {journal} {Phys. Rev. B}\ }\textbf {\bibinfo {volume} {103}},\ \bibinfo
  {pages} {165109} (\bibinfo {year} {2021})}\BibitemShut {NoStop}%
\bibitem [{\citenamefont {P\'erez-Garc\'{\i}a}\ \emph
  {et~al.}(2008)\citenamefont {P\'erez-Garc\'{\i}a}, \citenamefont {Wolf},
  \citenamefont {Sanz}, \citenamefont {Verstraete},\ and\ \citenamefont
  {Cirac}}]{PerezGarcia2008}%
  \BibitemOpen
  \bibfield  {author} {\bibinfo {author} {\bibfnamefont {D.}~\bibnamefont
  {P\'erez-Garc\'{\i}a}}, \bibinfo {author} {\bibfnamefont {M.~M.}\
  \bibnamefont {Wolf}}, \bibinfo {author} {\bibfnamefont {M.}~\bibnamefont
  {Sanz}}, \bibinfo {author} {\bibfnamefont {F.}~\bibnamefont {Verstraete}},\
  and\ \bibinfo {author} {\bibfnamefont {J.~I.}\ \bibnamefont {Cirac}},\
  }\bibfield  {title} {\bibinfo {title} {String order and symmetries in quantum
  spin lattices},\ }\href {https://doi.org/10.1103/PhysRevLett.100.167202}
  {\bibfield  {journal} {\bibinfo  {journal} {Phys. Rev. Lett.}\ }\textbf
  {\bibinfo {volume} {100}},\ \bibinfo {pages} {167202} (\bibinfo {year}
  {2008})}\BibitemShut {NoStop}%
\bibitem [{\citenamefont {Pollmann}\ \emph {et~al.}(2012)\citenamefont
  {Pollmann}, \citenamefont {Berg}, \citenamefont {Turner},\ and\ \citenamefont
  {Oshikawa}}]{Pollmann2012a}%
  \BibitemOpen
  \bibfield  {author} {\bibinfo {author} {\bibfnamefont {F.}~\bibnamefont
  {Pollmann}}, \bibinfo {author} {\bibfnamefont {E.}~\bibnamefont {Berg}},
  \bibinfo {author} {\bibfnamefont {A.~M.}\ \bibnamefont {Turner}},\ and\
  \bibinfo {author} {\bibfnamefont {M.}~\bibnamefont {Oshikawa}},\ }\bibfield
  {title} {\bibinfo {title} {Symmetry protection of topological phases in
  one-dimensional quantum spin systems},\ }\href
  {https://doi.org/10.1103/PhysRevB.85.075125} {\bibfield  {journal} {\bibinfo
  {journal} {Phys. Rev. B}\ }\textbf {\bibinfo {volume} {85}},\ \bibinfo
  {pages} {075125} (\bibinfo {year} {2012})}\BibitemShut {NoStop}%
\bibitem [{\citenamefont {Pollmann}\ and\ \citenamefont
  {Turner}(2012)}]{Pollmann2012b}%
  \BibitemOpen
  \bibfield  {author} {\bibinfo {author} {\bibfnamefont {F.}~\bibnamefont
  {Pollmann}}\ and\ \bibinfo {author} {\bibfnamefont {A.~M.}\ \bibnamefont
  {Turner}},\ }\bibfield  {title} {\bibinfo {title} {Detection of
  symmetry-protected topological phases in one dimension},\ }\href
  {https://doi.org/10.1103/PhysRevB.86.125441} {\bibfield  {journal} {\bibinfo
  {journal} {Phys. Rev. B}\ }\textbf {\bibinfo {volume} {86}},\ \bibinfo
  {pages} {125441} (\bibinfo {year} {2012})}\BibitemShut {NoStop}%
\bibitem [{\citenamefont {Shapourian}\ \emph {et~al.}(2017)\citenamefont
  {Shapourian}, \citenamefont {Shiozaki},\ and\ \citenamefont
  {Ryu}}]{Shapourian2017}%
  \BibitemOpen
  \bibfield  {author} {\bibinfo {author} {\bibfnamefont {H.}~\bibnamefont
  {Shapourian}}, \bibinfo {author} {\bibfnamefont {K.}~\bibnamefont
  {Shiozaki}},\ and\ \bibinfo {author} {\bibfnamefont {S.}~\bibnamefont
  {Ryu}},\ }\bibfield  {title} {\bibinfo {title} {Many-body topological
  invariants for fermionic symmetry-protected topological phases},\ }\href
  {https://doi.org/10.1103/PhysRevLett.118.216402} {\bibfield  {journal}
  {\bibinfo  {journal} {Phys. Rev. Lett.}\ }\textbf {\bibinfo {volume} {118}},\
  \bibinfo {pages} {216402} (\bibinfo {year} {2017})}\BibitemShut {NoStop}%
\bibitem [{\citenamefont {Shiozaki}\ \emph {et~al.}(2018)\citenamefont
  {Shiozaki}, \citenamefont {Shapourian}, \citenamefont {Gomi},\ and\
  \citenamefont {Ryu}}]{Shiozaki2018}%
  \BibitemOpen
  \bibfield  {author} {\bibinfo {author} {\bibfnamefont {K.}~\bibnamefont
  {Shiozaki}}, \bibinfo {author} {\bibfnamefont {H.}~\bibnamefont
  {Shapourian}}, \bibinfo {author} {\bibfnamefont {K.}~\bibnamefont {Gomi}},\
  and\ \bibinfo {author} {\bibfnamefont {S.}~\bibnamefont {Ryu}},\ }\bibfield
  {title} {\bibinfo {title}
  {\href{https://link.aps.org/doi/10.1103/PhysRevB.98.035151}{Many-body
  topological invariants for fermionic short-range entangled topological phases
  protected by antiunitary symmetries}},\ }\href
  {https://doi.org/10.1103/PhysRevB.98.035151} {\bibfield  {journal} {\bibinfo
  {journal} {Phys. Rev. B}\ }\textbf {\bibinfo {volume} {98}},\ \bibinfo
  {pages} {035151} (\bibinfo {year} {2018})}\BibitemShut {NoStop}%
\bibitem [{\citenamefont {Zhang}\ \emph {et~al.}(2023)\citenamefont {Zhang},
  \citenamefont {Manjunath}, \citenamefont {Kobayashi},\ and\ \citenamefont
  {Barkeshli}}]{zhang2023complete}%
  \BibitemOpen
  \bibfield  {author} {\bibinfo {author} {\bibfnamefont {Y.}~\bibnamefont
  {Zhang}}, \bibinfo {author} {\bibfnamefont {N.}~\bibnamefont {Manjunath}},
  \bibinfo {author} {\bibfnamefont {R.}~\bibnamefont {Kobayashi}},\ and\
  \bibinfo {author} {\bibfnamefont {M.}~\bibnamefont {Barkeshli}},\ }\href@noop
  {} {\bibinfo {title} {{Complete crystalline topological invariants from
  partial rotations in (2+1)D invertible fermionic states and Hofstadter's
  butterfly}}} (\bibinfo {year} {2023}),\ \Eprint
  {https://arxiv.org/abs/2303.16919} {arXiv:2303.16919 [cond-mat.str-el]}
  \BibitemShut {NoStop}%
\bibitem [{\citenamefont {Lieb}\ \emph {et~al.}(1961)\citenamefont {Lieb},
  \citenamefont {Schultz},\ and\ \citenamefont {Mattis}}]{Lieb1961}%
  \BibitemOpen
  \bibfield  {author} {\bibinfo {author} {\bibfnamefont {E.}~\bibnamefont
  {Lieb}}, \bibinfo {author} {\bibfnamefont {T.}~\bibnamefont {Schultz}},\ and\
  \bibinfo {author} {\bibfnamefont {D.}~\bibnamefont {Mattis}},\ }\bibfield
  {title} {\bibinfo {title} {Two soluble models of an antiferromagnetic
  chain},\ }\href@noop {} {\bibfield  {journal} {\bibinfo  {journal} {Annals of
  Physics}\ }\textbf {\bibinfo {volume} {16}},\ \bibinfo {pages} {407}
  (\bibinfo {year} {1961})}\BibitemShut {NoStop}%
\bibitem [{\citenamefont {Oshikawa}(2000)}]{Oshikawa2000}%
  \BibitemOpen
  \bibfield  {author} {\bibinfo {author} {\bibfnamefont {M.}~\bibnamefont
  {Oshikawa}},\ }\bibfield  {title} {\bibinfo {title} {Commensurability,
  excitation gap, and topology in quantum many-particle systems on a periodic
  lattice},\ }\href {https://doi.org/10.1103/PhysRevLett.84.1535} {\bibfield
  {journal} {\bibinfo  {journal} {Phys. Rev. Lett.}\ }\textbf {\bibinfo
  {volume} {84}},\ \bibinfo {pages} {1535} (\bibinfo {year}
  {2000})}\BibitemShut {NoStop}%
\bibitem [{\citenamefont {Hastings}(2004)}]{Hastings2004}%
  \BibitemOpen
  \bibfield  {author} {\bibinfo {author} {\bibfnamefont {M.~B.}\ \bibnamefont
  {Hastings}},\ }\bibfield  {title} {\bibinfo {title} {Lieb-schultz-mattis in
  higher dimensions},\ }\href {https://doi.org/10.1103/PhysRevB.69.104431}
  {\bibfield  {journal} {\bibinfo  {journal} {Phys. Rev. B}\ }\textbf {\bibinfo
  {volume} {69}},\ \bibinfo {pages} {104431} (\bibinfo {year}
  {2004})}\BibitemShut {NoStop}%
\bibitem [{\citenamefont {Hastings}(2005)}]{Hastings2005}%
  \BibitemOpen
  \bibfield  {author} {\bibinfo {author} {\bibfnamefont {M.~B.}\ \bibnamefont
  {Hastings}},\ }\bibfield  {title} {\bibinfo {title} {Sufficient conditions
  for topological order in insulators},\ }\href
  {https://doi.org/10.1209/epl/i2005-10046-x} {\bibfield  {journal} {\bibinfo
  {journal} {Europhysics Letters ({EPL})}\ }\textbf {\bibinfo {volume} {70}},\
  \bibinfo {pages} {824} (\bibinfo {year} {2005})}\BibitemShut {NoStop}%
\bibitem [{\citenamefont {Cheng}\ \emph {et~al.}(2016)\citenamefont {Cheng},
  \citenamefont {Zaletel}, \citenamefont {Barkeshli}, \citenamefont
  {Vishwanath},\ and\ \citenamefont {Bonderson}}]{Cheng2016}%
  \BibitemOpen
  \bibfield  {author} {\bibinfo {author} {\bibfnamefont {M.}~\bibnamefont
  {Cheng}}, \bibinfo {author} {\bibfnamefont {M.}~\bibnamefont {Zaletel}},
  \bibinfo {author} {\bibfnamefont {M.}~\bibnamefont {Barkeshli}}, \bibinfo
  {author} {\bibfnamefont {A.}~\bibnamefont {Vishwanath}},\ and\ \bibinfo
  {author} {\bibfnamefont {P.}~\bibnamefont {Bonderson}},\ }\bibfield  {title}
  {\bibinfo {title} {Translational symmetry and microscopic constraints on
  symmetry-enriched topological phases: A view from the surface},\ }\href
  {https://doi.org/10.1103/PhysRevX.6.041068} {\bibfield  {journal} {\bibinfo
  {journal} {Phys. Rev. X}\ }\textbf {\bibinfo {volume} {6}},\ \bibinfo {pages}
  {041068} (\bibinfo {year} {2016})}\BibitemShut {NoStop}%
\bibitem [{\citenamefont {Cho}\ \emph {et~al.}(2017)\citenamefont {Cho},
  \citenamefont {Hsieh},\ and\ \citenamefont {Ryu}}]{Cho2017}%
  \BibitemOpen
  \bibfield  {author} {\bibinfo {author} {\bibfnamefont {G.~Y.}\ \bibnamefont
  {Cho}}, \bibinfo {author} {\bibfnamefont {C.-T.}\ \bibnamefont {Hsieh}},\
  and\ \bibinfo {author} {\bibfnamefont {S.}~\bibnamefont {Ryu}},\ }\bibfield
  {title} {\bibinfo {title} {Anomaly manifestation of lieb-schultz-mattis
  theorem and topological phases},\ }\href
  {https://doi.org/10.1103/PhysRevB.96.195105} {\bibfield  {journal} {\bibinfo
  {journal} {Phys. Rev. B}\ }\textbf {\bibinfo {volume} {96}},\ \bibinfo
  {pages} {195105} (\bibinfo {year} {2017})}\BibitemShut {NoStop}%
\bibitem [{\citenamefont {Tasaki}(2018)}]{Tasaki2018}%
  \BibitemOpen
  \bibfield  {author} {\bibinfo {author} {\bibfnamefont {H.}~\bibnamefont
  {Tasaki}},\ }\bibfield  {title} {\bibinfo {title} {Lieb--schultz--mattis
  theorem with a local twist for general one-dimensional quantum systems},\
  }\href@noop {} {\bibfield  {journal} {\bibinfo  {journal} {Journal of
  Statistical Physics}\ }\textbf {\bibinfo {volume} {170}},\ \bibinfo {pages}
  {653} (\bibinfo {year} {2018})}\BibitemShut {NoStop}%
\bibitem [{\citenamefont {Jian}\ \emph {et~al.}(2018)\citenamefont {Jian},
  \citenamefont {Bi},\ and\ \citenamefont {Xu}}]{Jian2018}%
  \BibitemOpen
  \bibfield  {author} {\bibinfo {author} {\bibfnamefont {C.-M.}\ \bibnamefont
  {Jian}}, \bibinfo {author} {\bibfnamefont {Z.}~\bibnamefont {Bi}},\ and\
  \bibinfo {author} {\bibfnamefont {C.}~\bibnamefont {Xu}},\ }\bibfield
  {title} {\bibinfo {title} {Lieb-schultz-mattis theorem and its
  generalizations from the perspective of the symmetry-protected topological
  phase},\ }\href {https://doi.org/10.1103/PhysRevB.97.054412} {\bibfield
  {journal} {\bibinfo  {journal} {Phys. Rev. B}\ }\textbf {\bibinfo {volume}
  {97}},\ \bibinfo {pages} {054412} (\bibinfo {year} {2018})}\BibitemShut
  {NoStop}%
\bibitem [{\citenamefont {Ogata}\ and\ \citenamefont
  {Tasaki}(2019)}]{Ogata2019}%
  \BibitemOpen
  \bibfield  {author} {\bibinfo {author} {\bibfnamefont {Y.}~\bibnamefont
  {Ogata}}\ and\ \bibinfo {author} {\bibfnamefont {H.}~\bibnamefont {Tasaki}},\
  }\bibfield  {title} {\bibinfo {title} {{Lieb--Schultz--Mattis type theorems
  for quantum spin chains without continuous symmetry}},\ }\href
  {https://doi.org/10.1007/s00220-019-03343-5} {\bibfield  {journal} {\bibinfo
  {journal} {Communications in Mathematical Physics}\ }\textbf {\bibinfo
  {volume} {372}},\ \bibinfo {pages} {951} (\bibinfo {year}
  {2019})}\BibitemShut {NoStop}%
\bibitem [{\citenamefont {Ogata}\ \emph {et~al.}(2021)\citenamefont {Ogata},
  \citenamefont {Tachikawa},\ and\ \citenamefont {Tasaki}}]{Ogata2021}%
  \BibitemOpen
  \bibfield  {author} {\bibinfo {author} {\bibfnamefont {Y.}~\bibnamefont
  {Ogata}}, \bibinfo {author} {\bibfnamefont {Y.}~\bibnamefont {Tachikawa}},\
  and\ \bibinfo {author} {\bibfnamefont {H.}~\bibnamefont {Tasaki}},\
  }\bibfield  {title} {\bibinfo {title} {{General Lieb--Schultz--Mattis Type
  Theorems for Quantum Spin Chains}},\ }\href
  {https://doi.org/10.1007/s00220-021-04116-9} {\bibfield  {journal} {\bibinfo
  {journal} {Communications in Mathematical Physics}\ }\textbf {\bibinfo
  {volume} {385}},\ \bibinfo {pages} {79} (\bibinfo {year} {2021})}\BibitemShut
  {NoStop}%
\bibitem [{\citenamefont {Yao}\ and\ \citenamefont {Oshikawa}(2021)}]{Yao2021}%
  \BibitemOpen
  \bibfield  {author} {\bibinfo {author} {\bibfnamefont {Y.}~\bibnamefont
  {Yao}}\ and\ \bibinfo {author} {\bibfnamefont {M.}~\bibnamefont {Oshikawa}},\
  }\bibfield  {title} {\bibinfo {title} {{Twisted Boundary Condition and
  Lieb-Schultz-Mattis Ingappability for Discrete Symmetries}},\ }\href
  {https://doi.org/10.1103/PhysRevLett.126.217201} {\bibfield  {journal}
  {\bibinfo  {journal} {Phys. Rev. Lett.}\ }\textbf {\bibinfo {volume} {126}},\
  \bibinfo {pages} {217201} (\bibinfo {year} {2021})}\BibitemShut {NoStop}%
\bibitem [{\citenamefont {Aksoy}\ \emph
  {et~al.}(2021{\natexlab{b}})\citenamefont {Aksoy}, \citenamefont {Tiwari},\
  and\ \citenamefont {Mudry}}]{Aksoy2021b}%
  \BibitemOpen
  \bibfield  {author} {\bibinfo {author} {\bibfnamefont {{\"O}.~M.}\
  \bibnamefont {Aksoy}}, \bibinfo {author} {\bibfnamefont {A.}~\bibnamefont
  {Tiwari}},\ and\ \bibinfo {author} {\bibfnamefont {C.}~\bibnamefont
  {Mudry}},\ }\bibfield  {title} {\bibinfo {title} {{Lieb-Schultz-Mattis type
  theorems for Majorana models with discrete symmetries}},\ }\href
  {https://doi.org/10.1103/PhysRevB.104.075146} {\bibfield  {journal} {\bibinfo
   {journal} {Phys. Rev. B}\ }\textbf {\bibinfo {volume} {104}},\ \bibinfo
  {pages} {075146} (\bibinfo {year} {2021}{\natexlab{b}})}\BibitemShut
  {NoStop}%
\bibitem [{\citenamefont {Po}\ \emph {et~al.}(2018)\citenamefont {Po},
  \citenamefont {Watanabe},\ and\ \citenamefont {Vishwanath}}]{Po2018}%
  \BibitemOpen
  \bibfield  {author} {\bibinfo {author} {\bibfnamefont {H.~C.}\ \bibnamefont
  {Po}}, \bibinfo {author} {\bibfnamefont {H.}~\bibnamefont {Watanabe}},\ and\
  \bibinfo {author} {\bibfnamefont {A.}~\bibnamefont {Vishwanath}},\ }\bibfield
   {title} {\bibinfo {title} {{Fragile Topology and Wannier Obstructions}},\
  }\href {https://doi.org/10.1103/PhysRevLett.121.126402} {\bibfield  {journal}
  {\bibinfo  {journal} {Phys. Rev. Lett.}\ }\textbf {\bibinfo {volume} {121}},\
  \bibinfo {pages} {126402} (\bibinfo {year} {2018})}\BibitemShut {NoStop}%
\bibitem [{\citenamefont {Bradlyn}\ \emph {et~al.}(2019)\citenamefont
  {Bradlyn}, \citenamefont {Wang}, \citenamefont {Cano},\ and\ \citenamefont
  {Bernevig}}]{Bradlyn2019}%
  \BibitemOpen
  \bibfield  {author} {\bibinfo {author} {\bibfnamefont {B.}~\bibnamefont
  {Bradlyn}}, \bibinfo {author} {\bibfnamefont {Z.}~\bibnamefont {Wang}},
  \bibinfo {author} {\bibfnamefont {J.}~\bibnamefont {Cano}},\ and\ \bibinfo
  {author} {\bibfnamefont {B.~A.}\ \bibnamefont {Bernevig}},\ }\bibfield
  {title} {\bibinfo {title} {{Disconnected elementary band representations,
  fragile topology, and Wilson loops as topological indices: An example on the
  triangular lattice}},\ }\href {https://doi.org/10.1103/PhysRevB.99.045140}
  {\bibfield  {journal} {\bibinfo  {journal} {Phys. Rev. B}\ }\textbf {\bibinfo
  {volume} {99}},\ \bibinfo {pages} {045140} (\bibinfo {year}
  {2019})}\BibitemShut {NoStop}%
\bibitem [{\citenamefont {Bouhon}\ \emph {et~al.}(2019)\citenamefont {Bouhon},
  \citenamefont {Black-Schaffer},\ and\ \citenamefont {Slager}}]{Bouhon2019}%
  \BibitemOpen
  \bibfield  {author} {\bibinfo {author} {\bibfnamefont {A.}~\bibnamefont
  {Bouhon}}, \bibinfo {author} {\bibfnamefont {A.~M.}\ \bibnamefont
  {Black-Schaffer}},\ and\ \bibinfo {author} {\bibfnamefont {R.-J.}\
  \bibnamefont {Slager}},\ }\bibfield  {title} {\bibinfo {title} {Wilson loop
  approach to fragile topology of split elementary band representations and
  topological crystalline insulators with time-reversal symmetry},\ }\href
  {https://doi.org/10.1103/PhysRevB.100.195135} {\bibfield  {journal} {\bibinfo
   {journal} {Phys. Rev. B}\ }\textbf {\bibinfo {volume} {100}},\ \bibinfo
  {pages} {195135} (\bibinfo {year} {2019})}\BibitemShut {NoStop}%
\bibitem [{\citenamefont {Else}\ \emph {et~al.}(2019)\citenamefont {Else},
  \citenamefont {Po},\ and\ \citenamefont {Watanabe}}]{Else2019b}%
  \BibitemOpen
  \bibfield  {author} {\bibinfo {author} {\bibfnamefont {D.~V.}\ \bibnamefont
  {Else}}, \bibinfo {author} {\bibfnamefont {H.~C.}\ \bibnamefont {Po}},\ and\
  \bibinfo {author} {\bibfnamefont {H.}~\bibnamefont {Watanabe}},\ }\bibfield
  {title} {\bibinfo {title} {{Fragile topological phases in interacting
  systems}},\ }\href {https://doi.org/10.1103/PhysRevB.99.125122} {\bibfield
  {journal} {\bibinfo  {journal} {Phys. Rev. B}\ }\textbf {\bibinfo {volume}
  {99}},\ \bibinfo {pages} {125122} (\bibinfo {year} {2019})}\BibitemShut
  {NoStop}%
\bibitem [{\citenamefont {Aroyo}\ \emph
  {et~al.}(2006{\natexlab{a}})\citenamefont {Aroyo}, \citenamefont
  {Perez-Mato}, \citenamefont {Capillas}, \citenamefont {Kroumova},
  \citenamefont {Ivantchev}, \citenamefont {Madariaga}, \citenamefont {Kirov},\
  and\ \citenamefont {Wondratschek}}]{Aroyo2006}%
  \BibitemOpen
  \bibfield  {author} {\bibinfo {author} {\bibfnamefont {M.~I.}\ \bibnamefont
  {Aroyo}}, \bibinfo {author} {\bibfnamefont {J.~M.}\ \bibnamefont
  {Perez-Mato}}, \bibinfo {author} {\bibfnamefont {C.}~\bibnamefont
  {Capillas}}, \bibinfo {author} {\bibfnamefont {E.}~\bibnamefont {Kroumova}},
  \bibinfo {author} {\bibfnamefont {S.}~\bibnamefont {Ivantchev}}, \bibinfo
  {author} {\bibfnamefont {G.}~\bibnamefont {Madariaga}}, \bibinfo {author}
  {\bibfnamefont {A.}~\bibnamefont {Kirov}},\ and\ \bibinfo {author}
  {\bibfnamefont {H.}~\bibnamefont {Wondratschek}},\ }\bibfield  {title}
  {\bibinfo {title} {{Bilbao Crystallographic Server: I. Databases and
  crystallographic computing programs}},\ }\href
  {https://doi.org/doi:10.1524/zkri.2006.221.1.15} {\bibfield  {journal}
  {\bibinfo  {journal} {Zeitschrift für Kristallographie - Crystalline
  Materials}\ }\textbf {\bibinfo {volume} {221}},\ \bibinfo {pages} {15}
  (\bibinfo {year} {2006}{\natexlab{a}})}\BibitemShut {NoStop}%
\bibitem [{\citenamefont {Aroyo}\ \emph
  {et~al.}(2006{\natexlab{b}})\citenamefont {Aroyo}, \citenamefont {Kirov},
  \citenamefont {Capillas}, \citenamefont {Perez-Mato},\ and\ \citenamefont
  {Wondratschek}}]{Aroyo2006b}%
  \BibitemOpen
  \bibfield  {author} {\bibinfo {author} {\bibfnamefont {M.~I.}\ \bibnamefont
  {Aroyo}}, \bibinfo {author} {\bibfnamefont {A.}~\bibnamefont {Kirov}},
  \bibinfo {author} {\bibfnamefont {C.}~\bibnamefont {Capillas}}, \bibinfo
  {author} {\bibfnamefont {J.~M.}\ \bibnamefont {Perez-Mato}},\ and\ \bibinfo
  {author} {\bibfnamefont {H.}~\bibnamefont {Wondratschek}},\ }\bibfield
  {title} {\bibinfo {title} {{Bilbao Crystallographic Server. II.
  Representations of crystallographic point groups and space groups}},\ }\href
  {https://doi.org/10.1107/S0108767305040286} {\bibfield  {journal} {\bibinfo
  {journal} {Acta Crystallographica Section A}\ }\textbf {\bibinfo {volume}
  {62}},\ \bibinfo {pages} {115} (\bibinfo {year}
  {2006}{\natexlab{b}})}\BibitemShut {NoStop}%
\bibitem [{\citenamefont {Aroyo}\ \emph {et~al.}(2011)\citenamefont {Aroyo},
  \citenamefont {Perez-Mato}, \citenamefont {Orobengoa}, \citenamefont {Tasci},
  \citenamefont {De~La~Flor},\ and\ \citenamefont {Kirov}}]{Aroyo2011}%
  \BibitemOpen
  \bibfield  {author} {\bibinfo {author} {\bibfnamefont {M.}~\bibnamefont
  {Aroyo}}, \bibinfo {author} {\bibfnamefont {J.}~\bibnamefont {Perez-Mato}},
  \bibinfo {author} {\bibfnamefont {D.}~\bibnamefont {Orobengoa}}, \bibinfo
  {author} {\bibfnamefont {E.}~\bibnamefont {Tasci}}, \bibinfo {author}
  {\bibfnamefont {G.}~\bibnamefont {De~La~Flor}},\ and\ \bibinfo {author}
  {\bibfnamefont {A.}~\bibnamefont {Kirov}},\ }\bibfield  {title} {\bibinfo
  {title} {{Crystallography online: Bilbao crystallographic server}},\ }\href
  {https://www.scopus.com/inward/record.uri?eid=2-s2.0-80955140447&partnerID=40&md5=488772b9e21d2636a3952f66ae80ae84}
  {\bibfield  {journal} {\bibinfo  {journal} {Bulgarian Chemical
  Communications}\ }\textbf {\bibinfo {volume} {43}},\ \bibinfo {pages} {183
  – 197} (\bibinfo {year} {2011})}\BibitemShut {NoStop}%
\end{thebibliography}%

\end{document}